\documentclass[12pt,letterpaper]{article}
\usepackage{epsfig}
\usepackage{amsmath}
\usepackage{amssymb}
\usepackage{axodraw}
\usepackage{feynmp}
\usepackage{color} 
\usepackage{booktabs} 
\usepackage{cite} 

\newcommand{\mrm}[1]{\mathrm{#1}}
\newcommand{\mbf}[1]{\mathbf{#1}}

\newcommand{\ttt}[1]{\texttt{#1}}

\newlength{\tmplen}
\newcommand{\clab}[1]{\tiny\settowidth{\tmplen}{\scriptsize#1}%
\colorbox{white}{\textcolor{white}{#1}}\hspace*{-1.27\tmplen}\scriptsize#1}

\def\lsim{\mathrel{\rlap{\lower4pt\hbox{\hskip1pt$\sim$}}
    \raise1pt\hbox{$<$}}}                
\def\gsim{\mathrel{\rlap{\lower4pt\hbox{\hskip1pt$\sim$}}
    \raise1pt\hbox{$>$}}}                
 
\newcommand{\alphas}{\alpha_{\mathrm{s}}}

\newcommand{\pT}{\ensuremath{p_{\perp}}}
\newcommand{\kT}{\ensuremath{k_{\perp}}}
\newcommand{\pTmin}{p_{\perp\mathrm{min}}}
\newcommand{\pTo}{p_{\perp 0}}

\newcommand{\GeV}{\ensuremath{\!\ \mathrm{GeV}}}
\newcommand{\TeV}{\ensuremath{\!\ \mathrm{TeV}}}
\newcommand{\rem}{\ensuremath{\mathrm{rem}}}
 
\renewcommand{\c}{\mathrm{c}}
\renewcommand{\d}{\mathrm{d}}
\newcommand{\e}{\mathrm{e}}

\newcommand{\g}{\mathrm{g}}

\newcommand{\J}{\mathrm{J}}
\newcommand{\hrm}{\mathrm{h}}
\newcommand{\n}{\mathrm{n}}
\newcommand{\p}{\mathrm{p}}
\newcommand{\q}{\mathrm{q}}
\newcommand{\s}{\mathrm{s}}

\renewcommand{\u}{\mathrm{u}}

\newcommand{\W}{\mathrm{W}}
\newcommand{\Z}{\mathrm{Z}}

\newcommand{\nbar}{\overline{\mathrm{n}}}
\newcommand{\pbar}{\overline{\mathrm{p}}}
\newcommand{\qbar}{\overline{\mathrm{q}}}

\newcommand{\qsea}{\ensuremath{\q_{\mrm{s}}}}
\newcommand{\qcmp}{\ensuremath{\q_{\mrm{c}}}}
\newcommand{\val}{\ensuremath{{\mrm{v}}}}
\newcommand{\sea}{\ensuremath{{\mrm{s}}}}
\newcommand{\cmp}{\ensuremath{{\mrm{c}}}}
 
\newenvironment{Itemize}{\begin{list}{$\bullet$}%
{\setlength{\topsep}{0.2mm}\setlength{\partopsep}{0.2mm}%
\setlength{\itemsep}{0.2mm}\setlength{\parsep}{0.2mm}}}%
{\end{list}}
\newcounter{enumct}

\newlength{\abstwidth}
\newlength{\captivewidth}
\newcommand{\captive}[1]{\rule{5mm}{0mm}%
\begin{minipage}{\captivewidth}%
\caption[small]{#1}\end{minipage}}
 
\begin{document}
\sloppy
 
\pagestyle{empty}
 
\begin{flushright}
LU TP 04--08\\
hep-ph/0402078\\
March 2004
\end{flushright}
 
\vspace{\fill}
 
\begin{center}
{\LARGE\bf Multiple Interactions}\\[3mm]
{\LARGE\bf and the Structure of Beam Remnants}\\[10mm]
{\Large T. Sj\"ostrand\footnote{torbjorn@thep.lu.se} and %
P. Z. Skands\footnote{peter.skands@thep.lu.se}} \\[3mm]
{\it Department of Theoretical Physics,}\\[1mm]
{\it Lund University,}\\[1mm]
{\it S\"olvegatan 14A,}\\[1mm]
{\it S-223 62 Lund, Sweden}
\end{center}
 
\vspace{\fill}
 
\begin{center}
{\bf Abstract}\\[2ex]
\begin{minipage}{\abstwidth}
Recent experimental data have established some of the 
basic features of multiple interactions in hadron--hadron collisions. 
The emphasis 
is therefore now shifting, to one of exploring more 
detailed aspects. Starting from a brief review of the
current situation, a next-generation model is developed, 
wherein a detailed account is given of correlated
flavour, colour, longitudinal and transverse momentum 
distributions, encompassing both the partons initiating
perturbative interactions and the partons left in the
beam remnants. Some of the main features are illustrated
for the Tevatron and the LHC.
\end{minipage}
\end{center}
 
\vspace{\fill}
 
\clearpage
\pagestyle{plain}
\setcounter{page}{1}
 
\section{Introduction}
 
The physics of high-energy hadron--hadron interactions has become 
a topic of increasing interest in recent years. With the Tevatron 
Run II well under way and with the startup of the LHC drawing closer, 
huge data samples are becoming available that will challenge our 
current understanding of this physics. From the point of view of QCD, 
many interesting questions remain to be answered, and we shall take up 
some of these in detail below. Moreover, for new physics searches and 
precision measurements, it is important that these questions can be 
given meaningful and trustworthy answers, since ever-present yet
poorly-understood aspects of QCD can have a significant impact.

Much of the complexity involved in describing these phenomena --- 
specifically the
underlying event and minimum-bias collisions --- derives from the 
composite nature of hadrons; we are dealing with objects which possess 
a rich internal structure that is not calculable from perturbation
theory. This, however, does \emph{not} imply that the physics of the 
underlying event as such has to be an inherently non-perturbative 
quagmire. 

\begin{figure}[t]
\begin{center}
\begin{picture}(150,110)(0,0)
\SetScale{0.7}
\SetWidth{2}
\ArrowLine(10,130)(50,130)
\ArrowLine(10,20)(50,20)
\SetWidth{2}
\ArrowLine(50,121)(100,90)
\ArrowLine(50,29)(100,60)
\Gluon(100,60)(100,90){5}{2}
\ArrowLine(100,90)(200,120)
\ArrowLine(100,60)(200,30)
\Gluon(50,127)(150,90){5}{10}
\Gluon(50,23)(150,60){5}{10}
\Gluon(150,60)(150,90){5}{2}
\Gluon(150,90)(200,100){5}{4}
\Gluon(150,60)(200,50){5}{4}
\ArrowLine(50,133)(200,133)
\ArrowLine(50,139)(200,139)
\ArrowLine(50,17)(200,17)
\ArrowLine(50,11)(200,11)
\GOval(50,130)(15,8)(0){0.5}
\GOval(50,20)(15,8)(0){0.5}
\end{picture}
\end{center}
\captive{Schematic illustration of an event with two
$2 \to 2$ perturbative interactions.
\label{fig:doublescat}}
\end{figure}
 
Viewing hadrons as `bunches' of incoming partons, it is
apparent that when two hadrons collide it is possible that several distinct
pairs of partons collide with each other, as depicted in
Fig.~\ref{fig:doublescat}. Thus 
multiple interactions (also known as multiple
scatterings) in hadronic collisions is a phenomenon which is a direct
consequence of the composite nature of hadrons and which \textit{must} exist,
at some level. In fact, by extending simple
perturbation theory to rather low $\pT$ values, though still some
distance above $\Lambda_{\mrm{QCD}}$, most inelastic events in high-energy
hadronic collisions are guaranteed to contain several \emph{perturbatively
calculable} interactions \cite{Zijl}. Furthermore, such
interactions --- even when soft --- can be highly important, 
causing non-trivial changes to the colour
topology of the colliding system as a whole, with potentially drastic
consequences for the particle multiplicity in the final state.

Nevertheless, traditionally the exploration of multiple interactions
has not attracted much interest. For studies concentrating on 
high-$\pT$ jets, perturbative QCD emission is a more important source of
multijets than separate multiple interactions. The underlying event, on the
other hand, has in this context often been viewed as a mess of soft QCD 
interactions, that cannot be described from first principles but is better
simply parametrized. 

However, such parametrizations, even while reasonably successful in 
describing the average underlying activity, are not sophisticated enough 
to adequately describe correlations and fluctuations. This relates for 
instance to jet profiles and jet pedestals, and to systematic as well as 
random shifts in jet energies. The lack of sophistication implies that,
even when tuned to describe a few distributions well, they could not
be trusted for extrapolations beyond the fit region. Hence, a sound 
understanding of multiple interactions is prerequisite for precision 
physics involving jets and/or the underlying event.

It is interesting to note that this can also
impact physics studies in areas well beyond the
conventional QCD ones. As an example, consider the search for a Higgs
particle in the $\hrm^0 \to \gamma\gamma$ channel at the LHC, 
where the Higgs mass resolution at high luminosity depends on picking the
correct vertex between maybe 30 different $\p\p$ events. If the
event that contained the Higgs is special, by typically containing
more charged particles (in general or in some regions of phase space),
we would want to use that information to clean up the signal
\cite{higgsvertex}. 

The crucial leap of imagination is to postulate that \textit{all} 
particle production in inelastic hadronic collisions derives from the
multiple-interactions mechanism. This is not to say that 
many nonperturbative and poorly known phenomena will not 
force their entrance on the stage, in going from the perturbative
interactions to the observable hadrons, but the starting point is 
perturbative. If correct, this hypothesis implies that the typical
Tevatron hadron--hadron collision may contain something like 2--6 
interactions, and the LHC one 4--10.

A few models based on this picture were presented several years ago
\cite{Zijl}, and compared with the data then available. Though these models
may still be tuned to give a reasonable description of the underlying event at
various collider energies, several shortcuts had to be taken, particularly in
the description of the nonperturbative aspects alluded to above. For
instance, it was not possible to consider beam remnants with more than one
valence quark kicked out.
  
The increased interest and the new data now prompts
us to develop a more realistic framework for multiple interactions
than the one in ref.~\cite{Zijl}, while making use of many of the
same underlying ideas. This may not necessarily result in significant
improvements for fits made to only a few distributions at a time, 
but we hope it will enhance our ability to simultaneously 
describe many different observations inside one framework, thereby
improving the confidence with which we can make extrapolations from known
measurements to new distributions and to higher energies. 

One of the building blocks for the new model comes from our recent 
study of baryon-number-violating processes \cite{BNV}.
We then had to address the hadronization of colour topologies
of the same kind as found in baryons. Specifically, as an extension
of the standard Lund string fragmentation framework \cite{string},
we explored the concept of a junction in which three string pieces
meet, with a quark at the end of each string. This also opens the way
to a more realistic description of multiple interaction events.
 
The resulting improvements, relative to the framework in \cite{Zijl}, 
are especially notable in the description of the
structure of the incoming hadrons, i.e.\
how flavours, colours, transverse and longitudinal momenta are
correlated between all the partons involved, both those that undergo
interactions and those that are left behind in the remnants. (Brief
descriptions of some of these aspects can also be 
found in \cite{briefdesc}.) To give one specific example, we introduce
parton densities that are modified according to the flavours already affected
by interactions. 
 
Clearly, the model we present here is not the final word. 
For instance, we defer for the future
any discussions of processes involving photons in the initial state and 
whether and how diffractive topologies could arise
naturally from several interactions with a net colour singlet exchange.
More generally, the whole issue of colour correlations will require further
studies. The model also allows some options in a few places. A reasonable range
of possibilities can then be explored, and (eventually) experimental
tests should teach us more about the course taken by nature.
 
This article is organized as follows.
In Section~2 we give an introduction to multiple interactions in
general, to the existing multiple interactions machinery, to other
theoretical models, and to experimental data of relevance.
Sections~3--5 then describe the improvements introduced by the current
study: 3 a new option for impact-parameter dependence, 4 the main work
on flavour and momentum space correlations, and 5 the very difficult
topics of colour correlations and junction hadronization. Finally
Section~6 provides some further examples of the resulting physics
distributions and important tests, while Section~7 contains a
summary and outlook.
 
\section{Multiple Interactions Minireview}
 
\subsection{Basic Concepts}
 
The cross section for QCD hard $2 \to 2$ processes,
as a function of the $\pT^2$ scale, is given by
\begin{equation}
\frac{\d \sigma_{\mrm{int}}}{\d \pT^2} =
\sum_{i,j,k} \int \d x_1 \int \d x_2 \int \d \hat{t} \, \,
f_i(x_1,Q^2) \, f_j(x_2,Q^2) \,
\frac{\d \hat{\sigma}_{ij \to kl}}{\d \hat{t}} \,
\delta \! \left( \pT^2 - \frac{\hat{t}\hat{u}}{\hat{s}} \right) ~,
\label{eq:sigmapt}
\end{equation}
where $\hat{s} = x_1 x_2 s$. The jet cross section is twice as large,
$\sigma_{\mrm{jet}} = 2 \sigma_{\mrm{int}}$, since each interaction
gives rise to two jets, to first approximation. In the following,
we will always refer to the interaction rather than the jet cross
section, unless otherwise specified. We will also assume that the
`hardness' of processes is given by the $\pT$ scale of the
interaction, i.e. $Q^2 = \pT^2$.
 
The cross section for QCD $2 \to 2$ processes, the sum of
$\q\q'\to \q\q'$, $\q\qbar \to \q'\qbar'$, $\q\qbar \to \g\g$,
$\q\g \to \q\g$, $\g\g \to \g\g$ and $\g\g \to \q\qbar$, is
dominated by $t$-channel gluon exchange contributions. (This
justifies the use of `interaction' and `scattering' as almost
synonymous.) In the $|\hat{t}| \ll \hat{s}$ limit, where
$\pT^2 = \hat{t}\hat{u}/\hat{s} \approx |\hat{t}|$, quark and gluon
interactions just differ by the colour factors, so approximately we
may write
\begin{equation}
\frac{\d\sigma_{\mrm{int}}}{\d\pT^2} \approx \int \!\!\! \int
\frac{\d x_1}{x_1} \,  \frac{\d x_2}{x_2} \,
F(x_1, \pT^2) \, F(x_2, \pT^2) \, \frac{\d \hat{\sigma}}{\d \pT^2}~,
\end{equation}
where
\begin{equation}
\frac{\d \hat{\sigma}}{\d \pT^2} = \frac{8 \pi \alphas^2(\pT^2)}{9 \pT^4}~,
\end{equation}
and
\begin{equation}
F(x, Q^2) = \sum_q \left( x\,q(x, Q^2) +  x\,\overline{q}(x, Q^2) \right)
           +  {\textstyle\frac{9}{4}} x\,g(x, Q^2)~.
\end{equation}
 
\begin{figure}[tp]
\begin{center}
\mbox{\epsfig{file=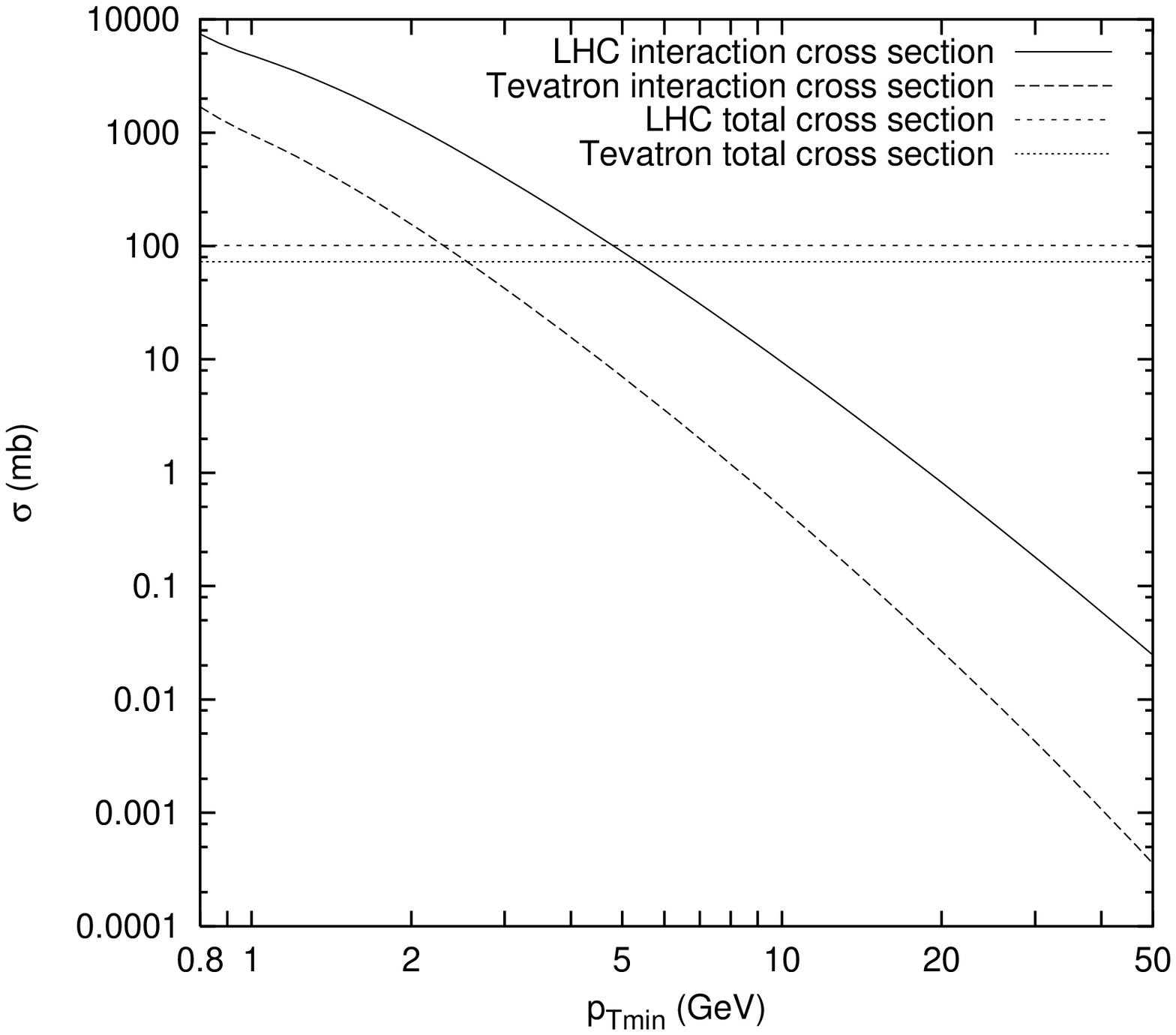,width=15cm,clip=}}
\end{center}
\captive{The integrated interaction cross section $\sigma_{\mrm{int}}$
above $\pTmin$ for the Tevatron, with 1.8~TeV $\p\pbar$ collisions, and
the LHC, with 14~TeV $\p\p$ ones. For comparison, the flat lines
represent the respective total cross section.
\label{fig:sigmapt}}
\end{figure}
 
Thus, for constant $\alphas$ and neglecting the $x$ integrals, the
integrated cross section above some $\pTmin$ is divergent in the
limit $\pT \to 0$:
\begin{equation}
\sigma_{\mrm{int}}(\pTmin) =
\int_{\pTmin}^{\sqrt{s}/2} \frac{\d\sigma}{\d\pT} \, \d\pT
\propto \frac{1}{\pTmin^2}~.
\end{equation}
A numerical illustration of this divergence is given in
Fig.~\ref{fig:sigmapt}. Note that the actual fall-off is everywhere
steeper than $1/\pT^2$. We have here used full $2\to 2$ matrix
elements and the CTEQ~5L parton density parametrizations
\cite{CTEQ5}, which are valid for $Q > 1.1$~GeV and $x > 10^{-6}$;
therefore results at the lowest $\pT$ values are not to be taken
too literally. For the studies in this article we are basing
ourselves on leading-order cross sections and parton densities, with
nontrivial higher-order corrections only approximately taken into
account by the addition of parton showers. Nevertheless, the trend
is quite clear, with an integrated cross section that exceeds
the total $\p\p$/$\pbar\p$ cross section $\sigma_{\mrm{tot}}$
(in the parametrization of ref.~\cite{DL}) for $\pTmin$ of the
order of a few GeV. As already mentioned in the introduction,
this is well above $\Lambda_{\mrm{QCD}}$, so one cannot postulate
a breakdown of perturbation theory in the conventional sense.
 
The resolution of the $\sigma_{\mrm{int}} > \sigma_{\mrm{tot}}$
paradox probably comes in two steps.
 
Firstly, the interaction cross section is an inclusive number.
Thus, if an event contains two interactions it counts twice in
$\sigma_{\mrm{int}}$ but only once in $\sigma_{\mrm{tot}}$,
and so on for higher multiplicities. Thereby we may identify
$\langle n \rangle(\pTmin) = \sigma_{\mrm{int}}(\pTmin) /
\sigma_{\mrm{tot}}$ with the average number of interactions
above $\pTmin$ per event, and that number may well be above unity.
 
One of the problems we will consider further in this article is
that this simple calculation of $\langle n \rangle(\pTmin)$ does
not take into account energy-momentum conservation effects.
Specifically, the average $\hat{s}$ of a scattering decreases
slower with $\pTmin$ than the number of interactions increases,
so naively the total amount of scattered partonic energy
becomes infinite. Thus corrections reduce the
$\langle n \rangle(\pTmin)$ number, but not sufficiently strongly:
one is lead to a picture with too little of the incoming energy
remaining in the small-angle beam jet region \cite{Zijl}.
 
Secondly, a more credible reason for taming the rise of
$\langle n \rangle(\pTmin)$ is that the incoming hadrons are
colour singlet objects. Therefore,
when the $\pT$ of an exchanged gluon is made small and the transverse
wavelength correspondingly large, the gluon can no longer resolve the
individual colour charges, and the effective coupling is decreased.
Note that perturbative QCD calculations are always performed assuming
free incoming and outgoing quark and gluon states, rather than partons
inside hadrons, and thus do not address this kind of nonperturbative
screening effects.
 
A naive estimate of an effective lower cutoff would be
\begin{equation}
\pTmin \simeq \frac{\hbar}{r_{\p}} \approx
\frac{0.2~\mrm{GeV}\cdot\mrm{fm}}{0.7~\mrm{fm}} \approx
0.3~\mrm{GeV} \simeq \Lambda_{\mrm{QCD}} ~, \label{eq:cutoff}
\end{equation}
but this again appears too low. The proton radius $r_{\p}$ has to be
replaced by the typical colour screening distance $d$, i.e.\ the average
size of the region within which the net compensation of a given colour
charge occurs. This number is not known from first principles, so
effectively one is forced to introduce some kind of cutoff parameter,
which can then just as well be put in transverse momentum space.
The simplest choice is to introduce a step function
$\theta(\pT - \pTmin)$, such that the perturbative cross section
completely vanishes below the $\pTmin$ scale. A more
realistic alternative is to note that the jet cross section is
divergent like $\alphas^2(\pT^2)/\pT^4$, and that therefore a
factor
\begin{equation}
\frac{\alphas^2(\pTo^2 + \pT^2)}{\alphas^2(\pT^2)} \,
\frac{\pT^4}{(\pTo^2 + \pT^2)^2}
\end{equation}
would smoothly regularize the divergences, now with $\pTo$ as the
free parameter to be tuned to data. Empirically the two procedures
give similar numbers, $\pTmin \approx \pTo$, and both of the order
of 2~GeV.
 
The parameters do not have to be energy-independent, however. 
Higher energies imply that parton densities can be probed at smaller
$x$ values, where the number of partons rapidly increases. Partons
then become closer packed and the colour screening distance $d$
decreases. Just like the small-$x$ rise goes like some power of $x$
one could therefore expect the energy dependence of $\pTmin$ and
$\pTo$ to go like some power of CM energy. Explicit toy simulations
\cite{Dischler} lend some credence to such an ansatz, although with
large uncertainties. Alternatively, one could let the cutoff increase
with decreasing $x$; this would lead to a similar phenomenology since
larger energies probe smaller $x$ values.
 
\subsection{Our Existing Models}
 
The models developed in ref.~\cite{Zijl} have been implemented
and are available in the \textsc{Pythia} event generator. They form
the starting point for the refinements we will discuss further on,
so we here review some of the main features.
 
The approach is not intended to cover elastic or diffractive physics,
so the $\sigma_{\mrm{int}}(\pTmin,s)$ or $\sigma_{\mrm{int}}(\pTo,s)$
interaction cross section is distributed among the
$\sigma_{\mrm{nd}}(s)$ nondiffractive inelastic one \cite{DL, Schuler}.
The average number of interactions per such event is then the ratio
$\langle n \rangle = \sigma_{\mrm{int}} / \sigma_{\mrm{nd}}$.
As a starting point we will assume that all hadron collisions are
equivalent, i.e.\ without an impact parameter dependence, and that
the different parton--parton interactions take place independently of
each other. The number of interactions per event is then distributed
according to a Poisson distribution with mean $\langle n \rangle$,
$\mathcal{P}_n = \langle n \rangle^n \exp( - \langle n \rangle) / n!$.
 
One (not used) approach would be, for each new event, to pick the actual
number of interactions $n$ according to the Poissonian, and select
the $n$ $\pT$ values independently according to eq.~(\ref{eq:sigmapt}).
One disadvantage is that this does not take into account correlations,
even such basic ones as energy--momentum conservation: the sum of
interaction energies may well exceed the total CM energy.
 
In an event with several interactions, it is therefore convenient to
impose an ordering. The logical choice is to arrange the scatterings
in falling sequence of $\pT$ values. The `first' scattering is thus
the hardest one, with the `subsequent' (`second', `third', etc.)
successively softer. This terminology is not primarily related to
any picture in physical time although, by the uncertainty relation,
large momentum transfers implies short timescales. When averaging over
all configurations of soft partons, one should effectively obtain
the standard QCD phenomenology for a hard scattering, e.g.\ in terms of
parton distributions. Correlation effects, known or estimated, can be
introduced in the choice of subsequent scatterings, given that the
`preceding' (harder) ones are already known. This will be developed
further in Section~4.
 
The generation of a sequence $\sqrt{s}/2 > p_{\perp 1} >
p_{\perp 2} > \ldots > p_{\perp n} > \pTmin$ now becomes one of
determining $p_{\perp i}$ from a known $p_{\perp i-1}$,
according to the probability distribution
\begin{equation}
\frac{\d\mathcal{P}}{\d p_{\perp i}} = \frac{1}{\sigma_{\mrm{nd}}}
\frac{\d\sigma}{\d\pT} \exp\left[ - \int_{\pT}^{p_{\perp i-1}}
\frac{1}{\sigma_{\mrm{nd}}} \frac{\d\sigma}{\d\pT'} \d\pT' \right] ~.
\label{eq:formfactor}
\end{equation}
The exponential expression is the `form factor' from the requirement
that no interactions occur between $p_{\perp i-1}$ and $p_{\perp i}$,
cf.\ radioactive decays or the Sudakov form factor \cite{Sudakov}
of parton showers.
 
When used with the standard differential cross section
$\d\sigma/\d\pT$, eq.~(\ref{eq:formfactor})
gives the same Poisson distribution as above. This time $n$ is not
known beforehand, but is defined by the termination of the iterative
procedure. Now, however, $\d\sigma/\d\pT$ can be modified to take into
account the effects of the $i-1$ preceding interactions. Specifically,
parton distributions are not evaluated at $x_i$ for the $i$'th
scattered parton from a hadron, but at the rescaled value
\begin{equation}
x'_i = \frac{x_i}{1 - \sum_{j=1}^{i-1} x_j} ~,
\label{eq:xresc}
\end{equation}
so that it becomes impossible to scatter more energy than initially
available in the incoming beam. This also dynamically suppresses the
high-multiplicity tail of the Poissonian and thereby reduces the average
number of interactions.
 
In a fraction of the events studied, there will be no hard scattering
at all above $\pTmin$. Such events are associated with nonperturbative
low-$\pT$ physics, and are simulated by exchanging a very soft gluon
between the two colliding hadrons, making the hadron remnants colour-octet
objects rather than colour-singlet ones. If only valence quarks
are considered, the colour-octet state of a baryon can be decomposed
into a colour triplet quark and an antitriplet diquark. In a
baryon--baryon collision, one would then obtain a two-string picture,
with each string stretched from the quark of one baryon to the diquark
of the other. A baryon--antibaryon collision would give one string
between a quark and an antiquark and another one between a diquark and
an antidiquark.
 
In a hard interaction, the number of possible string drawings are many
more, and the overall situation can become quite complex. In the studies
preceding this work, several simplifications were made. The hardest
interaction was selected with full freedom of flavour choice and colour
topology, but for the subsequent ones only three simple recipes were
available:
\begin{Itemize}
\item Interactions of the $\g\g \to \g\g $ type, with
the two gluons in a colour-singlet state, such that a double string is
stretched directly between the two outgoing gluons, decoupled from the
rest of the system.
\item Interactions $\g\g \to \g\g$, but colour correlations
assumed to be such that each of the gluons is connected to one
of the strings `already' present. Among the different possibilities of
connecting the colours of the gluons, the one which minimizes the total
increase in string length is chosen. This is in contrast to the
previous alternative, which roughly corresponds to a maximization
(within reason) of the extra string length.
\item Interactions $\g\g \to \q\qbar$, with the final pair again
in a colour-singlet state, such that a single string is stretched
between the outgoing $\q$ and $\qbar$.
\end{Itemize}
The three possibilities can be combined in suitable fractions.
 
Many further approximations were also required in the old framework, 
e.g.~the addition of initial- and final-state parton showers was feasible
only for the hardest interaction, and we address several of those in the
following.  

The model also includes several options for the impact-parameter
dependence. This offers an additional element of variability:
central collisions on the average will contain more interactions
than peripheral ones. Even if a Poisson distribution in the
number of interactions would be assumed for each impact parameter
separately, the net result would be a broader-than-Poisson
distribution. The amount of broadening can be `tuned' by the
choice of impact-parameter profile. We discuss this further in
section 3, where a new set of profiles is studied.

The above framework was originally formulated for $\p\pbar/\p\p$ collisions,
but has also been extended to $\gamma\p$ and $\gamma\gamma$ interactions
\cite{pythiagamma}. In these latter processes, however, the nature of the
photon needs to be modelled in detail, and this introduces many further
uncertainties. A study of such aspects is beyond the scope of the current
article.

\subsection{Other Models}
 
While the models of ref.~\cite{Zijl} may well be the ones most
frequently studied, owing to their implementation in \textsc{Pythia}
\cite{Pythia}, a number of other models also exist. Many of the
basic concepts have also been studied separately. We here give a
few examples, without any claim of completeness.
 
In Dual Topological Unitarization (DTU) language \cite{DTU},
and the Dual Parton Model based on it \cite{DPM}, or other
similar techniques \cite{diagrammatic}, inelastic
events are understood in terms of cut pomerons \cite{AKG}.
Translated into modern terminology, each cut pomeron corresponds
to the exchange of a soft gluon, which results in two `strings'
being drawn between the two beam remnants. Uncut pomerons give
virtual corrections that help preserve unitarity. A variable number
of cut pomerons are allowed. This approach has been the basis for
the simulation of underlying events in \textsc{Isajet} \cite{Isajet},
and was the starting point for \textsc{Dtujet} \cite{Dtujet}. However,
note that cut pomerons were originally viewed as purely soft objects,
and so did not generate any transverse momentum, unlike the multiple
interactions considered in this article. In \textsc{Dtujet} and its
\textsc{Phojet} \cite{Phojet} and \textsc{Dpmjet} \cite{Dpmjet}
relatives, however, also hard interactions have been included, so
that the picture now is one of both hard and soft pomerons, ideally
with a smooth transition between the two. Since the three related
programs make use of the \textsc{Pythia} hadronization description,
the differences relative to our scenarios is more a matter of details
(but ``the devil is in the details'') than of any basic incompatibility.
 
The possibility of observing two separate hard interactions has
been proposed since long \cite{twoint}, and from that has also
developed a line of studies on the physics framework for having
several hard interactions \cite{manyint}, also involving e.g.\
electroweak processes \cite{manyintdiff}. Again this is similar to
what we do, except that lower $\pT$ values and the transition to
nonperturbative physics are not normally emphasized.
 
The possibility of multiple interactions has also been implicit
\cite{sigmaeikonal} or explicit \cite{sigmaminijet} in many
calculations of total cross sections for hadron--hadron,
hadron--$\gamma$ and $\gamma\gamma$ events. The increase of
$\sigma_{\mrm{int}}$ with CM energy is here directly driving an
increase also of $\sigma_{\mrm{tot}}$; that the latter is rising
slower than the former comes out of an eikonalization procedure
that implies also an increasing $\langle n_{\mrm{int}} \rangle$.
 
Multiple interactions require an ansatz for the structure of the
incoming beams, i.e.\ correlations between the constituent partons.
Some of these issues have been studied, e.g.\ with respect to
longitudinal momenta \cite{hadronpcorr,DPM}, colours
\cite{hadronccorr} or impact parameter \cite{hadronbcorr}, but
very little of this has been tested experimentally. Dense-packing
of partons could become an issue \cite{diagrammatic}, of unknown
importance, but not necessarily a major one \cite{saturation}.
 
The \textsc{Herwig} \cite{Herwig} event generator does not contain
any physics simulation of multiple interactions. Instead a
parametrization procedure originally suggested by UA5 \cite{UA5gen}
is used, without any underlying physics scenario. It thus does
parametrize multiplicity and rapidity distributions, but does not
contain any minijet activity in the underlying event. The add-on
\textsc{Jimmy} package \cite{Jimmy} offers a multiple-interaction
component (both for $\p\p$, $\gamma\p$ and $\gamma\gamma$ events), 
which has recently been extended to include also a model
of soft interactions \cite{Ivan}.
 
The introduction of unintegrated parton densities, as used in
the BFKL/CCFM/LDC approaches to initial-state radiation
\cite{BFKL,CCFM,LDC}, allows the possibility to replace our
$\pTmin/\pTo$ cutoff by parton densities that explicitly vanish
in the $\pT \to 0$ limit \cite{Miukt}. This opens up the
possibility of an alternative implementation of multiple
interactions \cite{Miu}.
 
In heavy-ion collisions the multiple interactions rate can become
huge \cite{heavyion}. For small impact parameters, a major fraction 
of the energy of the two incoming nuclei is carried by partons
undergoing perturbative interactions, and which therefore define a
`resolved' partonic content. This suggests a mechanism for the 
construction of an `initial state' for the continued formation 
and thermalization (or not) of a quark--gluon plasma.

\subsection{Experimental Tests}
 
Experimental input to the understanding of multiple interactions
comes in essentially three categories: direct observation of double
parton scattering, event properties that directly and strongly
correlate with multiple interactions, and event properties that
do not point to multiple interactions by themselves but still
constrain multiple interactions models.
 
If an event contains two uncorrelated $2 \to 2$ interactions,
we expect to find four jets, grouped in two pairs that each
internally have roughly opposite and compensating transverse
momenta, and where the relative azimuthal angle between the
scattering planes is isotropic. Neither of these properties
are expected in a $2 \to 4$ event, where two of the partons can be
thought of as bremsstrahlung off a basic $2 \to 2$ process.
The problem is that $2 \to 4$ processes win out at large $\pT$,
so there is a delicate balance between having large enough
jet $\pT$ that the jets can be well measured and still not so
large that the signal drowns.
 
When the $\pTmin$ of the jets is sufficiently large that
$\exp(-\langle n \rangle) \approx 1$, Poisson statistics implies
that $\mathcal{P}_2 = \mathcal{P}_1^2/2$, where $\mathcal{P}_i$ is
the probability to have $i$ interactions. Traditionally this is
rewritten as
\begin{equation}
\sigma_2 = \sigma_{\mrm{nd}} \,
\left( \frac{\sigma_1}{\sigma_{\mrm{nd}}} \right)^2 \,
\frac{1}{2} \, \frac{\sigma_{\mrm{nd}}}{\sigma_{\mrm{eff}}} =
\frac{1}{2} \, \frac{\sigma_1^2}{\sigma_{\mrm{eff}}}~,
\label{eq:dps}
\end{equation}
where the ratio $\sigma_{\mrm{nd}}/\sigma_{\mrm{eff}}$
gauges deviations from the Poissonian ansatz. Values above unity,
i.e.~$\sigma_{\mrm{eff}}<\sigma_{\mrm{nd}}$,
arise naturally in models with variable impact parameter.
 
The first observation of double parton scattering is by AFS
\cite{AFSmiev}. The subsequent UA2 study \cite{UA2miev} ends up
quoting an upper limit, but has a best fit that requires them.
A CDF study \cite{CDFmiev} also found them. These experiments
all had to contend with limited statistics and uncertain
background estimates. The strongest signal has been obtained with a
CDF study involving three jets and a hard photon \cite{CDFmievnew};
here $\sigma_2 = \sigma_A \sigma_B/ \sigma_{\mrm{eff}}$,
without a factor $1/2$, since the two $2\to 2$ processes $A$ and
$B$ are inequivalent. In all cases, including the UA2 best fit,
$\sigma_{\mrm{eff}}$ comes out smaller than $\sigma_{\mrm{nd}}$;
typically the double parton scattering cross section is a factor
of three to four larger than the Poissonian prediction. For instance,
the CDF number is $\sigma_{\mrm{eff}} = 14.5 \pm 1.7^{+1.7}_{-2.3}$~mb.
More recently, ZEUS has observed a signal in $\gamma\p$ events
\cite{ZEUSmiev}. The D0 four-jet study shows the need to include
multiple interactions, but this is not quantified \cite{D0miev}.
 
So far, no direct tests of triple or more parton scattering exist.
However, the UA1 minijet rates \cite{UA1minijet}, going up to 5 jets,
are difficult to understand without such events.
 
Tests involving jets at reasonably large $\pT$ values do not probe
the total rate of multiple interactions, since the bulk of the
interactions occur at so small $\pT$ values that they cannot be
identified as separate jets. By the way colours are drawn across the
event, soft partons can drive the production of particles quite out of
proportion to the $\pT$ values involved, however. The multiplicity
distribution of multiple interactions thereby strongly influences
the multiplicity distribution of charged hadrons, $n_{\mrm{ch}}$.
 
A notable aspect here is that the measured $n_{\mrm{ch}}$
distribution, when expressed in the KNO variable
$z = n_{\mrm{ch}} / \langle n_{\mrm{ch}} \rangle$ \cite{KNO}, is
getting broader with increasing CM energy \cite{multdistu,multdiste}. 
This is contrary to the essentially Poissonian hadronization 
mechanism of the string model, where the KNO distribution becomes 
narrower. As an example, consider the UA5 measurements at 900 GeV 
\cite{multdistu}, where $\langle n_{\mrm{ch}} \rangle = 35.6$ and
$\sigma(n_{\mrm{ch}}) = 19.5$, while the Poissonian prediction would be
$\sigma(n_{\mrm{ch}}) = \sqrt{\langle n_{\mrm{ch}} \rangle} = 6.0$.
It is possible to derive approximate KNO scaling in $\e^+\e^-$
annihilation \cite{eeKNO} , but this then rests on having a perturbative
shower that involves the whole CM energy. However, allowing for at most
one interaction in $\p\pbar$ events and assuming that hadronization is
universal (so it can be tuned to $\e^+\e^-$ data), there is no (known)
way to accommodate the experimental multiplicity distributions, neither
the rapid increase with energy of the average nor the large width.
Either hadronization is \textit{very} different in hadronic events
from $\e^+\e^-$ ones, or one must accept multiple interactions as
a reality.(Unfortunately it is difficult to test the `hadronization
universality' hypothesis completely separated from the multiple
interactions and other assumptions. To give two examples, the relative
particle flavour composition appears to be almost but not quite
universal \cite{flavnonuniv}, and low-mass diffractive events display
`string-like' flavour correlations \cite{Schleinstring}.)
 
Further support is provided by the study of forward--backward
multiplicity correlations. For instance, UA5 and E735 define a `forward'
$n_F$ and a `backward' $n_B$ multiplicity in pseudorapidity windows
of one unit each, separated by a $\Delta\eta$ variable-width gap in
the middle \cite {fbcorr}. A forward--backward correlation strength
is now defined by
\begin{equation}
b =
\frac{\langle \left( n_F - \langle n_F \rangle \right) \,%
\left( n_B - \langle n_B \rangle \right) \rangle}%
{\sigma(n_F) \, \sigma(n_B)} =
\frac{\langle n_F n_B \rangle - \langle n_F \rangle^2}%
{\langle n_F^2 \rangle - \langle n_F \rangle^2} ~,
\end{equation}
where the last equality holds for a symmetric $\eta$ distribution,
i.e.\ for $\p\p/\p\pbar$ but not for $\gamma\p$. Measurements give
a positive and surprisingly large $b$, also for $\Delta\eta$ of
several units of rapidity. So it appears that there is some global
quantity, different for each event, that strongly influences particle
production in the full phase space. Again known fragmentation
mechanisms are too local, and effects of a single hard interaction
not strong enough. But the number of multiple interactions is
indeed a global quantity of the desired kind, and
multiple-interaction models can describe the data quite well.
 
It is a matter of taste which evidence is valued highest. The
direct observation of double parton scattering is easily recognized
as evidence for the multiple-interactions concept, but only affects
a tiny fraction of the cross section. By comparison, the broadening
multiplicity distribution and the strong forward--backward
correlations offer more indirect evidence, but ones strongly
suggesting that the bulk of events have several interactions.
We are not aware of any realistic alternative explanations for either
of the observables.
 
Another interesting phenomenon is the pedestal effect: events
with high-$\pT$ jets on the average contain more underlying activity
than minimum-bias ones, also well away from jets themselves. It has
been observed by several collaborations, like UA1 \cite{UA1ped},
CDF \cite{CDFpedjet,CDFped} and H1 \cite{H1ped}.
When the jet energy is varied from next-to-nothing to the highest
possible, the underlying activity initially increases, but then
flattens out around $p_{\perp\mrm{jet}} = 10$~GeV (details depending
on the jet algorithm used and the CM energy). This fits very nicely
with an impact-parameter-dependent multiple-interactions scenario:
the presence of a higher $\pT$ scale biases events towards a smaller
impact parameter and thereby a higher additional activity, but once
$\sigma_{\mrm{int}}(p_{\perp\mrm{jet}}) \ll \sigma_{\mrm{nd}}$
the bias saturates \cite{Zijl}. The height of the pedestal depends on
the form of the overlap function $\mathcal{O}(b)$, and can thus be
adjusted, while the $p_{\perp\mrm{jet}}$ at which saturation occurs
is rather model-insensitive, and in good agreement with the data.
 
The presence of pedestals also affects all measurements of jet profiles
\cite{jetshape,CDFpedjet}. It can lead to seemingly broader jets,
when the full underlying event cannot be subtracted, and enrich the
jet substructure, when a multiple-interactions jet is mistaken for
radiation off the hard subprocess. It can also affect
(anti)correlations inside a jet and with respect to the rest of the
event \cite{H1ped}.
 
Many further observables influence the modeling and understanding
of multiple interactions, without having an immediate interpretation
in those terms. A notable example here is the
$\langle \pT \rangle (n_{\mrm{ch}})$ distribution, i.e.\ how the
average transverse momentum of charged particles varies as a function
of the total charged multiplicity. The observed increasing trend
\cite{meanptfornch}
is consistent with multiple interactions: large multiplicity implies
many interactions and therefore more perturbatively generated $\pT$
to be shared between the hadrons. For it to work, however, each new
interaction should add proportionately less to the total
$n_{\mrm{ch}}$ than to the total $\pT$. Whether this is the case
strongly depends on the colour connections between the interactions,
i.e.\ whether strings tend to connect nearest neighbours in momentum
space or run criss-cross in the event. A rising trend can easily be
obtained, but it is a major challenge to get the quantitative behaviour
right, as we shall see.
 
Finally, one should mention that global fits to hadron collider data
\cite{oldglobal,CDFglobal,D0global,ATLASglobal,Butterworth:2002ts}
 clearly point to the
importance of a correct understanding of multiple interactions, and
constrains models down to rather fine details. This brings together
many of the aspects raised above, plus some further ones. A convenient
reference for our continued discussion is Tune A, produced by R.D. Field,
which is known to describe a large set of CDF minimum bias and jet data
\cite{CDFglobal}. Relative to the defaults of the old scenario,
it assumes $\pTo = 2.0$ GeV (\texttt{PARP(82)=2.0}) at the reference
energy 1.8 TeV (\texttt{PARP(89)=1800.0}), with an energy rescaling
proportional to $E_{\mrm{cm}}^{1/4}$ (\texttt{PARP(90)=0.25}). It is
based on a double Gaussian matter distribution (\texttt{MSTP(82)=4}),
with half of the matter (\texttt{PARP(83)=0.5}) in a central core
of radius 40\% of the rest (\texttt{PARP(84)=0.4}). Almost all of the
subsequent interactions are assumed to be of the type
$\g \g \to \g \g$ with minimal string length (\texttt{PARP(85)=0.9},
\texttt{PARP(86)=0.95}). Finally the matching of the initial-state
showers to the hard scattering is done at a scale
$Q^2_{\mrm{shower}} = 4 p^2_{\perp\mrm{hard}}$ (\texttt{PARP(67)=4.0}).

The above parameter set is sensible, within the framework of the model,
although by no means obvious. The matter distribution is intermediate 
between the extremes already considered in \cite{Zijl}, while the
string drawing is more biased towards small string lengths than 
foreseen there. The $\pTo$ energy dependence is steeper than previously
used, but in a sensible range, as follows. In reggeon theory, a
Pomeron intercept of $1 + \epsilon$ implies a total cross section 
rising like $s^{\epsilon}$, and a small-$x$ gluon density like 
$x g(x) \propto x^{-\epsilon}$ (at small $Q^2$). A $\pTo$ rising (at 
most) like $s^{\epsilon}$ would then be acceptable, while one rising 
significantly faster would imply a decreasing interaction cross section
$\sigma_{\mrm{int}}(\pTo)$, and by implication a decreasing
$\sigma_{\mrm{tot}}$, in contradiction with data. The DL fit to 
$\sigma_{\mrm{tot}}$ \cite{DL} gives $\epsilon \approx 0.08$, which 
would imply (at most) a $\pTo$ dependence like 
$s^{0.08} = E_{\mrm{cm}}^{0.16}$. However, $\sigma_{\mrm{tot}}$ already 
represents the unitarization of multiple-pomeron exchanges, and the 
`bare' pomeron intercept should be larger than this, exactly by how much 
being a matter of some debate \cite{Landdeb}. A value like 
$\epsilon_{\mrm{bare}} \approx 0.12$ is here near the lower end of 
the sensible range; the $x g(x)$ shape is consistent with a
rather larger value. Since it is actually the bare pomeron that 
corresponds to a single interaction, an $E_{\mrm{cm}}^{0.25}$ 
behaviour is thereby acceptable.
 
\section{Impact-Parameter Dependence}
 
In the simplest multiple-interactions scenarios, it is assumed that
the initial state is the same for all hadron collisions. More
realistically, one should include the possibility that each collision
also could be characterized by a varying impact parameter $b$
\cite{Zijl}. Within the classical framework we use here, $b$ is to
be thought of as a distance of closest approach, not as the Fourier
transform of the momentum transfer. A small $b$ value corresponds to
a large overlap between the two colliding hadrons, and hence an
enhanced probability for multiple interactions. A large $b$, on the
other hand, corresponds to a grazing collision, with a large
probability that no parton--parton interactions at all take place.
 
In order to quantify the concept of hadronic matter overlap, one may
assume a spherically symmetric distribution of matter inside a
hadron at rest, $\rho(\mbf{x}) \, \d^3 x = \rho(r) \, \d^3 x$.
For simplicity, the same spatial distribution is taken to apply
for all parton species and momenta. Several different matter
distributions have been tried. A Gaussian ansatz makes the
subsequent calculations especially transparent, but there is no
reason why this should be the correct form. Indeed, it appears to
lead to a somewhat too narrow multiplicity distribution and too
little of a pedestal effect. The next simplest choice, that does
provide more fluctuations, is a double Gaussian
\begin{equation}
\rho(r) \propto \frac{1 - \beta}{a_1^3} \, \exp \left\{
- \frac{r^2}{a_1^2} \right\} + \frac{\beta}{a_2^3} \,
\exp \left\{ - \frac{r^2}{a_2^2} \right\} ~.
\label{eq:doubleGauss}
\end{equation}
This corresponds to a distribution with a small core region, of radius
$a_2$ and containing a fraction $\beta$ of the total hadronic matter,
embedded in a larger hadron of radius $a_1$. If we want to give a
deeper meaning to this ansatz, beyond it containing two more
adjustable parameters, we could imagine it as an intermediate
step towards a hadron with three disjoint core regions (`hot spots'),
reflecting the presence of three valence quarks, together carrying
the fraction $\beta$ of the proton momentum. One could alternatively
imagine a hard hadronic core surrounded by a pion cloud. Such details
would affect e.g.\ the predictions for the $t$ distribution in elastic
scattering, but are not of any consequence for the current topics.
 
For a collision with impact parameter $b$, the time-integrated
overlap $\mathcal{O}(b)$ between the matter distributions of the
colliding hadrons is given by
\begin{eqnarray}
& & \mathcal{O}(b) \propto \int \d t \int \d^3 x \, \,
\rho(x,y,z) \, \rho(x+b,y,z+t)      \nonumber \\
& & \propto \frac{(1 - \beta)^2}{2a_1^2} \exp \left\{
- \frac{b^2}{2a_1^2} \right\} +
\frac{2 \beta (1-\beta)}{a_1^2+a_2^2} \exp \left\{
- \frac{b^2}{a_1^2+ a_2^2} \right\} +
\frac{\beta^2}{2a_2^2} \exp \left\{ - \frac{b^2}{2a_2^2} \right\} ~.
\end{eqnarray}
The necessity to use boosted $\rho(\mbf{x})$ distributions
has been circumvented by a suitable scale transformation of the $z$
and $t$ coordinates. The overlap function $\mathcal{O}(b)$ is closely
related to the $\Omega(b)$ of eikonal models \cite{sigmaeikonal}, but
is somewhat simpler in spirit.
 
The larger the overlap $\mathcal{O}(b)$ is, the more likely it is to
have interactions between partons in the two colliding hadrons.
In fact, to first approximation, there should be a linear relationship
\begin{equation}
\langle \tilde{n}(b) \rangle = k \mathcal{O}(b) ~,
\label{eq:bdepend}
\end{equation}
where $\tilde{n} = 0, 1, 2, \ldots$ counts the number of interactions
when two hadrons pass each other with an impact parameter $b$. At this
stage $k$ is an undefined constant of proportionality, to be specified
below.
 
For each given impact parameter, the number of interactions $\tilde{n}$
is assumed to be distributed according to a Poissonian, 
$\mathcal{P}_{\tilde{n}} = \langle \tilde{n} \rangle^{\tilde{n}} 
\exp ( - \langle \tilde{n} \rangle ) / \tilde{n} !$, before
energy--momentum and other constraints are included. If the matter
distribution has a tail to infinity (as the Gaussians do),
events may be obtained with arbitrarily large $b$ values. In order
to obtain finite total cross sections, it is necessary to assume
that each event contains at least one semi-hard interaction.
(Unlike the simpler, impact-parameter-independent approach above,
where $\pT = 0$ no-interaction events are allowed as a separate class.)
The probability that two hadrons, passing each other with an impact
parameter $b$, will produce a real event is then given by
\begin{equation}
\mathcal{P}_{\mrm{int}}(b) = 
\sum_{\tilde{n} = 1}^{\infty} \mathcal{P}_{\tilde{n}}(b) = 
1 - \mathcal{P}_0(b) = 
1 - \exp ( - \langle \tilde{n}(b) \rangle )
= 1 - \exp (- k \mathcal{O}(b) ) ~,
\end{equation}
according to Poisson statistics. The average number of
interactions per event at impact parameter $b$ is now
$\langle n(b) \rangle = \langle \tilde{n}(b) \rangle /
\mathcal{P}_{\mrm{int}}(b)$, where the denominator comes from the
removal of hadron pairs that pass without interaction, i.e.\ which do 
not produce any events. While the removal of $\tilde{n}=0$ from the potential 
event sample gives a narrower-than-Poisson interaction distribution
at each fixed $b$, the variation of $\langle n(b) \rangle$ with $b$
gives a $b$-integrated broader-than-Poisson interaction multiplicity
distribution.
 
Averaged over all $b$ the relationship
$\langle n \rangle = \sigma_{\mrm{int}}/\sigma_{\mrm{nd}}$
should still hold. Here, as before, $\sigma_{\mrm{int}}$ is the 
integrated interaction cross section for a given regularization 
prescription at small $\pT$, while the inelastic nondiffractive cross 
section $\sigma_{\mrm{nd}}$ is taken from parametrizations  
\cite{DL, Schuler}. This relation can be used to solve for the 
proportionality factor $k$ in eq.~(\ref{eq:bdepend}). Note that, 
since now each event has to have at least one interaction, 
$\langle n \rangle > 1$, one must ensure that
$\sigma_{\mrm{int}} > \sigma_{\mrm{nd}}$. The $\pTo$
parameter has to be chosen accordingly small --- since now the concept
of no-interaction low-$\pT$ events is gone, aesthetically it is
more appealing to use the smooth $\pTo$ turnoff than the sharp
$\pTmin$ cutoff, and thereby populate interactions continuously all
the way down to $\pT = 0$. The whole approach can be questioned at
low energies, since then very small $\pTo$ values would be required,
so that many of the interactions would end up in the truly
nonperturbative $\pT$ region.
 
Technically, the combined selection of $b$ and a set of scattering
$\p_{\perp i}$ values now becomes more complicated
\cite{Zijl, Pythia}. It can be reduced to a combined choice of
$b$ and $p_{\perp 1}$, according to a generalization of
eq.~(\ref{eq:formfactor})
\begin{equation}
\frac{\d\mathcal{P}}{\d p_{\perp 1} \, \d^2 b} =
\frac{\mathcal{O}(b)}{\langle \mathcal{O} \rangle} \,
\frac{1}{\sigma_{\mrm{nd}}} \frac{\d\sigma}{\d\pT}
\exp\left[ - \frac{\mathcal{O}(b)}{\langle \mathcal{O} \rangle} \,
\int_{\pT}^{\sqrt{s}/2}
\frac{1}{\sigma_{\mrm{nd}}} \frac{\d\sigma}{\d\pT'} \d\pT' \right] ~.
\label{eq:bandptsel}
\end{equation}
The removal of the $\tilde{n}=0$ non-events leads to a somewhat special 
definition of the average \cite{Zijl}
\begin{equation}
\langle \mathcal{O} \rangle = \frac{\int \mathcal{O}(b) \, \d^2 b}%
{\int \mathcal{P}_{\mrm{int}}(b) \, \d^2 b} =
\frac{1}{k} \, \frac{\sigma_{\mrm{int}}}{\sigma_{\mrm{nd}}} ~.
\end{equation}
The subsequent interactions can be generated sequentially in falling
$\pT$ as before, with the only difference that $\d\sigma/\d\pT^2$ now
is multiplied by $\mathcal{O}(b)/\langle \mathcal{O} \rangle$,
where $b$ is fixed at the value chosen above.
 
Note that this lengthy procedure, via $\rho(r)$ and $\mathcal{O}(b)$,
is not strictly necessary: the probability $\mathcal{P}_n$ for having
$n$ interactions could be chosen according to any desired distribution.
However, with only $\mathcal{P}_n$ known and an $n$ selected from this
distribution, there is no obvious way to order the interactions in
$\pT$ during the generation stage, with lower-$\pT$ interactions
modified by the flavours, energies and momenta of higher-$\pT$ ones.
(This problem is partly addressed in ref.~\cite{Miu}, by a post-facto
ordering of interactions and a subsequent rejection of some of the
generated interactions, but flavour issues are not easily solved
that way.)
 
There is also another issue, the parton-level pedestal effect,
related to the transition from hard
events to soft ones. To first approximation, the likelihood that an
event contains a very hard interaction is proportional to
$n \, \mathcal{P}_n$, since $n$ interactions in an event means $n$
chances for a hard one. If the requested hardest $\pT$ is gradually
reduced, the bias towards large $n$ dies away and turns into its
opposite: for events with the hardest $\pT \to 0$ the likelihood of
further interactions vanishes. The interpolation between these two
extremes can be covered if an impact parameter is chosen, and thereby
an $\mathcal{O}(b)$, such that one can calculate the probability
of \textit{not} having an interaction harder than the requested
hardest one, i.e.\ the exponential in eq.~(\ref{eq:bandptsel}).
 
\begin{figure}[tp]
\begin{center}
\mbox{\epsfig{file=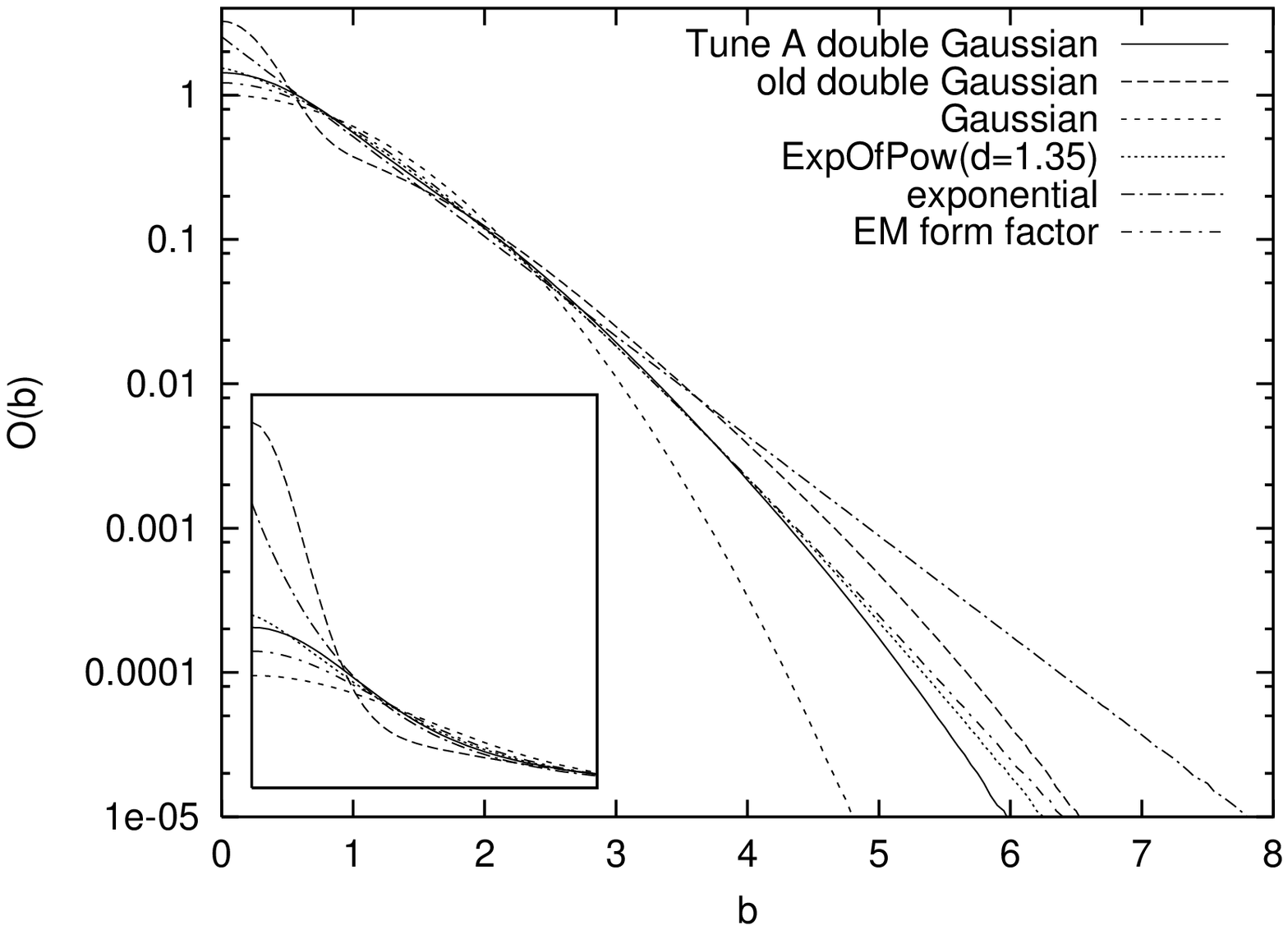,width=15cm,clip=}}
\end{center}
\captive{Overlap profile $\mathcal{O}(b)$ for a few different choices.
Somewhat arbitrarily the different parametrizations have been
normalized to the same area and average $b$, i.e. same
$\int \mathcal{O}(b) \, \d^2 b$ and $\int b \mathcal{O}(b) \, \d^2 b$.
(Recall that we have not specified $b$ in terms of any absolute units,
so both a vertical and a horizontal scale factor have to be fixed
for each distribution separately.) 
Insert shows the region $b <2$ on a linear scale.
\label{fig:bshape}}
\end{figure}
 
If the Gaussian matter distribution is the simplest possible choice,
the double Gaussian in some respects is the next-simplest one.
It does introduce two new parameters, however, where we might have
preferred to start with only one, to see how far that goes.
As an alternative, we will here explore an exponential of a power
of $b$
\begin{equation}
\mathcal{O}(b) \propto \exp \left\{ - b^d \right\}
\end{equation}
where $d \neq 2$ gives deviations from a Gaussian behaviour. We will
use the shorthand ExpOfPow($d=\ldots$) for such distributions.
Note that we do not present an ansatz for $\rho(r)$ from which the
$\mathcal{O}(b)$ is derived: in the general case the convolution of
two $\rho$ is nontrivial. A peaking of $\mathcal{O}(b)$ at $b=0$ is
related to one of $\rho(r)$ at $r=0$, however.
 
A lower $d$ corresponds to an overlap distribution more spiked at
$b=0$ and with a higher tail at large $b$, Fig.~\ref{fig:bshape},
i.e.\ leads to larger fluctuations. Specifically, the height of the
$b=0$ peak is related to the possibility of having fluctuations out
to high multiplicities. To give some feeling, an exponential,
ExpOfPow($d=1$), is not too dissimilar to the old \textsc{Pythia}
double Gaussian, with $\beta = 0.5$ and $a_2/a_1 = 0.2$.
Conveniently, the Tune A double Gaussian, still with $\beta = 0.5$
but now $a_2/a_1 = 0.4$, is well approximated in shape by an
ExpOfPow($d=1.35$). Another alternative, commonly used, is to assume 
the matter distribution to coincide with the charge distribution,
as gauged by the electric form factor 
$G_E(\pT^2) = (1+\pT^2/\mu^2)^{-2}$, with $\mu^2 = 0.71$~GeV$^2$. 
This gives an $\mathcal{O}(b) \propto (\mu b)^3 K_3(\mu b)$, which
also is close in form to Tune A, although somewhat less peaked at 
small $b$. 
 
\begin{figure}[p]
\begin{center}
\mbox{\epsfig{file=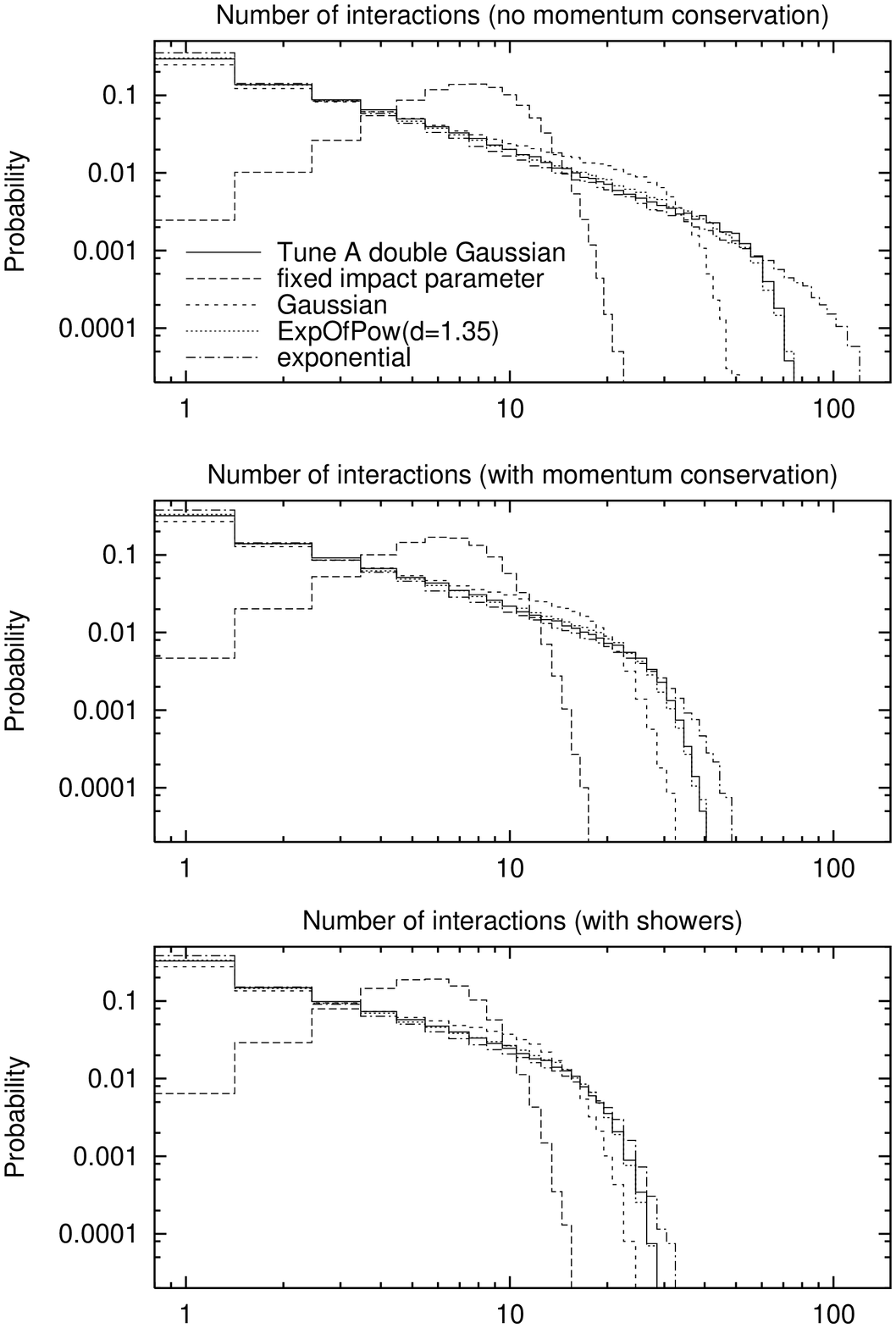,width=14.5cm,clip=}}
\end{center}
\captive{Distribution of the number of interactions for different
overlap profiles $\mathcal{O}(b)$, for $\p\pbar$ at 1.8~TeV, top
without momentum conservation constraints, middle with such
constraints included but without (initial-state) showers, and
bottom also with shower effects included. \label{fig:nintpartons}}
\end{figure}
 
\begin{figure}[p]
\begin{center}
\mbox{\epsfig{file=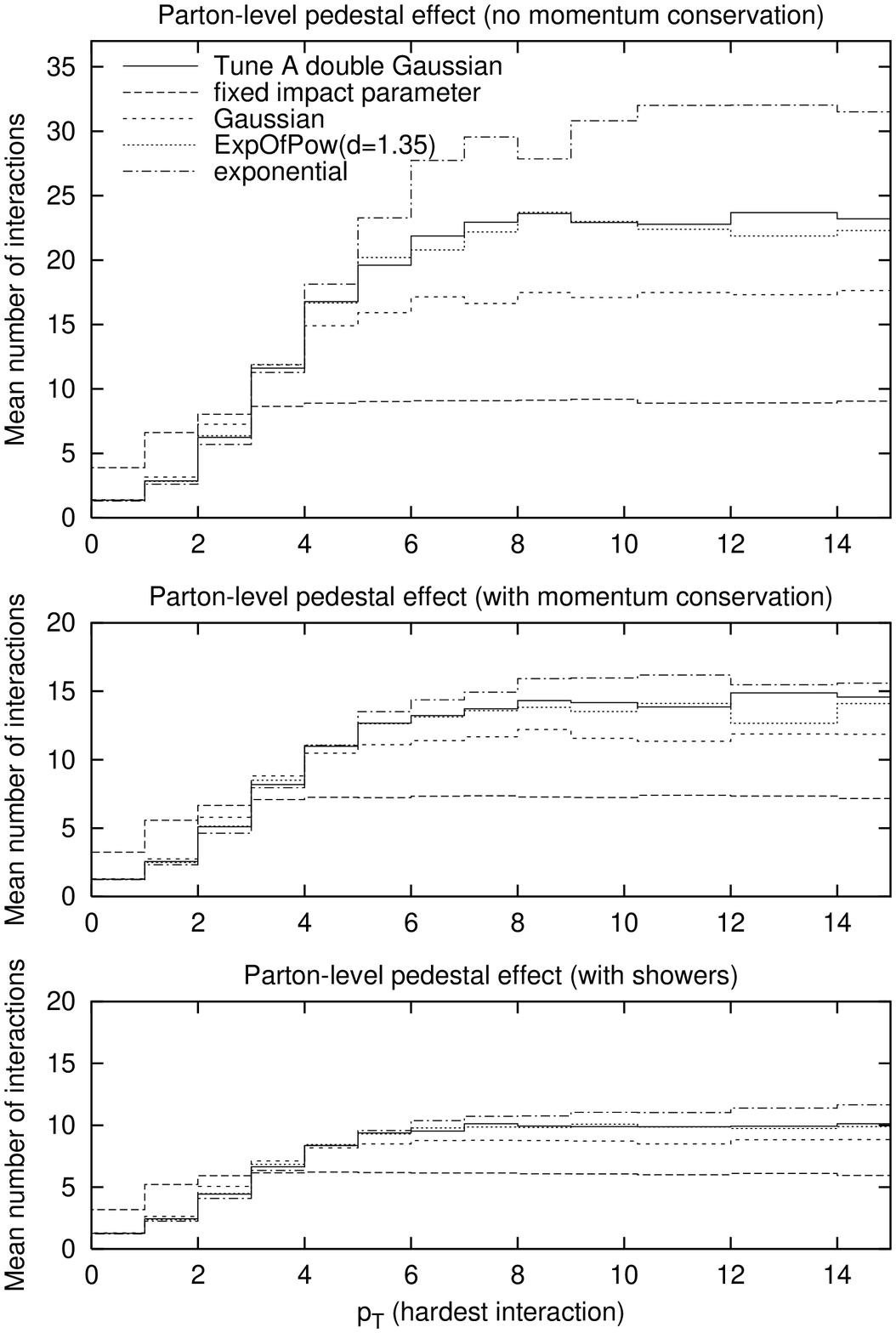,width=14.5cm,clip=}}
\end{center}
\captive{Average number of interactions as a function of the $\pT$
of the hardest interaction, for $\p\pbar$ at 1.8~TeV, top without
momentum conservation constraints, middle with such constraints
included but without (initial-state) showers, and bottom also with
shower effects included. \label{fig:pedestalpartons}}
\end{figure}
 
As indicated above, there are two key consequences of a an overlap
profile choice. One is the interaction multiplicity distribution
and the other the parton-level pedestal effect. These two are
illustrated in Figs.~\ref{fig:nintpartons} and
\ref{fig:pedestalpartons}, respectively, for $\p\pbar$ at 1.8~TeV, with
$\pTo = 2.0$~GeV as in Tune A. The three frames of each figure illustrate
how momentum conservation effects suppress the probability to have an
event with large multiplicity. This effect is even stronger now that each

interaction is allowed to undergo full shower evolution, so that it
carries away more of the available energy. In the figures, the default
lower shower cut-off of 1~GeV has been used; obviously a larger cut-off
would give results intermediate to the two lower frames. Further, the
possibility of two hard-scattering partons being part of the same
shower is not included. Note that the suppression of the high-multiplicity
tail implies that a distribution with large fluctuations in reality will
have fewer interactions on the average than a less-fluctuating one, if
they (as here) start with the same assumed average before the momentum
conservation effects are considered. This means that the choice of $\pTo$
is somewhat dependent on the one of overlap profile.
 
\begin{figure}[p]
\begin{center}
\mbox{\epsfig{file=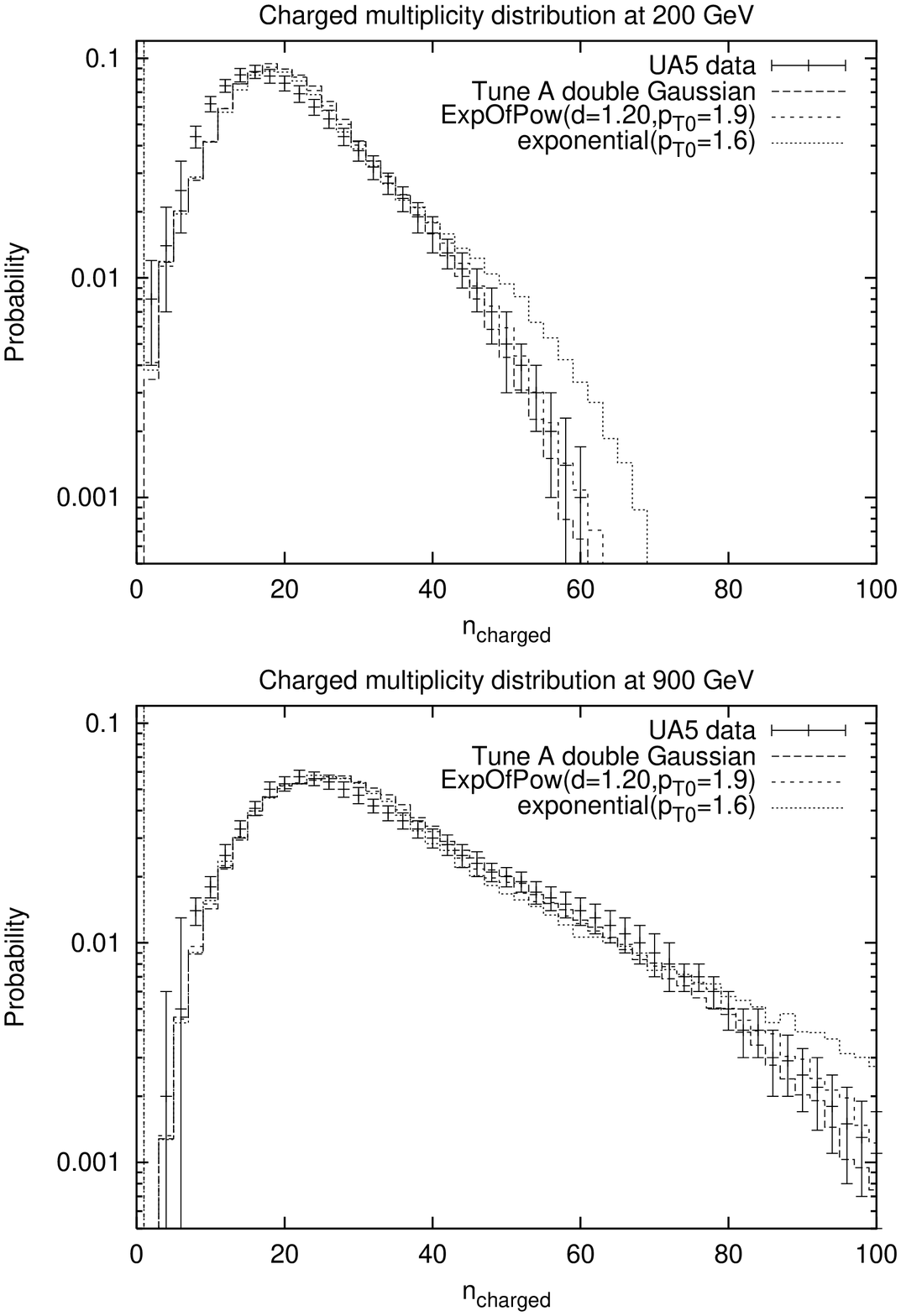,width=15cm}}
\end{center}
\captive{Charged multiplicity distribution at 200 and 900 GeV; different
overlap profiles compared with UA5 data \cite{multdistu}. 
\label{fig:ua5mult}}
\end{figure}
 
Let us now study the hadron-level multiplicity distribution, and 
begin with UA5 data at 200 and 900 GeV \cite{multdistu}. Tune A then 
does impressively well, Fig.~\ref{fig:ua5mult}, in spite of primarily 
having been tuned to pedestal effects rather than multiplicity 
distributions. In this comparison, we do not put too much emphasis 
on the low-multiplicity end, which is largely probing
diffractive physics. Here the \textsc{Pythia} description is known to
be too simple, with one or two strings stretched at low $\pT$ and no
hard interactions at all. More relevant is the mismatch in peak position,
which mainly is related to the multiplicity in events with only one
interaction. Assuming that most hadronization parameters are fixed by
$\e^+\e^-$ data, it is not simple to tune this position. The beam remnant
structure does offer some leeway, but actually the defaults are already
set towards the end of the sensible range that produces the lower peak
position, and still it comes out on the high side.
 
However, the main impression is of a very good description
of the fluctuations to higher multiplicities, better than obtained
with the old parameters explored in \cite{Zijl}. Of course, many
aspects have changed significantly since then, such as the shape
of parton densities at small $x$. One main difference is that Tune
A 90\% of the time picks subsequent interactions to be of the
$\g\g \to \g\g$ type with colour flow chosen to minimize the
string length. Since each further interaction thereby contributes
less additional multiplicity, the mean number of interactions can
be increased, and this obviates the need for the more extreme
double-Gaussian default parameters.
 
If, nevertheless, one should attempt to modify the Tune A parameters,
deviating from its ExpOfPow($d=1.35$) near equivalent, it would be
towards a smaller $d$, i.e. a slight enhancement of the tail towards
high multiplicities. An example is shown in Fig.~\ref{fig:ua5mult},
with ExpOfPow($d=1.2$) and $\pTo = 1.9$~GeV (at 1800~GeV, with the
Tune A energy rescaling).
 
\begin{figure}[t]
\begin{center}
\mbox{\epsfig{file=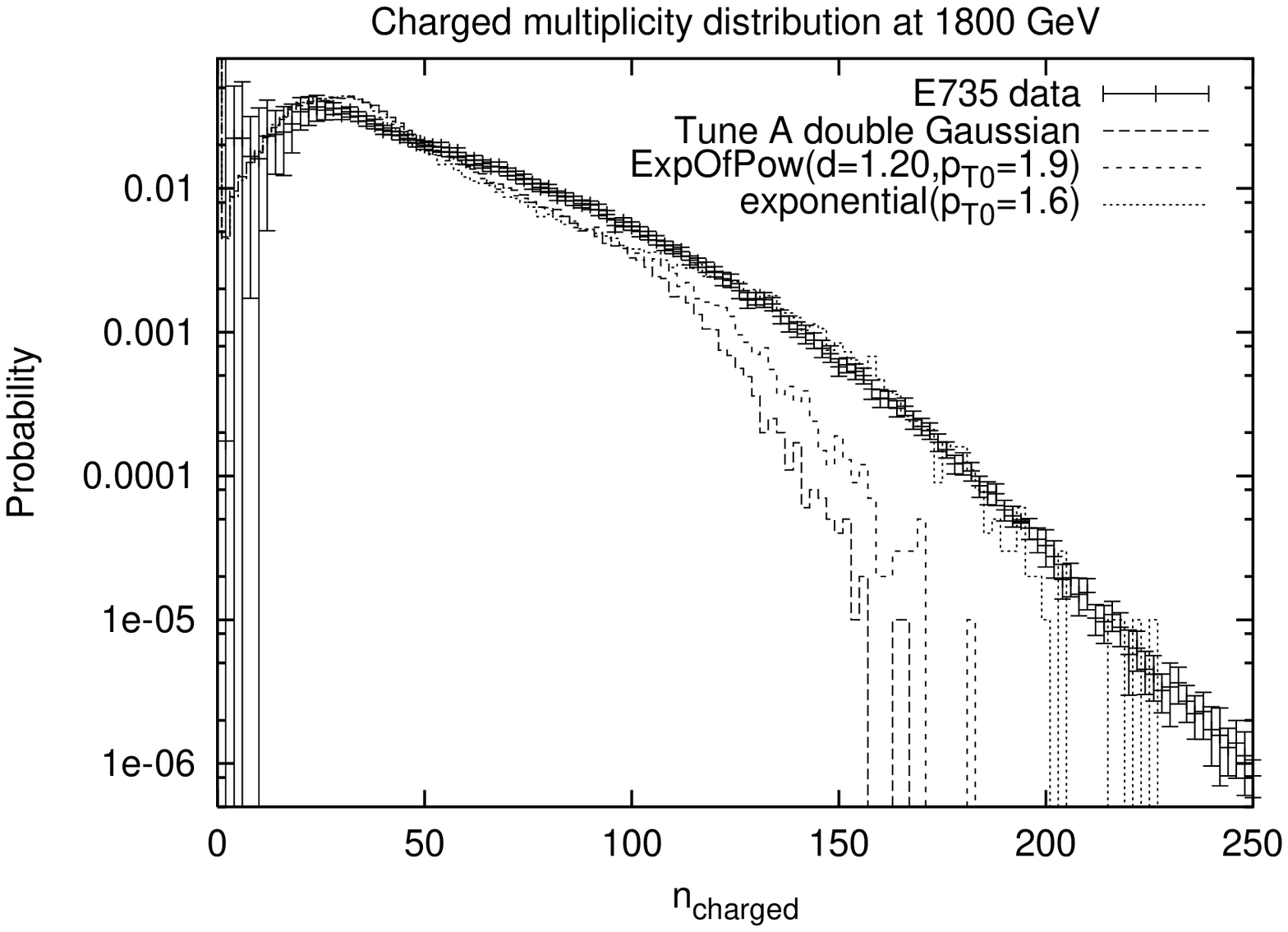,width=15cm}}
\end{center}
\captive{Charged multiplicity distribution at 1800 GeV; different
overlap profiles compared with E735 data \cite{multdiste}. 
\label{fig:e735mult}}
\end{figure}

However, the nice picture is shattered if one instead considers the E735 data 
at 1800~GeV \cite{multdiste}, Fig.~\ref{fig:e735mult}. 
Tune A gives a way too small tail out to large multiplicities, 
and also the ExpOfPow($d=1.2$) falls below the data. One would need 
something like an exponential with a rather low $\pTo = 1.6$~GeV 
to come near the E735 data, and that then disagrees with the 
lower-energy UA5 data, Fig.~\ref{fig:ua5mult}. The agreement could be 
improved, but not to the level of Tune A, by playing with the energy 
dependence of $\pTo$. However, the E735 collaboration itself notes 
that results from the two collaborations are incompatible over the 
whole UA5 energy range and especially at 546~GeV, where both have 
data \cite{multdiste}. Furthermore, we do not have the expertise to 
fully simulate E735 selection criteria, nor to assess the impact of 
the large acceptance corrections. E735 only covered the pseudorapidity 
range $|\eta| < 3.25$, so about half of the multiplicity is obtained 
by extrapolation from the measured region for the 1800 GeV data. 
UA5 extended further and observed 70\%--80\%, depending on energy, 
of its multiplicity. 

Obviously new experimental studies would be required to resolve 
the UA5--E735 ambiguity. As it stands, presumably a tune adjusted to
fit E735 would give disagreement with the CDF data that went 
into Tune A. In this particular case we suspect the differences 
to be of an experimental origin, but in other cases it could well be
that the \textsc{Pythia} model is incapable of fitting different 
(correct) distributions simultaneously, the model not 
being perfect. Indeed, speaking in general
terms, that is a main reason why we try to improve the model in this 
article. In this particular case and for the moment being, however, 
we choose to use the UA5-compatible Tune A as a convenient reference 
for a realistic multiplicity distribution at Tevatron energies.
 
\section{Correlations in Momentum and Flavour}
\label{s:momflav}
 
Consider a hadron undergoing multiple interactions in a collision.
Such an object should be described by multi-parton densities,
giving the joint probability of simultaneously finding $n$ partons with
flavours $f_1,\ldots,f_n$, carrying momentum fractions $x_1,\ldots,x_n$
inside the hadron, when probed by interactions at scales
$Q_1^2,\ldots,Q_n^2$. However, just like the standard
one-particle-inclusive parton densities, such distributions would
involve nonperturbative initial conditions that ultimately would have to
be pinned down by experiment. We are nowhere near such a situation:
the experimental information on double parton scattering, $n=2$, boils
down to one single number, the $\sigma_{\mrm{eff}}$ of eq.~(\ref{eq:dps}),
and for $n \geq 3$ there is no information whatsoever. Wishing to make
maximal use of the existing ($n=1$) information, we thus propose the
following strategy.
 
As described above, the interactions may be generated in an ordered
sequence of falling $\pT$. For the hardest interaction, all smaller $\pT$
scales may be effectively integrated out of the (unknown) fully correlated
distributions, leaving an object described by the standard one-parton
distributions, by definition. For the second and subsequent interactions,
again all lower-$\pT$ scales can be integrated out, but the correlations 
with the first cannot, and so on.
 
\begin{figure}
\begin{center}
\begin{fmffile}{fmfptordering}
\begin{fmfgraph*}(330,160)
\fmfstraight
\fmfright{r1,r2,r3,r4,r5}
\fmftop{t0,t1,t12,t2,t23,tddd,tn,t4}
\fmfbottom{b0,b1,b12,b2,b23,bddd,bn,b4}
\fmf{plain,tension=0.7}{t1,v11}
\fmf{plain,tension=0.85}{v11,v12}
\fmf{plain,tension=1}{v12,v13}
\fmf{plain,tension=1.3}{v13,b1}
\fmf{phantom,tension=0.7}{t12,e11}
\fmf{phantom,tension=0.85}{e11,e12}
\fmf{phantom,tension=1}{e12,e13}
\fmf{phantom,tension=1.3}{e13,b12}
\fmf{gluon,tension=0,label=\small ISR,l.sid=left}{v11,e11}
\fmf{gluon,tension=0}{v12,e12}
\fmf{gluon,tension=0}{v13,e13}
\fmf{phantom,tension=0.85}{t2,v21}
\fmf{plain,tension=0.9}{v21,v22}
\fmf{plain,tension=1.6}{v22,b2}
\fmf{phantom,tension=0.85}{t23,e21}
\fmf{phantom,tension=0.9}{e21,e22}
\fmf{phantom,tension=1.6}{e22,b23}
\fmf{gluon,tension=0}{v22,e22}
\fmf{phantom,tension=1.3}{tddd,vddd}
\fmf{phantom}{bddd,vddd,b23}
\fmfv{lab=\large$\cdots$,l.ang=-90}{bddd}
\fmf{phantom,tension=0.35}{tn,vn1}
\fmf{plain}{vn1,bn}
\fmfv{d.sh=cross,lab=\small\sc Interaction,l.ang=90,l.di=8}{t1}
\fmfv{d.sh=cross}{v11}
\fmfv{d.sh=cross}{v12}
\fmfv{d.sh=cross}{v13}
\fmfv{d.sh=cross}{v21}
\fmfv{d.sh=cross}{v22}
\fmfv{d.sh=cross}{vn1}
\fmfforce{0.0w,0.07h}{q01}
\fmfforce{1w,0.07h}{q02}
\fmfforce{0.0w,0.595h}{q11}
\fmfforce{1w,0.595h}{q12}
\fmfforce{0.0w,0.26h}{q21}
\fmfforce{1w,0.26h}{q22}
\fmf{dashes}{q11,q12}
\fmf{dashes}{q21,q22}
\fmfv{label=$p_{\perp2}$,l.ang=180}{q11}
\fmf{phantom,lab=$\vdots$,lab.sid=right,lab.dist=5}{q11,q21}
\fmfv{label=$p_{\perp n}$,l.ang=180}{q21}
\fmfforce{0.03w,0h}{y1}
\fmfforce{0.03w,1h}{y2}
\fmfforce{0w,0.02h}{x1}
\fmfforce{1w,0.02h}{x2}
\fmf{plain}{y1,y2}
\fmfv{d.sh=triangle,d.fil=full,d.siz=0.03w,lab=$\pT$}{y2}
\fmf{plain}{x1,x2}
\fmfv{d.sh=triangle,d.fil=full,d.siz=0.03w,d.ang=270,lab=$i$}{x2}
\fmfv{lab=$1$,l.ang=-90}{b1}
\fmfv{lab=$2$,l.ang=-90}{b2}
\fmfv{lab=$n$,l.ang=-90}{bn}
\end{fmfgraph*}\vspace*{3mm}
\end{fmffile}
\caption{Schematic representation of the evolution of parton shower
  initiators in a hadron collision with $n$ interactions (see text).
\label{fig:ptordering}}
\end{center}
\end{figure}
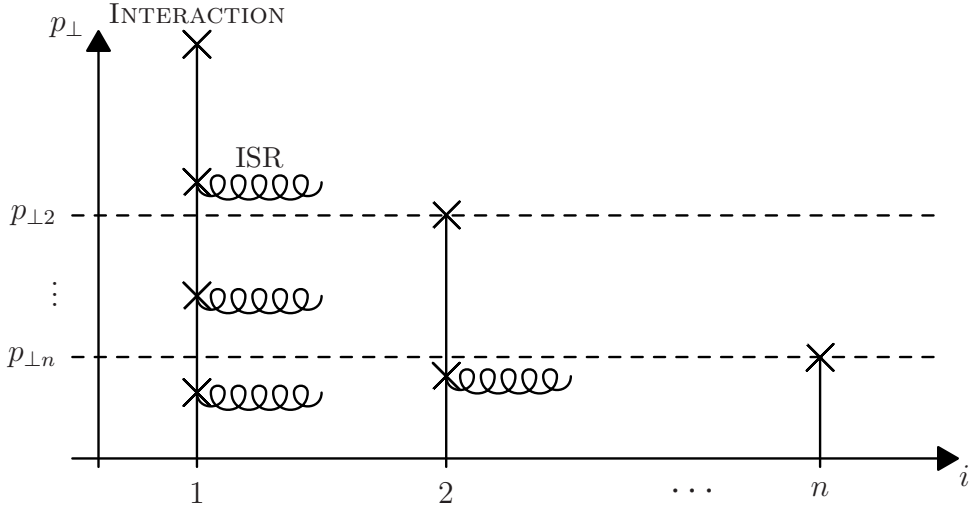

The general situation is
depicted in Fig.~\ref{fig:ptordering}.
This illustrates how, for the $i$'th interaction, only the correlations with
  the $i-1$ previous interactions need be taken into account, with all lower
  $\pT$ scales integrated out. Note, however, that this is only strictly
  true for the hard scatterings themselves. The initial-state shower
  evolution of, say, the first interaction, should exhibit
  correlations with the $i$'th at scales smaller than $p_{\perp i}$. Thus,
  the $\pT$ ordering (or equivalently, a virtuality ordering) is in some
  sense equivalent to a time ordering, with the harder physics being able to
  influence the softer physics, but not vice versa. For two interactions of
  comparable $\pT$ this order may appear quite arbitrary, and also should not
  matter much, but consider the case of one very hard and one very soft
  interaction. The soft one will then correspond to a long formation time
  (field regeneration time) \cite{DKMT}, $\sim p/\pT^2 \sim 1/\pT$, and
  indeed it is to be expected that the hard one can pre-empt or at least
  modify the soft one, whereas the influence in the other direction would be
  minor. This gives additional motivation to the choice of a $\pT$ ordering
  of interactions.
 
The possibility of intertwined shower evolution is not (yet) addressed.
Rather, we introduce modified parton densities, that correlate
the $i$'th interaction and its shower evolution to what happened in the $i-1$
previous ones, but we do not let the previous showers be affected by
what
happens in subsequent interactions. As partons are successively removed from
the hadron by hard scatterings
at smaller and smaller $\pT$ scales, the flavour, momentum and colour
structure of the remaining object changes. The colour structure in particular
is a thorny issue and will be discussed separately, in the next Section.
Here, we focus on deriving a set of parton distributions for a hadron
after an arbitrary number of interactions have occurred, on the need for
assigning a primordial transverse momentum to shower initiators, and on the
kinematics of the partons residing in the final beam remnants.
 
Our general strategy is thus to pick a succession of hard interactions and to
associate each interaction with initial- and final-state shower activity,
using the parton densities introduced below. The initial-state shower is
constructed by backwards evolution \cite{backw}, back to the shower initiators
at some low $Q_0$ scale, the parton shower cutoff scale.
Thus, even if the hard scattering does not
involve a valence quark, say, the possibility exists that the shower will
reconstruct back to one. This necessitates
dealing with quite complicated beam remnant structures. For instance, if two
valence quarks have been knocked out of the same baryon in different
directions, there will be three quarks, widely separated in momentum space, of
which no two may naturally be collapsed to form a diquark system.
 
In the old model, technical
limitations in the way the fragmentation was handled made it impossible
to address such remnant systems.
Consequently, it was not possible to associate
initial-state radiation with the interactions after the first, i.e.~the one
with the highest $p_\perp$ scale, and only a very limited set of $\q\qbar$
and $\g\g$ scatterings were allowed.
 
In a recent article \cite{BNV}, the Lund string model was augmented
to include string systems carrying non-zero baryon number, by the
introduction of `junction fragmentation'. In the context of
multiple interactions, this improvement means that almost arbitrarily
complicated beam remnants may now be dealt with. Thus,
a number of the restrictions that were present in the old
model may now be lifted.
 
\subsection{Parton Densities}
 
As mentioned above, we take the standard parton density functions as our
starting point in constructing parton distributions for the remnant
hadronic object after one or several interactions have occurred. Based
on considerations of momentum and flavour conservation we then introduce
successive modifications to these distributions.
 
The first and most trivial observation is that each interaction $i$ removes a
momentum fraction $x_i$ from the hadron remnant. This is the fraction carried
by the initiator of the initial-state shower, at the shower cutoff scale
$Q_0$, so that the two initiators of an interaction together carry all the
energy and momentum eventually carried by the hard scattering and
initial-state shower products combined.  To take into account that the total
momentum of the remaining object is thereby reduced, already in the old model
the parton densities were assumed to scale such that the point $x = 1$ would
correspond to the remaining momentum in the beam hadron, rather than the
total original beam momentum, cf.~eq.~(\ref{eq:xresc}). In addition to this
simple $x$ scaling ansatz we now introduce the possibility of genuine and
non-trivial changes in both shape and normalization of the distributions.
 
\subsubsection{Valence Quarks}
 
Whenever a valence quark is knocked out of an incoming hadron,
the number of remaining valence
quarks of that species should be reduced accordingly. Thus, for a proton, the
valence d distribution is completely removed if the valence $\d$ quark has
been kicked out, whereas the valence $\u$ distribution is halved when one of
the two is kicked out. In cases where the valence and sea $\u$ and $\d$ quark
distributions are not separately provided from the PDF libraries, we assume
that the sea is flavour-antiflavour symmetric, so that one can write e.g.
\begin{equation}
  u(x,Q^2) = u_{\val}(x,Q^2) + u_{\sea}(x,Q^2) =
  u_{\val}(x,Q^2) + \overline{u}(x,Q^2).
\end{equation}
Here and in the following, $q_{\val}$ ($q_{\sea}$) denotes the $\q$ valence
(sea) distribution.  The parametrized $u$ and $\overline{u}$ distributions
should then be used to find the relative probability for a kicked-out $\u$
quark to be either valence or sea. Explicitly, the quark valence distribution
of flavour $f$ after $n$ interactions, $q_{f\val n}(x,Q^2)$,
is given in terms of the initial distribution, $q_{f\val0}(x,Q^2)$,
and the ratio of remaining to
original $\q_f$ valence quarks, ${N_{f\val n}}/{N_{f\val 0}}$, as:
\begin{equation}\displaystyle
  q_{f\val n}(x,Q^2) = \frac{N_{f\val n}}{N_{f\val 0}}
  \frac{1}{X}q_{f\val0}\left(\frac{x}{X},Q^2\right) ~~~;~~X =
  1-\sum_{i=1}^{n}x_i, \label{eq:valdist}
\end{equation}
where $N_{u\val 0}=2$ and $N_{d\val 0}=1$ for the proton, and $x\in[0,X]$ is
the fraction of the original beam momentum ($\sum_{i=1}^{n}x_i$ is the total
momentum fraction already taken out of the incoming hadrons by the preceding
parton-shower initiators). The $Q^2$ dependence of $q_{f\val n}$ is
inherited from the standard parton densities $q_{f\val0}$, and this
dependence is reflected both in the choice of a hard scattering and in the
backwards evolution. The factor $1/X$ arises since we squeeze the
distribution in $x$
while maintaining its area equal to the number of $\q_f$ valence quarks
originally in the hadron, ${N_{f\val 0}}$, thereby ensuring that the
sum rule,
\begin{equation}
  \int_{0}^{X}q_{f\val n}(x,Q^2)~\d x = N_{f\val n}, \label{eq:valnumrule}
\end{equation}
is respected. There is also the total momentum sum rule,
\begin{equation}
  \int_0^{X}\!\left(\sum_f q_{fn}(x,Q^2) + g_n(x,Q^2)\right)x~\d x =
  X. \label{eq:totmomrule}
\end{equation}
Without any further change, this sum rule would not be respected since, by
removing a valence quark from the parton distributions in the above manner,
we also remove a total amount of momentum corresponding to
$\langle x_{f\val} \rangle$, the average momentum fraction carried by a
valence quark of flavour $f$:
\begin{equation}
\langle x_{f\val n}(Q^2) \rangle \equiv \frac{\int_0^X
q_{f\val n}(x,Q^2)~x~\d x}{\int_0^X q_{f\val n}(x,Q^2)~\d x}
= X \, \langle x_{f\val 0}(Q^2) \rangle ~.
\label{eq:avmom}
\end{equation}
 
The removal of $\sum_ix_i$, the total momentum carried by the
previously struck partons, has already been taken into account by
the `squeezing' in $x$ of the parton distributions (and expressed
in eq.~(\ref{eq:totmomrule}) by the RHS being equal to $X$ rather
than 1). By scaling down the $\q_\val$ distribution, we are
removing an \emph{additional} fraction, $\langle x_{f\val n}
\rangle$, which must be put back somewhere, in order to maintain
the validity of eq.~(\ref{eq:totmomrule}).
 
Strictly speaking, $\langle x_{f\val 0} \rangle$ of course depends on which
specific PDF set is used. Nevertheless, for the purpose at hand this
variation is negligible between most modern PDF sets.  Hence we make the
arbitrary choice of restricting our attention to the values obtained with the
CTEQ5L PDF set \cite{CTEQ5}.
 
More importantly, all the above parton densities depend on the
factorization scale $Q^2$. This dependence of course carries over to
$\langle x_{f\val 0} \rangle$, for which we assume the functional form
\begin{equation}
  \langle x_{f\val 0}(Q^2) \rangle = \frac{A_f}{1+B_f\log\left(\log
  (\mathrm{max}(Q^2,1~\mathrm{GeV}^2)/\Lambda_{\mathrm{QCD}}^2)\right)},
\end{equation}
inspired by the
$\d s = \d\log(\log Q^2/\Lambda^2) \propto \d Q^2/Q^2 \, \alphas(Q^2)$
pace of evolution, where
$\d \langle x \rangle / \d s \approx - B \langle x \rangle$
suggests a solution of the form
$\langle x \rangle \propto \exp(-B s) \approx 1/(1 + B s)$. Reasonable fits
to the CTEQ5L valence quark distributions in the proton are obtained for
$A_{\d}  =  0.385$, $B_{\d} = 1.60$, $A_{\u} = 0.48$ and $B_{\u} = 1.56$,
with the isospin conjugate for neutrons.
 
Essentially nothing is known about parton densities for other baryons,
such as the reasonably long-lived hyperons $\Lambda^0$, $\Sigma^{+,-}$,
$\Xi^{0,-}$ and $\Omega^-$, which can undergo secondary interactions that
one may wish to study. We here use essentially the same parton densities
and parameters as for protons. Thus the influence of the larger strange
quark mass is neglected, which ought to lead to harder $x$ spectra for
$\s$ quarks and softer for everything else.
The fact that the two proton valence $\u$ quarks have a harder distribution
than the single $\d$ one is carried over to other baryons with two equal
quarks, while the average (sum) of the $\u$ and $\d$ distributions are
used for baryons with three unequal (equal) quarks. For mesons and the
Vector Meson Dominance part of photons one could use a similar strategy,
with $\pi^+$ measurements as a starting point instead of protons, while
the anomalous part of the photon densities is perturbatively calculable.
There are still many further assumptions that would have to go into a
complete model of multiple interactions in $\gamma\p$ and $\gamma\gamma$
events \cite{pythiagamma}, however, and so far we did not pursue this further.
 
We now know how much momentum is `missing' in eq.~(\ref{eq:totmomrule}).
It is not possible to put this momentum back onto valence quarks without
changing the shape of the distributions (beyond the mere `$x$ squeezing')
or invalidating eq.~(\ref{eq:valnumrule}). Rather, we here assume that the
missing momentum is taken up by the sea+gluon distributions, which
thus are scaled up slightly when a valence quark is kicked out. This
enhancement of the sea+gluon momentum fraction may over- or undercompensate
the `$x$ squeezing' reduction, depending on whether the kicked-out valence
quark had a small or large $x$. However, before the procedure can be
discussed in more detail, we must consider another effect which affects
the normalization of the sea: changes in the content of the sea itself.
 
\subsubsection{Sea Quarks and their Companions}
 
When a sea quark is kicked out of a hadron,
it must leave behind a corresponding
antisea parton in the beam remnant, by flavour conservation. We call
this a companion quark.
In the perturbative approximation the sea quark $\qsea$ and its
companion $\qcmp$ come from a gluon branching
$\g \to \qsea + \qcmp$, where it is implicitly understood that if \qsea\ is
a quark, \qcmp\ is its antiquark, and vice versa.
This branching often would not be in the perturbative
regime, but we choose to make a perturbative ansatz, and also to
neglect subsequent perturbative evolution of the $q_{\cmp}$
distribution. Even if approximate, this procedure should catch
the key feature that a sea quark and its companion should not be expected too
far apart in $x$ (or, better, in $\ln x$).
 
With this approximation, we obtain the $q_{\cmp}$ distribution from
the probability that a sea quark $\qsea$, carrying a
momentum fraction $x_{\sea}$, is produced by the branching of a
gluon with momentum fraction $y$, so that the
companion has a momentum fraction $x=y-x_{\sea}$,
\begin{eqnarray}
q_{\cmp}(x;x_{\sea}) & = & C\int_0^1
g(y) \, P_{\g\to\qsea\qcmp}(z) \, \delta(x_{\s}-zy)~\d z
\nonumber \\
& = & C~g(y) \, P_{\g\to\qsea\qcmp}\left(\frac{x_{\s}}{y}\right)\frac{1}{y}
\nonumber \\
& = & C~\frac{g(x_{\s}+x)}{x_{\s}+x} \,
P_{\g\to\qsea\qcmp}\left(\frac{x_{\s}}{x_{\s}+x}\right),
\label{eq:qcmp}
\end{eqnarray}
with $C$ a normalization constant to be determined below, and
$P_{\g\to\qsea\qcmp}$ the DGLAP splitting kernel
\begin{eqnarray}
 P_{\g\to \qsea\qcmp}(z) &  = & \frac{1}{2} \left( z^2 + (1-z)^2 \right).
\label{eq:Pgqq}
\end{eqnarray}
In view of the approximate nature of the procedure, allowing a generic
$g(x)$ shape would give disproportionately complex expressions.
Instead, the following simple ansatz for the gluon distribution at low
$Q^2$ is used:
\begin{eqnarray}
 g(x) & \propto & \frac{(1-x)^p}{x} ~,
\label{eq:ga}
\end{eqnarray}
with the integer choices $p=0,1,2,3,4$ giving a range of
variability for the large-$x$ behaviour of the distribution and
the $1/x$ controlling the small-$x$ behaviour. Note that all the
above equations are defined assuming no previous energy loss and,
as for the valence quarks, should be `squeezed' by a factor $X$,
to ensure momentum conservation.
 
The overall normalization of a companion quark distribution is
obtained by imposing the sum rule:
\begin{equation}
\int_0^{1-x_{\sea}}\!q_{\cmp}(x;x_{\sea})~\d x = 1.
\end{equation}
Inserting eqs.~(\ref{eq:qcmp})--(\ref{eq:ga}) and inverting, one obtains the
normalization constants $C_p$:
\begin{eqnarray}
C_0 & = & \frac{3x_{\s}}{2-x_{\s}(3-3x_{\s}+2x_{\s})} \label{eq:C0}
~, \\
C_1 & = & \frac{3x_{\s}}{2-x_{\s}^2(3-x_{\s})+3x_{\s}\log(x_{\s})}
~, \\
C_2 & = & \frac{3x_{\s}}{2(1-x_{\s})(1+4x_{\s}+4x_{\s}^2)+
6x_{\s}(1+x_{\s})\log(x_{\s})} ~, \\
C_3 & = & \frac{6x_{\s}}{4+x_{\s}(27-31x_{\s}^2)+
  6x_{\s}(3+2x_{\s}(3+x_{\s}))\log(x_{\s})} ~, \\
C_4 & = &
\frac{3x_{\s}}{2(1+2x_{\s})((1-x_{\s})(1+x_{\s}(10+x_{\s}))+
6x_{\s}(1+x_{\s})\log(x_{\s}))}\label{eq:C4} ~.
\end{eqnarray}
 
\begin{figure}[t]
\begin{center}
\vspace*{-7mm}\includegraphics*[scale=0.75]{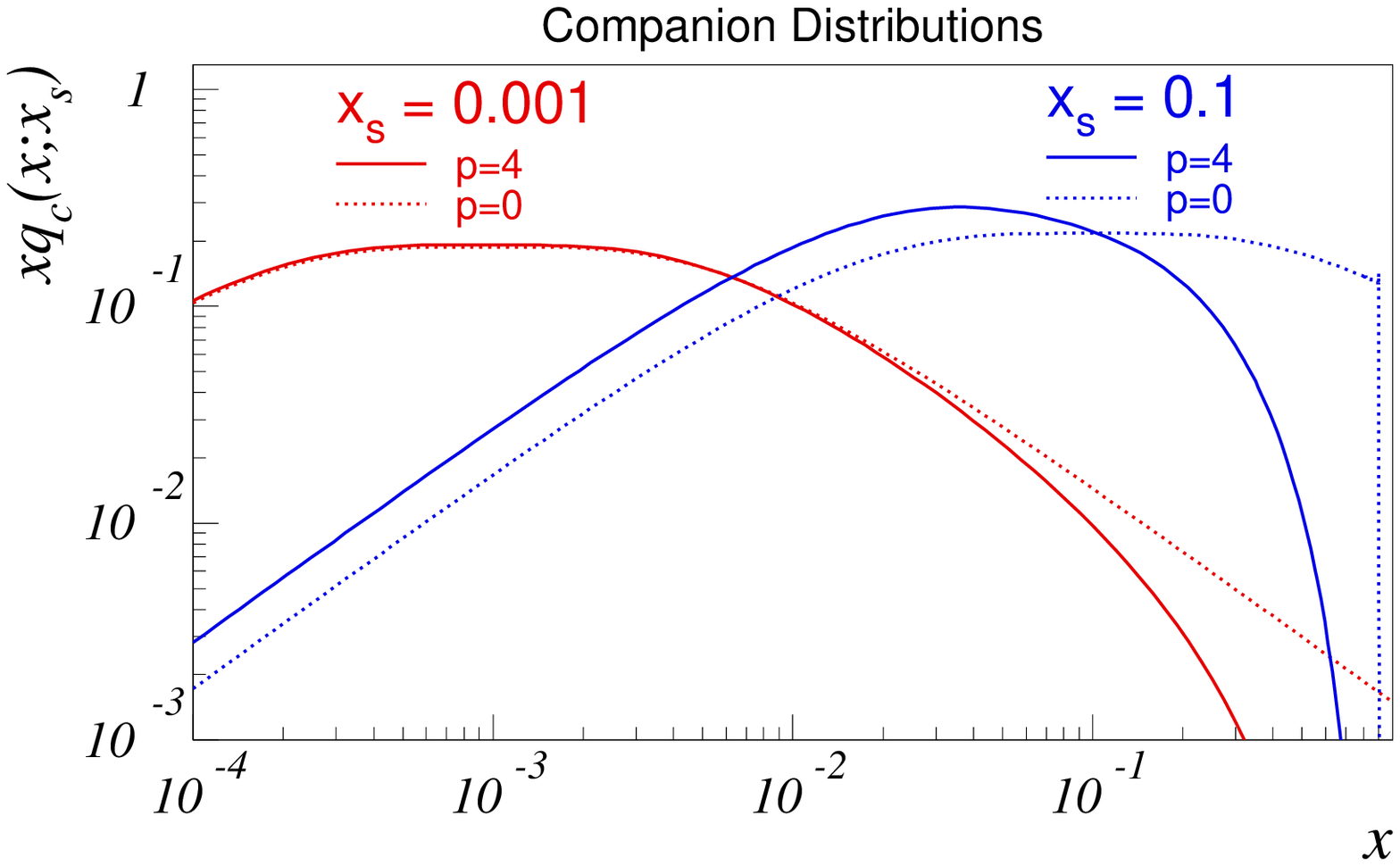}\vspace*{-1mm}
\caption{Companion distributions for $p=4$ (solid lines) and
$p=0$ (dotted lines), for two different values of $x_{\s}$.
\label{fig:compdist}}
\end{center}
\end{figure}
 
To illustrate, Fig.~\ref{fig:compdist} shows properly normalized
companion momentum distributions for $p=4$ and $p=0$, each for two
different values of $x_{\s}$. There are two noteworthy aspects
about these distributions. Firstly, for $p=0$ the discontinuity
at the point $x=1-x_{\s}$ is merely an artifact of our
parametrization of the gluon density, i.e.\ that $g(x)$ does not
vanish at $x=1$. That problem is absent for more realistic $p$
values; in our continued discussions we will use $p=4$ as default,
this being the closest to the CTEQ5L small-$Q^2$ gluon
distribution. Secondly, the falling gluon distribution convoluted
with the almost flat $\g\to\q\qbar$ splitting kernel give
distributions that roughly tend to a constant $q_{\cmp}(x;x_s)\sim
C_p/2x_{\s}^2$ below $x_{\s}$ and exhibit power-like fall-offs
$q_{\cmp}(x;x_s)\propto 1/x^2$ above it, with some modulation of
the latter depending on $p$. In order to display the probability
per $\log x$ interval, Fig.~\ref{fig:compdist} gives $x\,q_{\cmp}$
rather than $q_{\cmp}$ itself, and then a peaking occurs around $x
\approx x_{\sea}$, as should be expected from the symmetric
splitting kernel in eq.~(\ref{eq:Pgqq}).
 
Also here, the question arises of ensuring
that the total momentum sum rule, eq.~(\ref{eq:totmomrule}), is
respected, but now the difference has the opposite sign; by adding a
companion quark distribution, we are in effect bookkeeping a part of the
flavour and momentum content of the sea separately. One possibility is
that this momentum comes \emph{only} from the sea+gluons and that the
valence quarks are not affected, i.e.~that the rest of the sea+gluons
fluctuate down, in order to compensate.
 
The amount of momentum that will have to be compensated for each companion
quark, $\langle x_{\cmp n} \rangle = X \langle x_{\cmp 0} \rangle$, with
$\langle x_{\cmp n} \rangle$ defined in analogy with eq.~(\ref{eq:avmom}), is
straightforward to compute using the distribution in eq.~(\ref{eq:qcmp}) and
the normalizations given by eqs.~(\ref{eq:C0})--(\ref{eq:C4}):
\begin{eqnarray}
\langle x_{\cmp 0} \rangle_{p=0} \hspace*{-1mm} & = &
  \hspace*{-1mm}x_{\s}\frac{5-9x_{\s} +6 x_{\s}^2 - 2x_{\s}^3 +
3 \log x_{\s}}{(x_{\s}-1) (2-x_{\s}+2x_{\s}^2)} ~, \\[2mm]
\langle x_{\cmp 0} \rangle_{p=1} \hspace*{-1mm} & = &
  \hspace*{-1mm} -1-3 x_{\s}+\frac{2 {{(1-x_{\s})}^2} (1+x_{\s}+{x_{\s}^2})}%
  {2-3x_{\s}^2 +x_{\s}^3 +3 x_{\s}\log x_{\s}} ~, \\[2mm]
\langle x_{\cmp 0} \rangle_{p=2}\hspace*{-1mm} & = &
\hspace*{-1mm} \frac{x_{\s}}{4} \, \frac{
  19+24x_{\s} - 39x_{\s}^2-4x_{\s}^3+6 (1+6 x_{\s}+4 {x_{\s}^2})
  \log x_{\s}}{-1 -3 x_{\s} + 3x_{\s}^2 + x_{\s}^3-3 x_{\s} (1+x_{\s})
  \log x_{\s}} ~,\\[2mm]
\langle x_{\cmp 0} \rangle_{p=3}\hspace*{-1mm} & = & \hspace*{-1mm}
3 x_{\s} \frac{-7-21x_{\s} + 15 x_{\s}^2 + 13 x_{\s}^3
- 2 (1+ 9x_{\s}+12 x_{\s}^2
  + 2x_{\s}^3) \log x_{\s}}{
4+27 x_{\s}-31 {x_{\s}^3}+6 x_{\s} (3+6 x_{\s} +2 x_{\s}^2) \log x_{\s}}
 ~,\\[2mm]
\langle x_{\cmp 0} \rangle_{p=4}\hspace*{-1mm} &=&\hspace*{-1mm}3x_{\s}\frac{
3(5+24x_{\s} - 4x_{\s}^2-24x_{\s}^3-x^4)+
4 (1+12 x_{\s} +24 x_{\s}^2 +8 x_{\s}^3) \log x_{\s}}{
8 (1+2 x_{\s}) (-1-9x_{\s} +9x_{\s}^2+x_{\s}^3-6 x_{\s} (1+x_{\s})
\log x_{\s})} ~.
\end{eqnarray}
 
\subsubsection{Sea Quark and Gluon Density Normalizations}
As described above,
the normalization of valence and companion distributions is fixed
  by the respective number of quarks, i.e.\ the sum rules (for each flavour,
  $f$)
\begin{eqnarray}
\int_0^{X}\! q_{f\val n}(x,Q^2)~\d x & = & N_{f\val n},\\
\int_0^{X}\! q_{f\cmp_j n}(x;x_{\s_j})~\d x & = & 1 ~~~~
(\mrm{for~each}~j),
\end{eqnarray}
where $X$ is still the longitudinal momentum fraction left after the
$n$ previous interactions and $N_{f\val n}$ is
the number of $\q_f$ valence quarks remaining. The index $j$ on the companion
distribution, $q_{f\cmp_j n}$, counts different companion quarks of
the same flavour, $f$.
 
On the other hand, the sea+gluon distributions do not have fixed
multiplicities, hence no corresponding sum rules exist for their
normalizations.  We use this freedom to fulfill the last remaining sum rule,
eq.~(\ref{eq:totmomrule}), letting the sea+gluon normalizations fluctuate up
when we reduce a valence distribution and down when we add a companion
distribution. In addition, the requirement of a physical $x$ range is of
course maintained by still `squeezing' all distributions into the interval
$x\in[0,X]$.
 
For simplicity, and since eq.~(\ref{eq:totmomrule}) only
furnishes us with one equation of constraint, we assume the same scale factor
for all sea flavours as well as for the gluon, i.e.~
\begin{eqnarray}
q_{f\sea}(x,Q^2) & \to & a q_{f\sea}(x,Q^2), \\
g(x,Q^2) & \to & a g(x,Q^2) .
\end{eqnarray}
The momentum sum rule now reads:
\begin{eqnarray}
1 & = & \frac{1}{X}\int_0^X\!\left(\sum_f\left[
q_{f\val n}(x,Q^2)+\sum_j q_{f\cmp_jn}(x;x_j) +
aq_{f\sea}(x,Q^2)
\right] + ag_n(x,Q^2)\right)x~\d x\nonumber\\
& = & \int_0^1\!\left(\sum_f\left[
\frac{N_{f\val n}}{N_{f\val 0}}
q_{f\val0}(x,Q^2)
+\sum_j q_{f\cmp_j 0}(x;x_j)
 + aq_{f\sea 0}(x,Q^2)
\right] + ag_0(x,Q^2)\right) x~\d x\nonumber\\
& = & a + \sum_f\int_0^1\!\left[
\left(\frac{N_{f\val n}}{N_{f\val 0}}-a\right)
q_{f\val0}(x,Q^2)
+\sum_j q_{f\cmp_j 0}(x;x_j)\right] x~\d x \nonumber\\
& = &a\left(1- \sum_fN_{f\val 0}\langle x_{f\val 0} \rangle\right) +
\sum_fN_{f\val n}\langle x_{f\val 0} \rangle
+\sum_{f,j} \langle x_{f\cmp_j 0} \rangle~,
\end{eqnarray}
and hence,
\begin{equation}
a = \frac{1-\sum_fN_{f\val n}\langle x_{f\val 0} \rangle
-\sum_{f,j} \langle x_{f\cmp_j 0} \rangle}{1- \sum_fN_{f\val 0}\langle x_{f\val
    0} \rangle}.
\end{equation}
One easily checks that $a=1$ before the first interaction, as it should be,
and that $a$ is driven larger by $N_{f\val n}<N_{f\val 0}$, while introducing
companion quarks drives it the opposite way, also as expected.
 
\subsection{Beam Remnants}
\label{sec:remnants} 

The longitudinal momenta and flavours of the initiator partons are defined by
the sequence of $\pT$-ordered hard scatterings and their associated
initial-state showers, as described above. What is left in the beam remnant
is then a number of partons, with flavours given by the remaining valence
content plus the number of sea quarks required for overall flavour
conservation. That is, gluons in the remnant are not explicitly
accounted for, but are implicit as confinement clouds around the
quarks and as unresolved originators of sea quark pairs.
 
\begin{figure}[t]
\vspace*{0.5cm}
\begin{center}
\setlength{\unitlength}{0.3mm}
\begin{fmffile}{fmfcomposite}
\begin{tabular}{p{8cm}p{5.cm}}\hspace*{-2.3cm}%
\begin{fmfgraph*}(150,50)
\fmftop{t}
\fmfbottom{b}
\fmf{phantom}{t,v,b}
\fmfv{d.sh=circ,d.f=full,d.siz=10,lab=Parton in beam remnant,l.ang=0,l.dist=15}{t}
\fmfv{d.sh=circ,d.f=empty,d.siz=10,lab=Parton going to hard
  interaction,l.ang=0,l.dist=15}{v}
\fmfv{d.sh=circ,d.f=20,d.siz=15,lab=Composite object,l.ang=0,l.dist=15}{b}
\end{fmfgraph*}\hspace*{2cm} &\vspace*{-1.5cm}%
\begin{fmfgraph*}(150,100)
\fmfleft{d1,q1,q2,d2}
\fmfright{q3}
\fmftop{d3,d4,g1,g2}
\fmf{phantom}{q1,j,q2}
\fmf{phantom,tension=0.4}{j,v1,v2,q3}
\fmffreeze
\fmfv{d.sh=circ,d.f=20,d.siz=0.6h,lab=$\q\q$,lab.dist=30}{j}
\end{fmfgraph*}\hspace*{-150\unitlength}%
\begin{fmfgraph*}(150,100)
\fmfleft{d1,q1,q2,d2}
\fmfright{q3}
\fmftop{d3,d4,g1,g2}
\fmfbottom{b}
\fmf{plain}{q1,j,q2}
\fmf{plain,tension=0.4}{j,v1,v2,q3}
\fmffreeze
\fmf{gluon}{v1,g1}
\fmf{gluon}{v2,g2}
\fmfv{d.sh=circ,d.f=full,d.siz=0.1h,lab=$\q_{\val1}$,l.ang=0}{q1}
\fmfv{d.sh=circ,d.f=full,d.siz=0.1h,lab=$\q_{\val2}$,l.ang=0}{q2}
\fmfv{d.sh=circ,d.f=full,d.siz=0.1h,lab=$\q_{\val3}$,l.ang=0}{q3}
\fmfv{d.sh=circ,d.f=empty,d.siz=0.1h,lab=$\g_1$}{g1}
\fmfv{d.sh=circ,d.f=empty,d.siz=0.1h,lab=$\g_2$}{g2}
\fmfv{lab=\it a),l.dist=5,l.ang=90}{b}
\end{fmfgraph*}\vspace*{8mm}\\\vspace*{-125\unitlength}
\begin{fmfgraph*}(150,100)
\fmfleft{d1,q1,q2,d2}
\fmfright{q3}
\fmftop{d6,q5,d7,q4,g2}
\fmf{phantom}{q1,j,q2}
\fmf{phantom}{j,v1}
\fmf{phantom,tension=0.42}{v1,v2,q3}
\fmffreeze
\fmf{phantom,tension=0.4}{q5,vd2}
\fmf{phantom}{vd2,j}
\fmfv{d.sh=circ,d.f=20,d.siz=0.9h,lab=B,lab.dist=43}{vd2}
\end{fmfgraph*}\hspace*{-150\unitlength}%
\begin{fmfgraph*}(150,100)
\fmfleft{d1,q1,q2,d2}
\fmfright{q3}
\fmftop{d6,q5,d7,q4,g2}
\fmfbottom{b}
\fmfv{lab=\it b),l.dist=5,l.ang=90}{b}
\fmf{plain}{q1,j,q2}
\fmf{plain}{j,v1}
\fmf{plain,tension=0.4}{v1,v2,q3}
\fmffreeze
\fmf{gluon,tension=0.8}{v1,vqq}
\fmf{plain,tension=0.2}{vqq,q5}
\fmf{plain,tension=0.4}{vqq,q4}
\fmf{gluon}{v2,g2}
\fmfv{d.sh=circ,d.f=full,d.siz=0.1h,lab=$\q_{\val1}$,l.ang=0}{q1}
\fmfv{d.sh=circ,d.f=full,d.siz=0.1h,lab=$\q_{\val2}$,l.ang=0}{q2}
\fmfv{d.sh=circ,d.f=full,d.siz=0.1h,lab=$\q_{\val3}$,l.ang=0}{q3}
\fmfv{d.sh=circ,d.f=full,d.siz=0.1h,lab=$\q_{\cmp}$,l.ang=180}{q5}
\fmfv{d.sh=circ,d.f=empty,d.siz=0.1h,lab=$\g$}{g2}
\fmfv{d.sh=circ,d.f=empty,d.siz=0.1h,lab=$\qbar_{\sea}$}{q4}
\end{fmfgraph*} &
\begin{fmfgraph*}(150,100)
\fmfleft{d1,q1,q2,d2}
\fmfright{q3}
\fmftop{d6,g2,q5,d7,q4}
\fmf{phantom}{q1,j,q2}
\fmf{phantom}{j,v1}
\fmf{phantom,tension=0.4}{v1,vd6,v2,q3}
\fmffreeze
\fmf{phantom,tension=0.8}{v2,vqq}
\fmf{phantom,tension=0.2}{vqq,q5}
\fmf{phantom,tension=0.4}{vqq,q4}
\fmf{phantom}{q4,vd2}
\fmf{phantom,tension=1.25}{vd2,q3}
\fmf{phantom,tension=0.5}{vqq,vd2}
\fmfv{d.sh=circ,d.f=20,d.siz=0.75h,lab=M,lab.dist=35}{vd2}
\end{fmfgraph*}\hspace*{-150\unitlength}%
\begin{fmfgraph*}(150,100)
\fmfleft{d1,q1,q2,d2}
\fmfright{q3}
\fmftop{d6,g2,q5,q4}
\fmfbottom{b}
\fmfv{lab=\it c),l.dist=5,l.ang=90}{b}
\fmf{plain}{q1,j,q2}
\fmf{plain}{j,v1}
\fmf{plain,tension=0.75}{v1,vd6,v2,q3}
\fmffreeze
\fmf{gluon,tension=0.8}{v2,vqq}
\fmf{plain,tension=0.2}{vqq,q5}
\fmf{plain,tension=0.4}{vqq,q4}
\fmf{gluon}{v1,g2}
\fmfv{d.sh=circ,d.f=full,d.siz=0.1h,lab=$\q_{\val1}$,l.ang=0}{q1}
\fmfv{d.sh=circ,d.f=empty,d.siz=0.1h,lab=$\q_{\val2}$,l.ang=0}{q2}
\fmfv{d.sh=circ,d.f=full,d.siz=0.1h,lab=$\q_{\val3}$,l.ang=0}{q3}
\fmfv{d.sh=circ,d.f=empty,d.siz=0.1h,lab=$\q_{\sea}$,l.ang=180}{q5}
\fmfv{d.sh=circ,d.f=empty,d.siz=0.1h,lab=$\g$}{g2}
\fmfv{d.sh=circ,d.f=full,d.siz=0.1h,lab=$\qbar_{\cmp}$,l.ang=0}{q4}
\end{fmfgraph*}
\end{tabular}
\end{fmffile}\vspace*{-.4cm}
\caption{Examples of the formation of composite objects in a baryon beam
  remnant: (\textit{a}) diquark, (\textit{b}) baryon and (\textit{c})
           meson.\label{fig:composite}}
\end{center}
\end{figure}
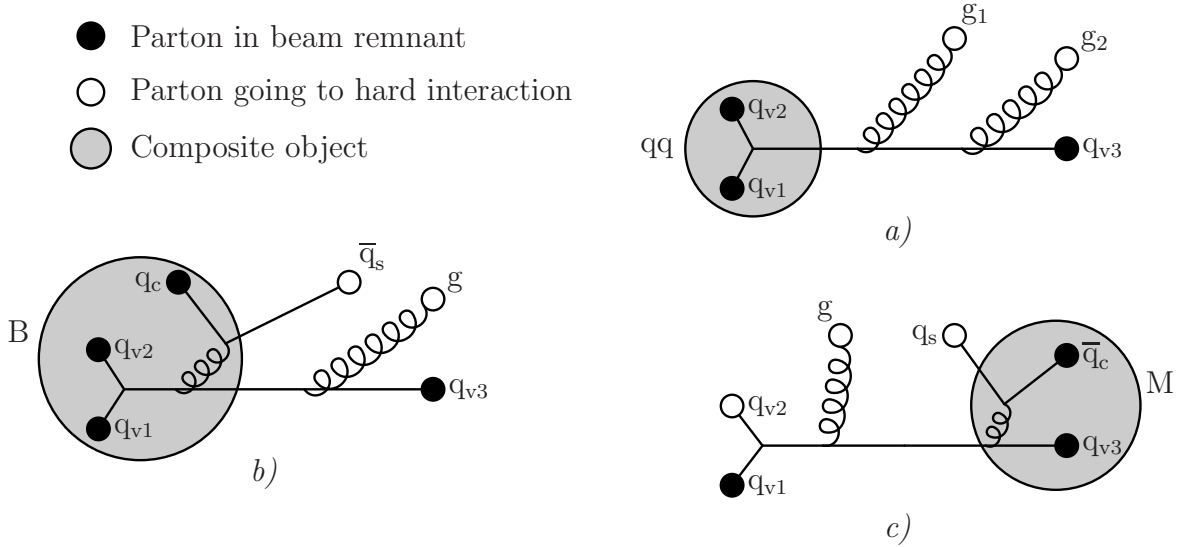
 
A remnant may thus contain several objects but, when the colour configuration
is studied, simplifications can occur. A colour antitriplet $\q\q$ pair in the
remnant can be associated with a diquark, a colour singlet $\q\q\q$ triplet
with a baryon, and a colour singlet $\q\qbar$ pair with a meson, see
Fig.~\ref{fig:composite}. When hadrons are formed, the standard string
fragmentation relative probabilities are used to select spin and other
quantum numbers, i.e.\ whether $\pi$ or $\rho$, etc.
 
It is here assumed that the respective pair/triplet has a sufficiently
small invariant mass that it can reasonably be projected onto a single
composite state. Thus a $\q\qbar$ system with large invariant mass would
define a string that could fragment into several mesons, rather than collapse
to a single meson. In principle this could be modeled dynamically, but it
would
require the introduction of some nonperturbative parameters, to describe the
partitioning of the proton into arbitrary--mass subsystems. At this point,
we consider it meaningful only to study a few specific scenarios for which
partons to allow in the formation of composite objects. We have chosen four
such: 
\begin{enumerate}
\item No composite objects are formed \emph{ab initio}. All partons act as
  single units, either as endpoints (quarks) or kinks (gluons) on strings
  that fragment in the normal way.
\item Composite objects may be formed, but only when all partons involved in
  the formation are valence quarks.
\item The formation of diquarks may involve both valence and sea quarks,
but the formation of colour singlet subsystems (i.e.~hadrons) is
still restricted to involve valence quarks only.
\item Sea quarks may also be used for colour singlet formation.
\end{enumerate}
The idea is thus that (spectator) valence quarks tend to have comparable
velocities, while sea quarks can be more spread out and therefore are less
likely to form low-mass systems.
 
\begin{figure}%
\begin{center}\hspace*{-0.9cm}
\includegraphics*[scale=0.73]{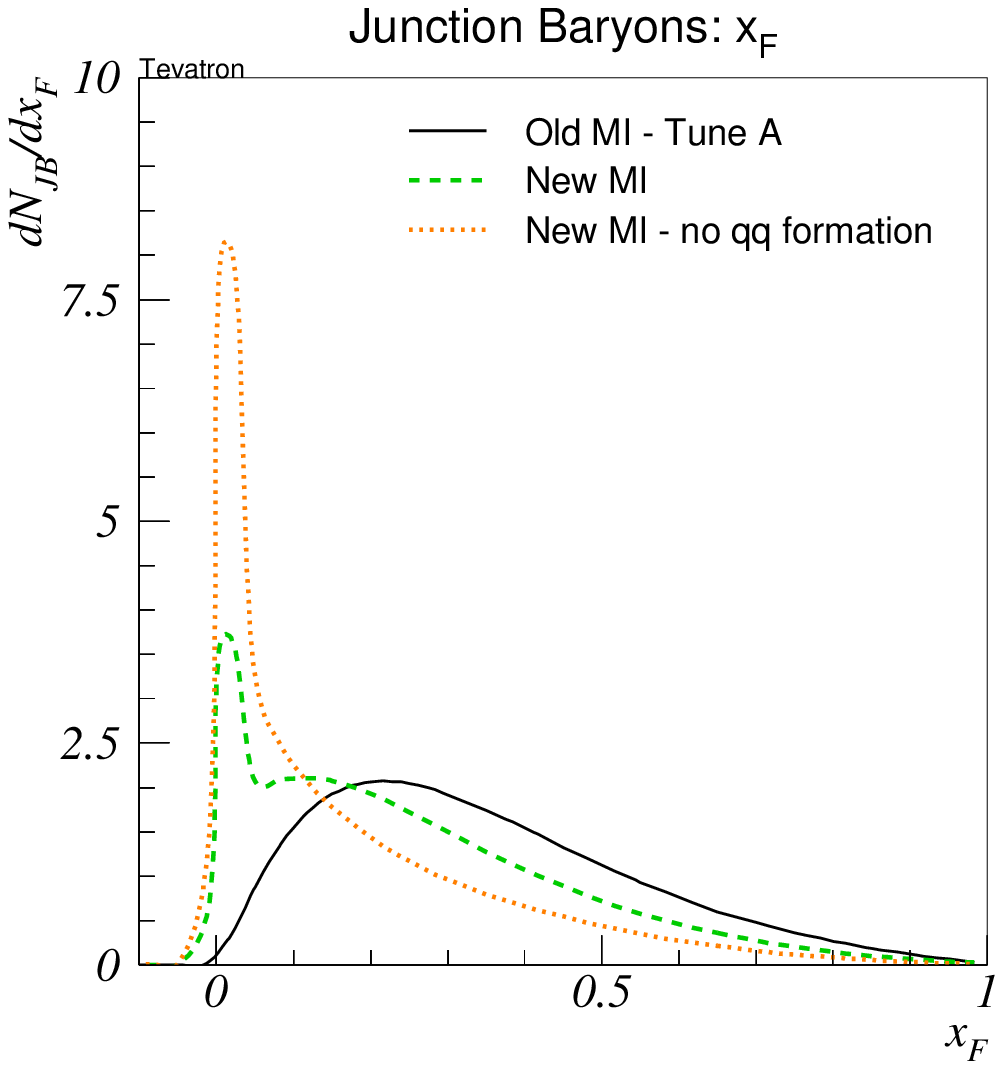}%
\includegraphics*[scale=0.73]{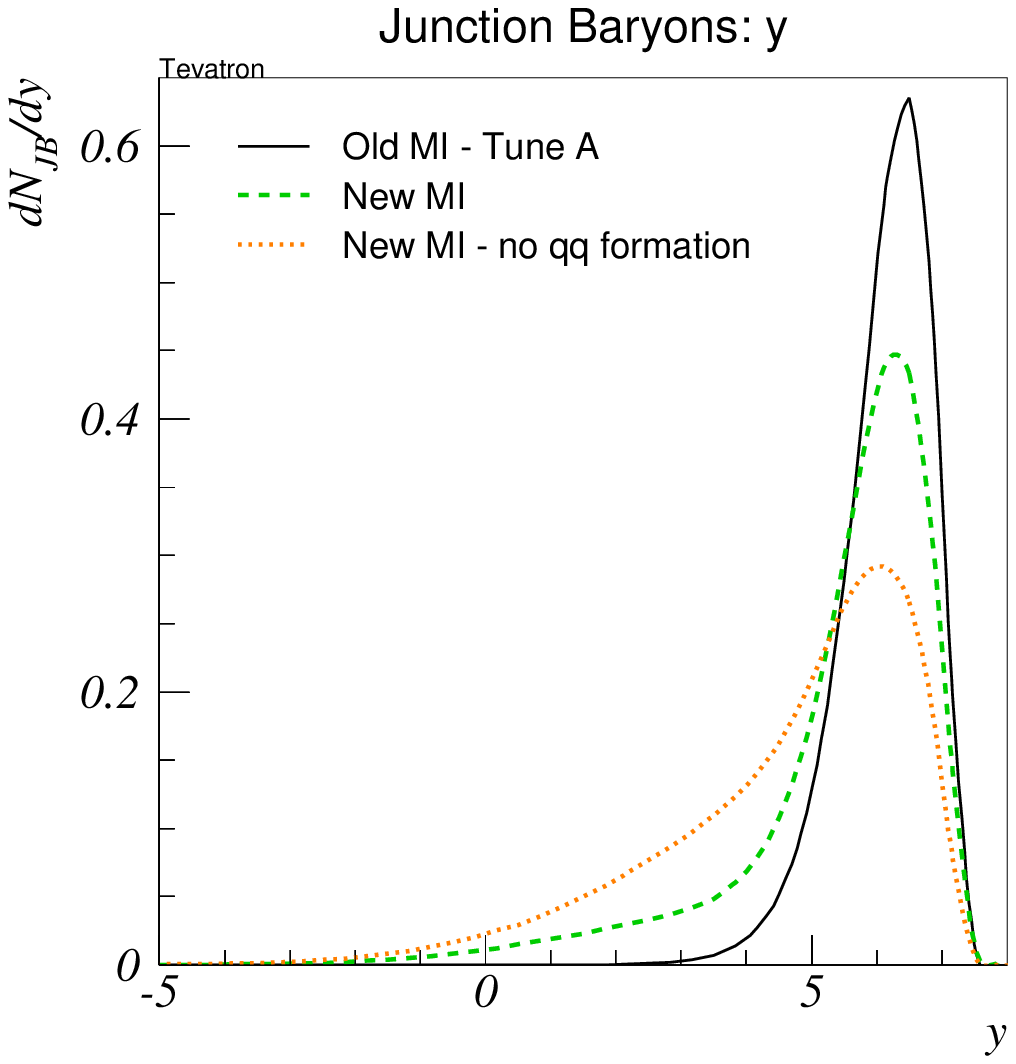}\vspace*{-.2cm}
\\
\caption{Feynman $x$ (left) and rapidity (right) 
distributions for junction baryons: distributions are
  shown for Tune A of the old MI scenario (solid lines), and for the new
  model with diquark formation in the beam remnant switched on (dashed)
  and off (dotted). \label{fig:compflow}}
\end{center}
\end{figure}

Whether composite systems in the beam remnant are formed or not has important
consequences for the baryon number flow. For $\p\pbar$ collisions at 1.8~TeV
CM energy, we show in Fig.~\ref{fig:compflow} the
Feynman $x$ (left plot) and rapidity (right plot) distributions
for the baryon which `inherits' the beam baryon number.
We denote this baryon the `junction baryon'. To better illustrate what
happens to each of the two initial beam baryon numbers separately, only
distributions for the junction baryon, not \emph{anti}baryon, are
shown. Possibilities 1 and 2 above are compared with 
the old multiple interactions model (Tune A). 
One immediately observes that the beam
baryon number migrates in a radically different way when diquark formation is
allowed or not (compare the dashed and dotted sets of curves). In
fact, in the new model it is not possible to reproduce the old
distribution (compare the solid curve). This comes about
since, even when all possible diquark formation is allowed in the new
model, it is not certain that the beam remnant actually contains the
necessary quark content, hence in some fraction of the
events the formation of a beam remnant diquark is simply not possible. 
Here is thus an example where the introduction of more physics into
the model has given rise to a qualitatively different expectation: the beam
baryon number appears to be stopped to a larger extent than would previously
have been expected.
 
One should note that, also at later stages, a small-mass string piece can
collapse to a single hadron, as part of the normal string fragmentation
procedure. There, however, it is intended only to cover a rare low-mass
tail of systems mainly defined by hard processes and perturbative shower
evolution, while the simplifications considered for the above kinematical
configurations are quite common and in a nonperturbative context. Beam
remnant quarks that were not collapsed in the nonperturbative first stage
could in the later stage be collapsed with other partons, e.g.\ from the
showers. Also collapses not allowed in the more restrictive scenarios
above could occur at this stage. Whether that happens or not depends on
the transverse and longitudinal momenta that will be defined below.
 
\subsection{Primordial $\kT$}
\label{sec:kT}
 
Until now, we have considered only the longitudinal part of parton
momenta. In reality, partons are also expected to have some non-zero $\kT$
values caused by Fermi motion inside the incoming hadrons. This kind of $\kT$
is denoted `primordial $\kT$', since it is not generated by the (DGLAP)
shower evolution nor from hard interactions, but rather represents an input
to the perturbative stages of the event. Based on Fermi motion alone, one
would expect values of the order of a few hundred MeV, just like in
eq.~(\ref{eq:cutoff}). But to reproduce
e.g.~the $\pT$ distributions of $\Z$ bosons produced in hadron--hadron
collisions, one notes a need for a significantly larger nonperturbative
input, either in parton showers or in resummation descriptions \cite{Huston}.
This problem is still awaiting a satisfactory explanation, although some ideas
have been explored that might alleviate it \cite{Erik}. Until such an
explanation has been found, we therefore have reason to consider an effective
`primordial $\kT$', at the level of the initiators, larger than the one
above. For simplicity, a parametrized $Q$-dependent width
\begin{equation}
\sigma(Q) = \mathrm{max}\left(\sigma_{\mathrm{min}},~
\sigma_\infty\frac{1}{1+Q_{\frac12}/Q}\right)  \label{eq:ktbroad}
\end{equation}
is introduced, where $\sigma$ is the width of the two-dimensional Gaussian
distribution of the initiator primordial $\kT$ (so that
$\langle \kT^2 \rangle = \sigma^2$), $Q$ is the scale of the hard interaction,
$\sigma_{\infty}$ is the value asymptotically approached as $Q\to\infty$, and
$Q_\frac12$ is the $Q$ scale where $\sigma= \frac12
\sigma_\infty$. A reasonable fit to the few available experimental `data points',
\begin{equation}\hspace*{1.3cm}
\begin{array}{lcll}
\sigma(Q \sim 1\GeV)\approx \sigma_{\mathrm{min}} & \approx &
0.36\GeV & \mathrm{(fragmentation)}\\[1mm]
\sigma(Q \sim 5-10\GeV) & \approx & 0.9\GeV &
\mathrm{(EMC)~{\mbox{\cite{EMC}}}}\\[1mm] 
\sigma(Q \sim M_Z) & \approx & 2\GeV & \mathrm{(Tevatron)~{\mbox{\cite{Huston}}}},
\end{array}
\end{equation}
is obtained with the values $\sigma_\infty = 2.1$~GeV and $Q_{\frac12} =
7$~GeV. 
The $\sigma_{\mathrm{min}}$ in eq.~(\ref{eq:ktbroad}) represents a minimum
broadening, at the level of Fermi motion, which we take to be the standard
fragmentation $\pT$ width. In addition to partons participating in relatively
soft interactions, this minimum broadening is also applicable to the remnant
partons, which by definition do not participate in hard interactions and
hence 
are not naturally associated with a particular $Q$ scale.
 
Apart from the selection of each individual $\kT$,
there is also the requirement that the total $\kT$ of the beam adds up to
zero. The question of how partons recoil off one another in transverse
momentum space inside hadrons is so far largely unaddressed in the
literature. We imagine a few different possibilities here:
\begin{enumerate}
\item The primordial transverse momenta are generated at a stage where the
  partons have low virtualities and hence large wavefunctions. Moreover, at
  least for that part which is due to Fermi motion, the dynamics responsible
  for the generation of primordial $\kT$ is that of a
  Fermi gas of partons in equilibrium, where each parton has received its
  total primordial $\kT$ through a sequence of many collisions with many
  different partons. Therefore, one possibility is to let the
  recoil of one parton be shared uniformly among all other initiator and
  remnant partons.
\item Since Fermi motion alone appears unable to account for the bulk of
  primordial $\kT$ in large-$Q$ interactions, there may exist a mechanism,
  involving presently unknown dynamics, which ties the generation of this
  $\kT$ to the presence of a large virtuality in the interaction, for
  instance by unresolved/unresummed bremsstrahlung radiation off
  the initiator parton and/or by the $\kT$ compensations in
  the shower being of a not strictly local nature (as happens e.g.~in the
  dipole description of parton showers \cite{LDC}). Regardless of the exact
  nature of the mechanism, recoils should in this picture
  primarily be taken up by initiators and beam remnant partons which are
  close in colour space.\\
  (\textit{a}) The extreme variant is here to let
  the recoil of a particular parton be taken up by its nearest colour
  neighbours only.\\
  (\textit{b}) A perhaps more realistic possibility is to let
  the compensation happen along a parton chain in colour space, with
  successive dampening of the compensation along the chain.
\end{enumerate}
These three possibilities are included in the present study. However, the
possibilities involving colour chains are
complicated by the fact that the colour connections between
initiator and remnant partons are very poorly known, since no perturbative
information is available. These problems will be discussed in  more detail in
Section \ref{sec:coltop} below.
 
Irrespective of which particular method is used to ensure
$\sum_i \vec{k}_{\perp i}=0$, the
question now arises how the kinematics of the initially collinear partons
should be reinterpreted to include non-zero $\kT$ assignments. This may be
done either by associating the generation of $\kT$ with the building up of
space-like virtualities among the partons, or by keeping the partons massless
while allowing (non-perturbatively small) longitudinal momentum transfers
between the beam remnants. (In the latter approximation also heavy quarks are
kept massless rather than assigned a spacelike virtuality; when initial-state
showers are  included no heavy quarks need be assigned to the beam remnants,
however.) In the first case, the invariant mass of initiator and beam
remnant partons combined in each hadron is maintained equal to the original
hadron mass, while in the second the mass can be significantly larger. The
difference, however, should be considered mostly technical, since the
momentum transfers involved are quantitatively small. For this study, the
second option is chosen, since this avoids potential technical
problems in dealing with string systems having negative mass squares.
 
For a specific interaction, consider a pair of massless initiator
partons in their rest frame, before $\kT$ is added:
\begin{eqnarray}
p_{1,2} = \frac{\sqrt{\hat{s}}}{2}(1,0,0,\pm1)~~~;~~\hat{s}=x_1x_2 s ~.
\end{eqnarray}
With primordial $\kT$ included, these momentum vectors should now be recast as
\begin{equation}
p_{1,2} =
(\sqrt{p_z^2+p_{\perp1,2}^2}~,~\vec{p}_{\perp1,2}~,~\pm p_z)~,
\label{eq:pnew}
\end{equation}
if the system should still be at longitudinal rest. Since we are
merely reinterpreting the kinematics of the initial-state partons,
the centre-of-mass energy, $\sqrt{\hat{s}}$, of the interaction
should be left unchanged. To ensure this, it is simple to solve
\begin{equation}
\hat{s}
 =  \left(\sqrt{p^2_z+p_{\perp1}^2}+ \sqrt{p^2_z+p_{\perp1}^2}\right)^2
- (\vec{p}_{\perp1}+\vec{p}_{\perp2})^2~,
\end{equation}
to obtain the required $p_z$ as a function of $\vec{p}_{\perp1}$,
$\vec{p}_{\perp2}$, and $\hat{s}$:
\begin{equation}
p_z^2
=\frac{\lambda(\hat{s}_\perp,p_{\perp1}^2,p_{\perp2}^2)}{4\hat{s}_\perp}~~~ ;
~\hat{s}_\perp \equiv \hat{s} + (\vec{p}_{\perp1}+\vec{p}_{\perp2})^2~,
\label{eq:pznew}
\end{equation}
with $\lambda$ the standard K\"all\'en function,
\begin{equation}
\lambda(a,b,c) = a^2+b^2+c^2-2ab-2bc-2ac~. \label{eq:lambdafunction}
\end{equation}
Naturally, only $\kT$ assignments which result in $p_z^2>0$ are acceptable.
 
\subsection{Beam Remnant Longitudinal Momenta}
 
In addition to flavours and transverse momenta, the beam remnants must also
together carry the remaining fraction, approximately $X$ (as defined by
eq.~(\ref{eq:valdist})) of longitudinal momentum. 
The sharing is based on the character of the remnant constituents.
First a fraction $x$ is defined for each constituent, and then these $x$
fractions are rescaled for overall energy and momentum conservation.
 
Thus a valence quark receives an $x$ picked at random according to a
small-$Q^2$ valence-like parton density, proportional to $(1-x)^a/\sqrt{x}$,
where $a = 2$ for a  $\u$ quark in a proton and $a = 3.5$ for a $\d$ quark.
A sea quark must be the companion of one of the initiator quarks, and can
have an $x$ picked according to the $q_{\c}(x ; x_{\s})$ distribution
introduced above. In the rare case that no valence quarks remain and no sea
quarks need be added for flavour conservation, the beam remnant is
represented by a gluon, carrying all of the beam remnant longitudinal
momentum.
 
Among composite objects, a diquark would na\"\i vely obtain an $x$
given by the sum of its constituent quarks, while baryons and
mesons would receive an $x$ equated with the $z$ value obtainable
from a fragmentation function, in this study the Lund symmetric
fragmentation function. However, earlier studies on quark--diquark
remnants \cite{Zijl} have shown that, within the multiple
interactions formalism, it is very difficult to accommodate
observed remnant multiplicity distributions if the composite
system (the diquark) does not take a much larger fraction than
implied by the na\"\i ve estimate above. Physically, this could
correspond to the momentum carried by a surrounding pion/gluon
cloud being larger for a composite object than for a single
parton. The possibility of enhancing the $x$ values picked for
composite objects is therefore retained in the present study.
 
Finally, once $x$ values (and primordial $\kT$) have been picked for each of
the remnants, an overall rescaling is performed such that the remnants together
carry the desired longitudinal momentum. The simplest way to accomplish
this would be to fix the normalization of the beam remnant $x$ values on
each side separately, by requiring conservation of longitudinal momentum,
$\sum_i x_{i_\mathrm{BR}}=X$ on each side. Unfortunately, the introduction
of non-zero $\kT$ values with massless partons and on-shell hadrons rules
out such a simple approach, since energy would then
not be conserved. Instead, small non-zero lightcone momentum fractions in
the direction opposite to the parent hadron direction must be
allowed. As already noted in Section~\ref{sec:kT}, this procedure should
be thought of merely as a technical trick, necessitated by insisting on a
description in terms of on-shell partons.
 
The amount of light-cone momentum removed from the remnant system by each
pair of initiators, $i$, is in the overall cm frame of the event
(omitting subscript $i$ to avoid cluttering the notation),
\begin{equation}
\begin{array}{lclcl}
w^+
& = &  E^{\mathrm{cm}} + p_z^{\mathrm{cm}}
& = &  \gamma(1+\beta_{z})(E_{1}'+E_{2}')~, \\[2mm]
w^-
& = & E^{\mathrm{cm}} - p_z^{\mathrm{cm}}
& = & \gamma(1-\beta_{z})(E_{1}'+E_{2}')~,
\end{array}\label{eq:lightconearbitrary}
\end{equation}
where subscripts 1 and 2 denote the initiator partons
of the hard scattering on each side respectively, and the primed frame is
chosen as the longitudinal rest frame, defined by eq.~(\ref{eq:pnew}).
Then the boost is only along the $z$ direction,
\begin{equation}
\beta_z=\beta = \frac{x_{1}-x_2}{x_1+x_2}~,
\end{equation}
and from equations (\ref{eq:pnew})--(\ref{eq:pznew})
\begin{equation}
E_1'+E_2' = \sqrt{\hat{s}_\perp}~.
\end{equation}
Inserting these results in eq.~(\ref{eq:lightconearbitrary}) above yields
\begin{equation}
\begin{array}{lcl}
w^+ & = & \displaystyle\sqrt{\frac{1+\beta}{1-\beta}}\sqrt{\hat{s}_{\perp}}
= \sqrt{\frac{x_1}{x_2}}\sqrt{\hat{s}_{\perp}}~, \\[6mm]
w^- & = & \displaystyle\sqrt{\frac{1-\beta}{1+\beta}}\sqrt{\hat{s}_{\perp}}
= \sqrt{\frac{x_2}{x_1}}\sqrt{\hat{s}_{\perp}}~.
\end{array}
\end{equation}
For vanishing $\pT$ this simplifies to the familiar
$\hat{s}_{\perp} = \hat{s} = x_1 x_2 s$, $w^+ = x_1 \sqrt{s}$ and
$w^- = x_2 \sqrt{s}$.
 
The light-cone momenta remaining for the combined beam remnant system are
thus:
\begin{eqnarray}
W^+_\rem & = & \sqrt{s} - \sum_i w^+_i = \sqrt{s} - \sum_i
\sqrt{\frac{x_{i1}}{x_{i2}}}\sqrt{\hat{s}_{\perp i}}\label{eq:Wrem+}\\
W^-_\rem & = & \sqrt{s} - \sum_i w^-_i =  \sqrt{s} - \sum_i
\sqrt{\frac{x_{i2}}{x_{i1}}}\sqrt{\hat{s}_{\perp i}}\label{eq:Wrem-}\\
W^2_\rem & = & W^+_\rem W^-_\rem.
\end{eqnarray}
In extreme cases, it may happen that
the hard interactions have removed so much energy and
momentum from the beam remnants that the remnant system nominally
becomes space-like, $W^2_\rem<0$, if large $\kT$ values have been
assigned. Though not strictly speaking unphysical, such a situation could
lead to problems at the fragmentation stage. The
requirement $W^2_\rem>0$ is therefore imposed as an additional constraint
when primordial $\kT$ values are assigned.
 
The $x$ values picked for the beam remnant partons are now interpreted as
fractions of the light-cone momenta, $W^+_\rem$ and $W^-_\rem$, of the beam
remnant system, modulo an overall rescaling on each
side, to leave room for overall momentum conservation. Using index $j$
to refer to beam remnant partons on side 1 and index $k$ for the ones on side
2, we thus make the identification
\begin{equation}
\begin{array}{lclclcl}
p_{j}^+ & = & \alpha x_j W^+_\rem & \implies & p_{j}^- & = & 
 \displaystyle\frac{m_{\perp j}^2}{p_j^+} \\[6mm]
p_{k}^- & = & \beta x_k W^-_\rem & \implies & p_{k}^+ & = & 
 \displaystyle\frac{m_{\perp k}^2}{p_k^-},
\end{array}
\end{equation}
where $m_\perp^2 = m^2 + \pT^2$ and $\alpha$ and $\beta$ are global
normalizations, to be determined from overall energy and
momentum conservation in the beam remnant system:
\begin{eqnarray}
W^+_\rem & = & \sum_j p^+_j + \sum_k p^+_k = \alpha W^+_\rem \sum_j x_j +
  \frac{1}{\beta W^-_\rem} \sum_k \frac{m_{\perp k}^2}{x_k} \\
W^-_\rem & = & \sum_j p^-_j + \sum_k p^-_k =
\beta W^-_\rem \sum_k x_k + \frac{1}{\alpha  W^+_\rem} \sum_j
  \frac{m_{\perp j}^2}{x_j}.
\end{eqnarray}
Equating these expressions with eqns.~(\ref{eq:Wrem+}) and (\ref{eq:Wrem-}),
one obtains for $\alpha$ and $\beta$
\begin{eqnarray}
\alpha & = & \frac{W_\rem^2 + W_1^2 - W_2^2 +
\lambda^{1/2}(W_\rem^2,W_1^2,W_2^2)}{2W_\rem^2\sum_j x_j } ~, \\
\beta & = & \displaystyle\frac{W_\rem^2 + W_2^2 - W_1^2 +
\lambda^{1/2}(W_\rem^2,W_1^2,W_2^2)}{2W_\rem^2\sum_k x_k } ~,
\end{eqnarray}
with the K\"all\'en function given by eq.~(\ref{eq:lambdafunction}) and $W_1$
and $W_2$ the total transverse masses of each of the two beam remnant
systems,
\begin{equation}
W_1^2  =  \left(\sum_jx_j\right)\displaystyle \left(\sum_j \frac{m_{\perp
      j}^2}{x_j}\right) ~~~~~~,~~~~~
W_2^2 = \left(\sum_kx_k\right)\displaystyle \left(\sum_k \frac{m_{\perp
      k}^2}{x_k}\right).
\end{equation}
Finally, for physical choices of $x_{j,k}$ and primordial $\kT$, the sum of the
individual remnant transverse masses must be smaller than that of the
total remnant system, $W_1 + W_2 < W_\rem$. If this is not the case, new
$\kT$ sets and/or new $x_{j,k}$ values are tried, until a physical set of
values is found.
 
\section{Colour Correlations and String Topologies}
\label{sec:coltop}
 
The formalism described in the previous Sections may be used to obtain a
sequence of hard 
scatterings with associated initial- and final-state showers and ordered by the
$\pT$ of the hard interactions they contain. Kinematics is completely
specified for all partons involved in the scatterings, in the associated
showers as well as in the left-behind beam remnants. However, 
at some point, the time evolution of this system results in inter--parton
distance scales larger than about a femtometer, where the perturbative
QCD description in terms of partons breaks down and must be supplanted by one
of hadrons. Now, if each parton hadronized independently of the rest
of the event, the information on kinematics and flavours alone would 
suffice to pass from the language of partons to that of hadrons, but
confinement is precisely \emph{not} a strictly local effect. Rather, it is 
a statement about a (colour singlet)
\textit{system} of partons, and hadronization is one concerning the evolution
when colour charges inside such a system have been imparted with large
momenta \textit{relative to each other}. It is therefore not meaningful
to study the hadronization of a single parton in isolation. Instead, it
becomes necessary to consider the interplay between colour charges and 
to take correlations into account when modeling the
hadronization process. 

In this Section, we begin by considering the hadronization of systems
containing non--zero baryon numbers, but where the colours of all
participating partons are known. We will use the planar approximation of QCD
as a starting point, with a junction picture for the baryons. 

Next, since the hard-scattering and parton-shower histories discussed above 
do not provide sufficient colour information --- specifically 
information about how the different scattering initiators 
are correlated in colour is completely absent --- 
we consider several possibilities for the assignment of correlated
colours to the parton-shower initiators of the scatterings. 
Combined with the hadronization model, this allows us to
study what is obtained under `minimal' assumptions. 
 
Finally, further issues are whether the original colour arrangement survives
all the way to the long-distance hadronization era, and whether nonlinear
effects arise in the hadronization process itself. That is, with several
partons and string pieces moving out from the collision process, these
partons and pieces will largely overlap in space and time. We do not know
whether such overlaps can lead to colour rearrangements or nontrivial
hadronization effects, e.g.\ of the Bose--Einstein kind. In principle
$\e^+\e^- \to \W^+\W^- \to \q_1\qbar_2\q_3\qbar_4$ offers a clean
environment to study such crosstalk, but experimental results are
inconclusive \cite{crosstalkexp}. In hadronic collisions, Bose--Einstein
studies by UA1 and E735 also give a splintered image \cite{ppbarBE}: the
strength parameter of BE effects drops with increasing particle density,
consistent with a picture where a higher multiplicity comes from having
several independently hadronizing strings, but the BE radius also
increases, which suggests correlations between the strings.
 
\subsection{Hadronization \label{sec:jufrag}}
Taking colour interference into account when modeling the hadronization
process could easily become an unmanageable task. One simplification
(disregarding baryons for the moment) is to go to the limit with infinitely
many colours, $N_C \to \infty$ \cite{NCinfinite}. In this limit the
confinement force acting on a gluon is twice that on a quark, i.e.\ the
gluon colour and anticolour charges decouple. Further, colour diagrams
are planar, so that final-state colour-anticolour pairs are always
uniquely matched, via an unbroken colour line through the diagram.

In the Lund string model \cite{string}, the two ends of such a colour line
define a string piece. The string piece can be viewed as a Lorentz covariant
and causal implementation of a linear confinement potential between the
two partons. Transverse degrees of freedom here play no dynamical role,
but one can visualize the colour field lines as compressed into a tube of a
typical hadronic width ($\sim 1$~fm). As the partons move apart and a
string piece is stretched out, it can break by the production of new
$\q\qbar$ pairs that screen the endpoint colour charges. The $\q$ of one
such break and the $\qbar$ of an adjacent break together define a meson,
which may be unstable and decay further.
 
The classical example is $\e^+\e^- \to \q\qbar\g$, where one may assign
a red colour to the $\q$, antired+green to the $\g$ and antigreen to the
$\qbar$, so that the string consists of two pieces, one $\q$--$\g$ and one
$\g$--$\qbar$. There is no piece directly between $\q$ and $\qbar$, with
observable consequences in the particle flow \cite{Jadestring}.
 
\begin{figure}
\begin{center}\vspace*{5mm}
\begin{fmffile}{fmfjtop}
\begin{tabular}{ccc}
\begin{fmfgraph*}(60,60)
\fmfleft{q1,q2}
\fmfright{q3}
\fmf{plain}{q1,j}
\fmf{plain}{q2,j}
\fmf{plain}{q3,j}
\fmfv{d.sh=circ,d.f=full,d.siz=7,lab=$\q_{\val1}$,l.ang=0}{q1}
\fmfv{d.sh=circ,d.f=full,d.siz=7,lab=$\q_{\val2}$,l.ang=0}{q2}
\fmfv{d.sh=circ,d.f=full,d.siz=7,lab=$\q_{\val3}$,l.ang=135}{q3}
\fmfv{lab=J,l.ang=180,l.dist=3.5}{j}
\end{fmfgraph*} & & 
\raisebox{0cm}{\begin{fmfgraph*}(60,50)
\fmftop{t}
\fmfbottom{b}
\fmf{phantom}{t,v,b}
\fmfv{d.sh=circ,d.f=full,d.siz=7,lab=Valence quark in beam
  hadron,l.ang=0,l.dist=8}{t} 
\end{fmfgraph*}}
\\[4mm]
\end{tabular}
\end{fmffile}
\caption{The initial state of a baryon, consisting of 3 valence quarks
  connected antisymmetrically in colour via a central `string junction', J.
\label{fig:jtop}}
\end{center}
\end{figure}
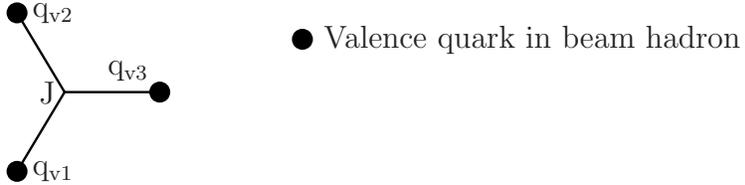

Turning now to baryon beams as the more interesting and difficult example, 
we picture the initial state of a baryon as
consisting of three valence quarks connected in colour via a
central `junction', cf.~Fig.\ref{fig:jtop}. 
At the most basic level, 
such a picture finds its motivation by
considering the simplest locally gauge invariant operator in
$SU(3)$ which carries non-zero baryon number \cite{mrv80}:
\begin{equation}
B_{f_1f_2f_3} = \epsilon^{\alpha_1\alpha_2\alpha_3}
\prod_{i=1}^{3} P\left[\exp\left(ig
\int_{\mathcal{P}(x,x_i)\hspace*{-4ex}}G_\mu~\d
x^\mu\right)q_{f_i}(x_i)\right]_{\alpha_i}.
\end{equation}
Physically, this operator assigns the space-time coordinates $x_i$
to three valence quarks (with flavours $f_i$ and colours $\alpha_i$)
and connects each of them via the gluon field $G_\mu$ along the
path $\mathcal{P}$ to the point $x$ (with $P$ representing the 
path-ordering operation), which we may identify as the
locus of the string junction. Such ideas were
already introduced in the early string model of hadrons
\cite{mrv80,artrustring,earlyjunc}, and 
have been used to construct baryon wavefunctions in confinement
studies \cite{confjunction}. In a recent article \cite{BNV}, we
argued that this picture also arises naturally from string energy
minimization considerations. 

In a collision, the fact that a gluon carries colour implies that the 
junction will, in general, be separated in colour space from the original 
valence quarks. As a simple example, consider a valence $\q\q\to\q\q$ 
scattering in a $\p\p$ collision. The exchanged gluon will flip colours, 
so that each junction becomes attached to a $\q$ from the other proton.

Hence the junction may well end up colour-connected to partons --- or
chains of partons --- which 
are widely separated in momentum space and of which no two may be 
naturally considered to form a diquark system. To describe the hadronization
of such systems, a model capable of addressing colour topologies containing 
explicit non--zero baryon numbers, here in the form of junctions, becomes
necessary. Such a model was first developed in \cite{BNV}, for dealing with
the colour topologies that arise in baryon number violating 
supersymmetric scenarios. In the following, we show how this approach 
may be applied in a multiple interactions context to describe the
physics of beam remnants.   

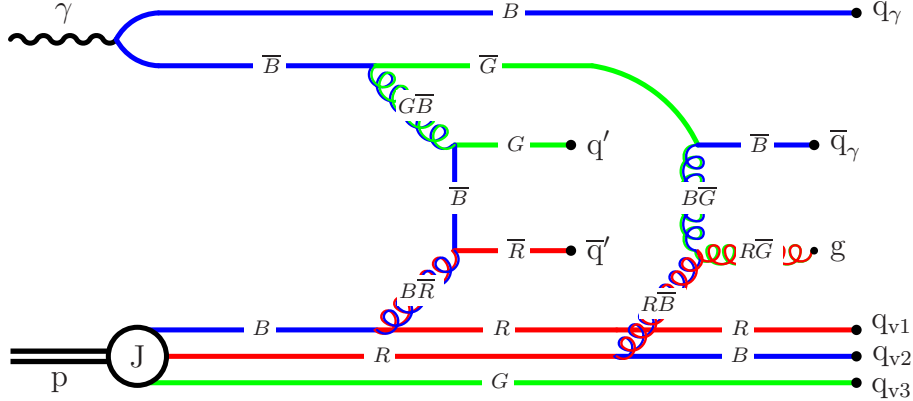
\begin{figure}
\vspace*{-5mm}%
\begin{center}
\begin{fmffile}{fmfjustring}
\begin{fmfgraph*}(400,200)\fmfpen{thick}
\fmfforce{0.1w,0.2h}{p_in}
\fmfforce{0.1w,0.8h}{f_in}
\fmfforce{0.22w,0.2h}{p_res}
\fmfforce{0.2w,0.8h}{f_res}
\fmf{photon,label=$\gamma$,label.s=left}{f_in,f_res}
\fmf{double,label=$\p$}{p_in,p_res}
\fmfv{d.sh=circ,d.siz=0.11h,label=J,label.dist=0,d.fill=empty}{p_res}
\fmfforce{0.24w,0.75h}{f_llo}
\fmfforce{0.24w,0.85h}{f_lhi}
\fmf{plain,right=0.27,fore=blue}{f_res,f_llo}
\fmf{plain,left=0.27,fore=blue}{f_res,f_lhi}
\fmfforce{0.9w,0.85h}{f_rhi}
\fmfforce{0.65w,0.75h}{f_rlo}
\fmf{plain,fore=blue,lab=\clab{$\overline{B}$},l.dist=0}{f_llo,f_1}
\fmf{plain,fore=green,lab=\clab{$\overline{G}$},l.dist=0}{f_1,f_rlo}
\fmf{plain,fore=blue,lab=\clab{$B$},l.dist=0,l.side=left}{f_lhi,f_rhi}
\fmfforce{0.22w,0.25h}{p_l1}
\fmfforce{0.22w,0.20h}{p_l2}
\fmfforce{0.22w,0.15h}{p_l3}
\fmfforce{0.9w,0.25h}{p_r1}
\fmfforce{0.9w,0.20h}{p_r2}
\fmfforce{0.9w,0.15h}{p_r3}
\fmf{plain,fore=blue,lab=\clab{${B}$},l.dist=0}{p_l1,p_1}
\fmf{plain,fore=red,lab=\clab{${R}$},l.dist=0}{p_1,p_2,p_r1}
\fmf{plain,fore=red}{p_l2,p_m1,p_m2}
\fmf{phantom,tension=0,lab=\clab{${R}$},l.dist=0}{p_l2,p_m2}
\fmf{plain,fore=blue,lab=\clab{${B}$},l.dist=0}{p_m2,p_r2}
\fmf{plain,fore=green,lab=\clab{${G}$},l.dist=0}{p_l3,p_r3}
\fmfforce{0.75w,0.6h}{f_i2}
\fmfforce{0.86w,0.6h}{f_e2}
\fmfforce{0.75w,0.4h}{p_i2}
\fmfforce{0.86w,0.4h}{p_e2}
\fmf{plain,fore=green,left=0.25}{f_rlo,f_i2}
\fmf{plain,fore=blue,lab=\clab{$\overline{B}$},l.dist=0}{f_i2,f_e2}
\fmfforce{0.52w,0.6h}{f_i1}
\fmfforce{0.52w,0.4h}{p_i1}
\fmfforce{0.63w,0.6h}{f_e1}
\fmfforce{0.63w,0.4h}{p_e1}
\fmf{plain,fore=green,lab=\clab{${G}$},l.dist=0}{f_i1,f_e1}
\fmf{plain,fore=blue,lab=\clab{$\overline{B}$},l.dist=0}{f_i1,p_i1}
\fmf{plain,fore=red,lab=\clab{$\overline{R}$},l.dist=0}{p_i1,p_e1}
\fmfv{lab=$\q'$,l.ang=0,d.sh=circ,d.fill=full,d.siz=2}{f_e1}
\fmfv{lab=$\qbar'$,l.ang=0,d.sh=circ,d.fill=full,d.siz=2}{p_e1}
\fmfv{lab=$\q_{\gamma}$,l.ang=0,d.sh=circ,d.fill=full,d.siz=2}{f_rhi}
\fmfv{lab=$\qbar_\gamma$,l.ang=0,d.sh=circ,d.fill=full,d.siz=2}{f_e2}
\fmfv{lab=$\q_{\val1}$,l.ang=20,d.sh=circ,d.fill=full,d.siz=2}{p_r1}
\fmfv{lab=$\q_{\val2}$,l.ang=0,d.sh=circ,d.fill=full,d.siz=2}{p_r2}
\fmfv{lab=$\q_{\val3}$,l.ang=-20,d.sh=circ,d.fill=full,d.siz=2}{p_r3}
\fmfv{lab=$\g$,l.ang=0}{p_e2}
\end{fmfgraph*}%
\vspace*{-1\unitlength}\hspace*{-401\unitlength}%
\begin{fmfgraph*}(400,200)\fmfpen{thin}
\fmfset{curly_len}{2.5mm}
\fmfforce{0.24w,0.75h}{f_llo}
\fmfforce{0.24w,0.85h}{f_lhi}
\fmfforce{0.65w,0.75h}{f_rlo}
\fmfforce{0.22w,0.25h}{p_l1}
\fmfforce{0.22w,0.20h}{p_l2}
\fmfforce{0.9w,0.25h}{p_r1}
\fmfforce{0.9w,0.20h}{p_r2}
\fmfforce{0.75w,0.6h}{f_i2}
\fmfforce{0.86w,0.6h}{f_e2}
\fmfforce{0.75w,0.4h}{p_i2}
\fmfforce{0.86w,0.4h}{p_e2}
\fmfforce{0.52w,0.6h}{f_i1}
\fmfforce{0.52w,0.4h}{p_i1}
\fmfforce{0.63w,0.6h}{f_e1}
\fmfforce{0.63w,0.4h}{p_e1}
\fmf{phantom}{p_l1,p_1,p_2,p_r1}
\fmf{phantom}{p_l2,p_m1,p_m2,p_r2}
\fmf{phantom}{f_llo,f_1}
\fmf{phantom}{f_1,f_rlo}
\fmffreeze
\fmf{gluon,fore=red}{p_1,p_i1}
\fmf{gluon,fore=blue}{f_1,f_i1}
\fmf{gluon,fore=blue}{p_m2,p_i2}
\fmf{gluon,fore=green}{p_i2,p_e2}
\fmf{gluon,fore=green}{f_i2,p_i2}
\end{fmfgraph*}%
{\hspace*{-399\unitlength}%
\begin{fmfgraph*}(400,200)\fmfpen{thin}
\fmfset{curly_len}{2.5mm}
\fmfforce{0.24w,0.75h}{f_llo}
\fmfforce{0.65w,0.75h}{f_rlo}
\fmfforce{0.22w,0.25h}{p_l1}
\fmfforce{0.22w,0.20h}{p_l2}
\fmfforce{0.9w,0.25h}{p_r1}
\fmfforce{0.9w,0.20h}{p_r2}
\fmfforce{0.75w,0.6h}{f_i2}
\fmfforce{0.86w,0.6h}{f_e2}
\fmfforce{0.75w,0.4h}{p_i2}
\fmfforce{0.86w,0.4h}{p_e2}
\fmfforce{0.52w,0.6h}{f_i1}
\fmfforce{0.52w,0.4h}{p_i1}
\fmfforce{0.63w,0.6h}{f_e1}
\fmfforce{0.63w,0.4h}{p_e1}
\fmf{phantom}{p_l1,p_1,p_2,p_r1}
\fmf{phantom}{p_l2,p_m1,p_m2,p_r2}
\fmf{phantom}{f_llo,f_1}
\fmf{phantom}{f_1,f_rlo}
\fmffreeze
\fmf{gluon,fore=blue,lab=\clab{${B\overline{R}}$},lab.s=left,l.dist=0}{p_1,p_i1}
\fmf{gluon,fore=green,lab=\clab{${G\overline{B}}$},l.dist=0}{f_1,f_i1}
\fmf{gluon,fore=red,l.side=left,lab=\clab{${R\overline{B}}$},l.dist=0}{p_m2,p_i2}
\fmf{gluon,fore=red,lab=\clab{${R\overline{G}}$},l.dist=0}{p_i2,p_e2}
\fmf{gluon,fore=blue,lab=\clab{${B\overline{G}}$},l.dist=0}{f_i2,p_i2}
\fmfv{d.sh=circ,d.fill=full,d.siz=2}{p_e2}
\end{fmfgraph*}}%
\end{fmffile}\vspace*{-5mm}
\caption{Example of colour assignments in a $\gamma\p$ collision with two
  interactions. Explicit colour labels are shown on each propagator line. In
  this example, the string system containing the junction is 
  spanned by $\J$ connected to $\q_\gamma$, to $\q_{\val3}$, and via $\g$ to
  $\q'$, as can be seen by tracing each of the three colour lines to the
  junction. 
  \label{fig:justring}} 
\end{center}
\end{figure}

We begin by considering a simplified situation where only one of the
initial beam particles is a baryon. Leaving the ambiguities in assigning
correlated colours aside for the moment, Fig.\  \ref{fig:justring} gives an
example of how the colour structure of a $\gamma\p$ collision might look. 
In this example, the final-state colour-singlet system containing the
junction consists of the three string pieces $\J$---$\q_\gamma$,
$\J$---$\q_{\val 3}$, and $\J$---$\g$---$\q'$. 
The two other string systems in the event, $\q_{\val1}$---$\qbar'$
and $\q_{\val 2}$---$\qbar_\gamma$ are standard and do not concern us here. 

To understand how the junction system hadronizes, the motion of the junction
must first be established. This can be inferred from noting that the opening
angle between any pair of the connected string pieces is 120$^\circ$ in the
rest frame of the junction, i.e.~in 
that frame the system consisting of the junction and its nearest 
colour-connected neighbours looks like a perfect Mercedes topology. This is
derivable \cite{BNV} from the action of the classical string
\cite{artrustring} (which has a linear potential and thus exerts a constant
force), but follows more directly from symmetry arguments.  

Note that the junction 
motion need not be uniform. In the example above, one of the string pieces
goes from the junction, via $\g$ to $\q'$. At early times, the junction
only experiences the pull of its immediate neighbour, $\g$, and the
direction of $\q'$ is irrelevant. However, as the gluon moves out from the
origin, it loses energy to the string traced out behind it. From 
the point when
\emph{all} its energy has been converted to potential energy of the string
and this information has propagated back to the junction,
it will be the direction of $\q'$ which determines the direction of the
`pull' exerted by this string piece on the junction, and not that of the 
gluon. In the general case, with arbitrarily many gluons, 
the junction will thus be `jittering around', being pulled in different
directions at different times. 

However, rather than trying to trace this jitter in detail --- which at any
rate is at or below what it is quantum mechanically meaningful to speak about
--- we choose to define an effective pull of each string on the junction, as
if from a single parton with four-momentum given by \cite{BNV}:
\begin{equation}
p_{\mathrm{pull}} = \sum_{i=1}^n \, p_i \, \exp \left( -
{\textstyle \sum}_{j=1}^{i-1}
E_j / E_{\mathrm{norm}} \right) ~,
\label{eq:ppull}
\end{equation}
where the outermost sum runs over the parton chain which defines the string
piece, from the junction outwards (in colour space), and where the sum inside
the exponent runs over all gluons closer to the junction than the one
considered (meaning it vanishes for $i=1$). The energy normalization
parameter $E_{\mrm{norm}}$ is by default associated with the characteristic
energy stored in the string at the time of breaking,
$E_{\mrm{norm}}\simeq1.5\GeV$. Naturally, the energies
$E_j$ should be evaluated in the junction rest frame, yet since this is not
known to begin with, we use an iterative sequence of successively improved
guesses. 

With the motion of the junction determined, the fragmentation of the system
as a whole may now be addressed. Since 
the string junction represents a localized topological feature
of the gluon/string field, we would not 
expect the presence of the junction in the string topology to significantly
affect the fragmentation in the regions close to the endpoint
quarks. Specifically, in an event where each of the three endpoint quarks
have large energies in the junction rest frame, the
energies of the leading and hence hardest particles 
of each jet should agree, on average, with that of an equivalent jet in an
ordinary two-jet event. 

The hadronization model developed in \cite{BNV} ensures this by fragmenting
each of the string pieces outwards--in, as for a normal $\q\qbar$ string (in
both cases opposite to the physical time ordering of the process). The
leading quark of a string piece is combined with a newly created
quark--antiquark pair to form a meson plus a new leftover quark, and so
on. Parton flavours and hadron spins are selected in a manner identical to
that of the ordinary string, as are fragmentation functions and the handling
of gluon kinks on the string pieces.

However, junctions were not included in the original string model, so here
a new procedure needs to be introduced. If all three string pieces were 
fragmented in the
above way until little energy was left in each to form more hadrons, then
it would be extremely unlikely that the resulting leftover system of three
unpaired quarks would just happen to have an invariant mass equal to that of
any on--shell baryon. While one could in principle amend this by shuffling
momentum and energy to other hadrons in the vicinity of the junction, such
a procedure would be arbitrary and
result in an undesirable and large systematic distortion of the junction baryon
spectrum. The way such systematic biases are avoided for ordinary
$\q\qbar$ strings is to alternate between fragmenting the system from the
$\q$ end and from the $\qbar$ end in a random way, so that the hadron pair
that is used to ensure overall energy--momentum conservation does not always
sit at the same location. Thus, while the distortion is still local in
each event, it is smeared out when considering a statistical sample
of events. 

In the case of a junction system, such a procedure is not immediately
applicable. Instead, we first
fragment two of the three string pieces, from their respective endpoint quarks
inwards. At the point where
more energy has been used up for the fragmentation than is available in the
piece, the last quark--antiquark pair formed is rejected and the
fragmentation is stopped. The two resulting unpaired quarks, one from each
fragmented string piece, are then combined into a single diquark, which
replaces the 
junction as the endpoint of the third string piece. Subsequently, this last
string piece is fragmented in the normal way, with overall energy and
momentum conservation ensured exactly as described for ordinary strings
above. In order to minimize the systematics
of the distortion and ensure that it is at all possible to produce
at least two hadrons from this final string system, we choose to always
select the highest energy string piece as the last to be fragmented. It was
shown in \cite{BNV,BNVsmall} that this asymmetry in the description does not
lead to large systematic effects.

\begin{figure}
\vspace*{-5mm}%
\begin{center}
\begin{fmffile}{fmfjujustring}
\begin{fmfgraph*}(300,200)\fmfpen{thick}
\fmfforce{0.1w,0.2h}{p_in}
\fmfforce{0.1w,0.8h}{pbar_in}
\fmfforce{0.22w,0.2h}{p_res}
\fmfforce{0.22w,0.8h}{pbar_res}
\fmf{double,label=$\pbar$,label.s=left}{pbar_in,pbar_res}
\fmf{double,label=$\p$}{p_in,p_res}
\fmfv{d.sh=circ,d.siz=0.11h,label=$\overline{\mrm{J}}$,%
label.dist=0,d.fill=empty}{pbar_res}
\fmfv{d.sh=circ,d.siz=0.11h,label=$\mrm{J}$,label.dist=0,d.fill=empty}{p_res}
\fmfforce{0.22w,0.75h}{pbar_l1}
\fmfforce{0.22w,0.8h}{pbar_l2}
\fmfforce{0.22w,0.85h}{pbar_l3}
\fmfforce{0.9w,0.75h}{pbar_r1}
\fmfforce{0.9w,0.80h}{pbar_r2}
\fmfforce{0.9w,0.85h}{pbar_r3}
\fmf{plain,fore=blue,lab=\clab{$\overline{B}$},l.dist=0}{pbar_l1,pbar_1}
\fmf{plain,fore=green}{pbar_1,pbar_2,pbar_r1}
\fmf{phantom,tension=0,lab=\clab{$\overline{G}$},l.dist=0}{pbar_1,pbar_r1}
\fmf{plain,fore=red,lab=\clab{$\overline{R}$},l.dist=0}{pbar_l2,pbar_r2}
\fmf{plain,fore=green,lab=\clab{$\overline{G}$},l.dist=0}{pbar_l3,pbar_r3}
\fmfforce{0.22w,0.25h}{p_l1}
\fmfforce{0.22w,0.20h}{p_l2}
\fmfforce{0.22w,0.15h}{p_l3}
\fmfforce{0.9w,0.25h}{p_r1}
\fmfforce{0.9w,0.20h}{p_r2}
\fmfforce{0.9w,0.15h}{p_r3}
\fmf{plain,fore=blue,lab=\clab{${B}$},l.dist=0}{p_l1,p_1}
\fmf{plain,fore=red}{p_1,p_2,p_r1}
\fmf{phantom,tension=0,lab=\clab{${R}$},l.dist=0}{p_1,p_r1}
\fmf{plain,fore=green,lab=\clab{${G}$},l.dist=0}{p_l2,p_r2}
\fmf{plain,fore=red,lab=\clab{${R}$},l.dist=0}{p_l3,p_r3}
\fmfforce{0.55w,0.5h}{i}
\fmfforce{0.7w,0.5h}{e}
\fmfforce{0.84w,0.66h}{eu}
\fmfforce{0.84w,0.34h}{ed}
\fmfv{lab=$\q_{\val1}$,l.ang=20,d.sh=circ,d.fill=full,d.siz=2}{p_r1}
\fmfv{lab=$\q_{\val2}$,l.ang=0,d.sh=circ,d.fill=full,d.siz=2}{p_r2}
\fmfv{lab=$\q_{\val3}$,l.ang=-20,d.sh=circ,d.fill=full,d.siz=2}{p_r3}
\fmfv{lab=$\qbar_{\val1}$,l.ang=-20,d.sh=circ,d.fill=full,d.siz=2}{pbar_r1}
\fmfv{lab=$\qbar_{\val2}$,l.ang=0,d.sh=circ,d.fill=full,d.siz=2}{pbar_r2}
\fmfv{lab=$\qbar_{\val3}$,l.ang=20,d.sh=circ,d.fill=full,d.siz=2}{pbar_r3}
\fmfv{lab=$\g_2$,l.ang=0}{ed}
\fmfv{lab=$\g_1$,l.ang=0}{eu}
\end{fmfgraph*}%
\vspace*{-1\unitlength}\hspace*{-301\unitlength}%
\begin{fmfgraph*}(300,200)\fmfpen{thin}
\fmfset{curly_len}{2.5mm}
\fmfforce{0.22w,0.75h}{pbar_l1}
\fmfforce{0.9w,0.75h}{pbar_r1}
\fmf{phantom}{pbar_l1,pbar_1}
\fmf{phantom}{pbar_1,pbar_2,pbar_r1}
\fmfforce{0.22w,0.25h}{p_l1}
\fmfforce{0.9w,0.25h}{p_r1}
\fmf{phantom}{p_l1,p_1}
\fmf{phantom}{p_1,p_2,p_r1}
\fmfforce{0.55w,0.5h}{i}
\fmfforce{0.7w,0.5h}{e}
\fmfforce{0.84w,0.66h}{eu}
\fmfforce{0.84w,0.34h}{ed}
\fmffreeze
\fmf{gluon,fore=red,l.dist=0}{p_1,i}
\fmf{gluon,fore=blue,l.dist=0}{pbar_1,i}
\fmf{gluon,fore=red,l.dist=0}{i,e}
\fmf{gluon,fore=blue}{e,eu}
\fmf{gluon,fore=red}{e,ed}
\end{fmfgraph*}%
{\hspace*{-299\unitlength}%
\begin{fmfgraph*}(300,200)\fmfpen{thin}
\fmfset{curly_len}{2.5mm}
\fmfforce{0.22w,0.75h}{pbar_l1}
\fmfforce{0.9w,0.75h}{pbar_r1}
\fmf{phantom}{pbar_l1,pbar_1}
\fmf{phantom}{pbar_1,pbar_2,pbar_r1}
\fmfforce{0.22w,0.25h}{p_l1}
\fmfforce{0.9w,0.25h}{p_r1}
\fmf{phantom}{p_l1,p_1}
\fmf{phantom}{p_1,p_2,p_r1}
\fmfforce{0.55w,0.5h}{i}
\fmfforce{0.7w,0.5h}{e}
\fmfforce{0.84w,0.66h}{eu}
\fmfforce{0.84w,0.34h}{ed}
\fmffreeze
\fmf{gluon,fore=blue,lab=\clab{$B\overline{R}$},l.dist=0}{p_1,i}
\fmf{gluon,fore=green,lab=\clab{$G\overline{B}$},l.dist=0}{pbar_1,i}
\fmf{gluon,fore=green,lab=\clab{$G\overline{R}$},l.dist=0}{i,e}
\fmf{gluon,fore=green,lab=\clab{$G\overline{B}$},l.dist=0}{e,eu}
\fmf{gluon,fore=blue,lab=\clab{$B\overline{R}$},l.dist=0}{e,ed}
\fmfv{d.sh=circ,d.fill=full,d.siz=2}{eu}
\fmfv{d.sh=circ,d.fill=full,d.siz=2}{ed}
\end{fmfgraph*}}%
\end{fmffile}\vspace*{-5mm}
\\
\caption{Example of colour assignments in a $\p\pbar$ collision, with 
explicit colour labels shown on each propagator line. Note that the blue
colour line starting on the junction $\mrm{J}$ is connected via the colour
flow of 
the hard scattering to the antiblue colour line of $\overline{\mrm{J}}$.
  \label{fig:jujustring}} 
\end{center}
\end{figure}
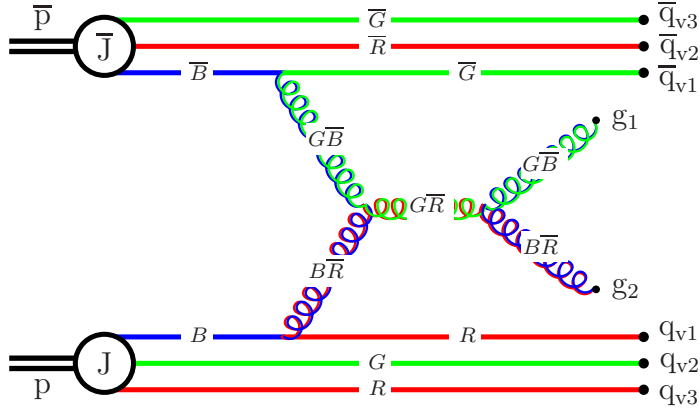

In proton--proton collisions, two junction systems will be present, but it is
physically impossible for these to be connected in colour. Hence, 
the hadronization of each system again proceeds exactly as described
above. However, in  
$\p\pbar$ collisions a new possibility arises, as depicted in Fig.\ 
\ref{fig:jujustring}. 
This simple example goes to illustrate that a junction and an `antijunction'
may become colour-connected by the colour exchanges taking place in a given
process. In such cases, the fragmentation of each of the junctions is no
longer disconnected from what happens to the other one; instead the
fragmentation of the system as a whole must be considered. 
The necessary generalization of the principles outlined above to the case of
connected junction--junction systems \cite{BNV} is not very complicated.

As before, two of the three strings from a junction are fragmented first,
outwards--in towards the junction, but in this case we always choose these
two string pieces to be the ones not connected to the other
junction. Diquarks are then formed around each junction exactly as before.  
What remains is a single string piece, spanned between a diquark at one end
and an antidiquark at the other, which can be
fragmented in the normal way. In fact, the only truly new 
question that arises at this point is how to generalize eq.~(\ref{eq:ppull})
to describe the pull of one junction on another. Here gluons on the string
between the two junctions are considered as normally, i.e.\ their momenta 
are added, with a suppression factor related to the energy of the 
intermediate gluons. The partons on the far side of the other junction also 
contribute their momenta, separately for each of the two strings, with an 
energy sum suppression now given by the intervening gluons on that particular 
string, plus the gluons on the junction--junction string.

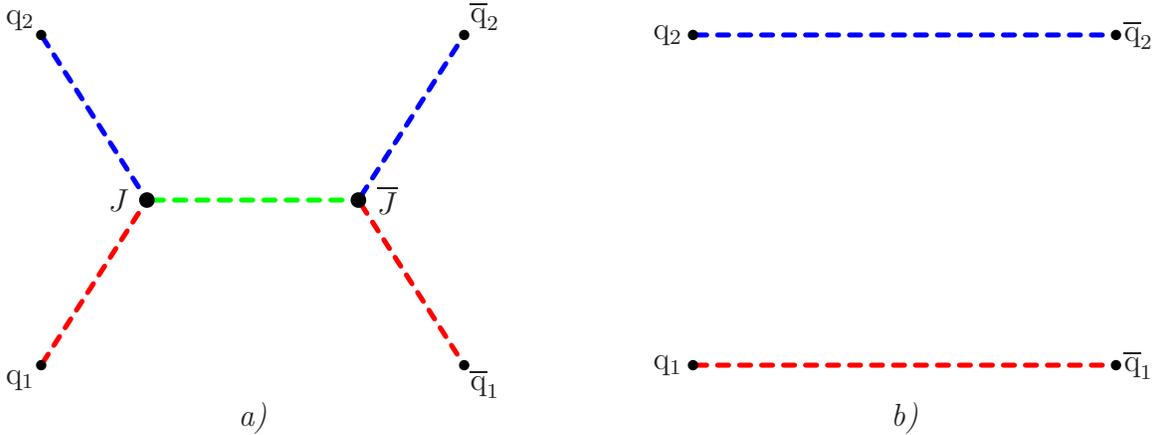
\begin{figure}[t]
\vspace*{7mm}
\begin{center}
\begin{fmffile}{fmfjjannihilation}
\begin{fmfgraph*}(200,125)
\fmfpen{thick}
\fmfbottom{t1}
\fmfleft{l1,l2}
\fmfright{r1,r2}
\fmfv{lab= {\it a)},l.dist=14}{t1}
\fmf{dashes,fore=red}{l1,jl}
\fmf{dashes,fore=blue}{l2,jl}
\fmf{dashes,fore=green}{jl,jr}
\fmf{dashes,fore=red}{jr,r1}
\fmf{dashes,fore=blue}{jr,r2}
\fmfdot{jl,jr}
\fmfv{lab=$J$,l.dist=7}{jl}
\fmfv{lab=$\overline{J}$,l.dist=7}{jr}
\fmfv{lab=$\q_1$,l.dist=3,d.sh=circ,d.fill=full,d.siz=2}{l1}
\fmfv{lab=$\q_2$,l.dist=3,d.sh=circ,d.fill=full,d.siz=2}{l2}
\fmfv{lab=$\qbar_1$,l.dist=3,d.sh=circ,d.fill=full,d.siz=2}{r1}
\fmfv{lab=$\qbar_2$,l.dist=3,d.sh=circ,d.fill=full,d.siz=2}{r2}
\end{fmfgraph*}\hspace*{1.5cm}
\begin{fmfgraph*}(200,125)
\fmfpen{thick}
\fmfbottom{t1}
\fmfv{lab= {\it b)},l.dist=14}{t1}
\fmfleft{l1,l2}
\fmfright{r1,r2}
\fmf{dashes,fore=red}{l1,r1}
\fmf{dashes,fore=blue}{l2,r2}
\fmfv{lab=$\q_{1}$,l.dist=4,l.ang=180,d.sh=circ,d.fill=full,d.siz=2}{l1}
\fmfv{lab=$\q_{2}$,l.dist=4,l.ang=180,d.sh=circ,d.fill=full,d.siz=2}{l2}
\fmfv{lab=$\qbar_1$,l.dist=3,l.ang=0,d.sh=circ,d.fill=full,d.siz=2}{r1}
\fmfv{lab=$\qbar_2$,l.dist=3,l.ang=0,d.sh=circ,d.fill=full,d.siz=2}{r2}
\end{fmfgraph*} 
\end{fmffile}\\[10mm]
\caption{%
{\it a)} A string system (dashed lines) spanned between four quarks and
containing a junction and an antijunction.  
{\it b)} The same parton configuration in colour space but with an
alternative string topology. In {\it a)} the beam baryon numbers will still be
present in the final state, while in {\it b)} they will have disappeared
through annihilation.
\label{fig:jjannihilation}}
\end{center}
\end{figure}

However, an alternative topology is also possible, 
where the junction
and the antijunction annihilate to produce two separate $\q\qbar$
systems \cite{BNV}, as illustrated in Fig.\  \ref{fig:jjannihilation}b. 
While it is
not clear from basic principles 
how often this should happen, it seems likely that, for a given
event, the topology which has the minimal string length is the one selected
dynamically. In this case, the string topology depicted in Fig.\ 
\ref{fig:jjannihilation}a would result when the
$\q_1\q_2$ and $\qbar_1\qbar_2$ opening angles are small, while the
topology in Fig.~\ref{fig:jjannihilation}b 
would result if the $\q_1\qbar_1$ and $\q_2\qbar_2$
opening angles are the small ones. Since, in the context we discuss here,
the quarks colour--connected to the junction will more often than not reside
in the beam remnants, we do not expect annihilation between the incoming
baryon numbers to be a large effect. Indeed, for the range of more realistic 
models that are investigated in Section \ref{sec:tests} below,
junction--junction annihilation is a feature of less than 1\% of the events
at Tevatron energies.

An ugly situation occurs in the rare events when the two junctions are 
connected by \emph{two} colour lines. If these lines contain intermediate 
gluons, it would be possible but difficult to fragment the system, in 
particular when the energy of these gluons becomes small. Without 
intermediate gluons, a first guess would be that the junction and 
antijunction annihilate to give a simple string spanned between a quark 
and an antiquark endpoint, so that the original baryon numbers are lost.
However, this assumes that the system starts out from a point in space 
and time, a commonly used approximation in the string language. Viewed 
in the transverse plane of the collision, the original positions of the 
junctions and of the hard scatterings involved could well be separated 
by distances up to a fm, i.e.\ the intervening strings could have 
energies up to a GeV. It may then be that the strings can break before 
the junctions annilate, so that a baryon--antibaryon pair nevertheless 
is produced. A detailed modeling would be required, beyond the scope 
of the current study, and possibly beyond the validity of the string
framework, so for now we choose to reject these rare events. 

\subsection{Initial-State Colour Correlations}
In the planar approximation, $N_C\to\infty$, 
a $2 \to 2$ process, such as $\g\g \to \g\g$,
can receive contributions from several possible colour flows, but the
cross section for a colour flow is uniquely defined, so that each flow
can be selected according to its appropriate weight \cite{HansUno}.
Furthermore, in our leading-order parton showers, the colour flow is
well-defined in each branching. Within each separate interaction and its
associated shower activity, the colour flow can thus be selected
unambiguously. In events with only one hard interaction, the colour of the
shower initiator can also uniquely be hooked up to that of the beam remnant.
Thus a knocked-out quark leaves behind a colour antitriplet beam remnant,
a gluon leaves a colour octet beam remnant, and so forth.
 
Unfortunately, once several interactions are allowed, there is no longer
a unique answer how to hook up the different shower initiators with the
beam-remnant partons. To illustrate, consider an incident meson 
out of which $n$ gluons are kicked out. These gluons may be ordered in colour
sequence such that the colour of the quark matches the anticolour of one
gluon, which then has another colour that matches the anticolour of
another gluon, and so on till the antiquark. Obviously there are $n!$
such possible arrangement of the $n$ gluons, each leading to a unique
colour topology for the hadronizing partonic system. Perturbation
theory has nothing to say about the relative probability for each of
these configurations; the colour correlations we now consider arise at
scales below the parton-shower cutoff $Q_0 \sim 1$~GeV. Further
arrangements would exist if we allowed some of the above gluons to form
a separate colour singlet, disconnected from the sequence between the
quark and antiquark ends; some of these could contribute to diffractive
topologies.
 
So far we only considered the planar colour topologies of the
$N_C \to \infty$ limit. The real-world $N_C = 3$ offers further complications.
First, interference terms of order $1/N_C^2$ --- modulated by
kinematical factors --- arise between different possible colour flows
in hard processes. The more partons are involved, the more assignments
need be made, and the larger the total uncertainty. Even at the
perturbative level it is thus no longer possible to speak of a unique
colour arrangement. Second, the situation is more complicated for baryon
beams. As described above, the initial state of a baryon, before
any scatterings occur, is represented by three valence quarks, connected
in a Y-shaped topology via a central junction which acts as a switchyard
for the colour flow and which carries the net baryon number. This situation is
illustrated in Fig.~\ref{fig:jtop}. 
Each of the gluons considered in the meson-beam example above may now be
arranged in colour on either of the three string pieces, leading to a
further multiplication of possibilities. 

We choose to address this question by determining 
a sequence of fictitious gluon emissions by which
this configuration evolves (in colour space)
to give rise to the parton shower initiators and beam
remnant partons actually present in a given event. We here assume that only
the minimal number of emissions required to obtain the given
set of initiators and remnants is dynamically relevant. Further, 
since sea quarks together with their
companion partners can pairwise be associated with a gluon branching below the
parton shower cutoff, only gluon emissions remain to be considered. 
(This also means that a sea quark, in our model, can
never form a colour singlet system together with its own companion.)
 
\subsubsection{Random Colour Correlations}
\begin{figure}
\begin{center}\vspace*{3mm}
\begin{fmffile}{fmfinitial}
\raisebox{2.2cm}{\hspace*{-5cm}\begin{fmfgraph*}(60,40)
\fmftop{t}
\fmfbottom{b}
\fmf{phantom}{t,v,b}
\fmfv{d.sh=circ,d.f=full,d.siz=7,lab=Parton in hadron
  remnant,l.ang=0,l.dist=8}{t}
\fmfv{d.sh=circ,d.f=empty,d.siz=7,lab=Parton shower initiator,%
l.ang=0,l.dist=8}{v}
\end{fmfgraph*}}\hspace*{5cm}
\begin{fmfgraph*}(120,80)
\fmftop{d1,g1,t2,t3}
\fmfbottom{d3,g2,g3,d4}
\fmfleft{d5,q1,q2,d6}
\fmfright{q3}
\fmf{plain}{q1,vq1,j}
\fmf{plain}{q2,vq2,j}
\fmf{plain}{q3,vq3,vq4,j}
\fmf{gluon}{vq1,g2}
\fmf{gluon}{vq2,g1}
\fmf{gluon}{vq3,vqq}
\fmf{plain}{t2,vqq,t3}
\fmf{gluon}{vq4,g3}
\fmfv{d.sh=circ,d.f=full,d.siz=7,lab=$\q_{\val1}$}{q1}
\fmfv{d.sh=circ,d.f=full,d.siz=7,lab=$\q_{\val2}$}{q2}
\fmfv{d.sh=circ,d.f=full,d.siz=7,lab=$\q_{\val3}$}{q3}
\fmfv{lab=J,l.ang=45,l.dist=3.5}{j}
\fmfv{d.sh=circ,d.f=empty,d.siz=7,lab=$\g_1$}{g1}
\fmfv{d.sh=circ,d.f=empty,d.siz=7,lab=$\g_2$}{g2}
\fmfv{d.sh=circ,d.f=empty,d.siz=7,lab=$\g_3$}{g3}
\fmfv{d.sh=circ,d.f=empty,d.siz=7,lab=$\qsea$}{t2}
\fmfv{d.sh=circ,d.f=full,d.siz=7,lab=$\qcmp$}{t3}
\end{fmfgraph*}
\\[4mm]
\end{fmffile}
\caption{Example of how a given set of parton shower initiators
could have
  been radiated off the initial baryon valence configuration, in the case of
  the `purely random' correlations discussed in 
  the text. In this example, the baryon number is disconnected
from the beam remnant. Instead, it is the final-state partons connected to the
colour lines of $\g_1$, $\g_2$, and $\g_3$ which determine how the junction
moves and hence how the baryon number flows in the event.
\label{fig:initialstate}}
\end{center}
\end{figure}
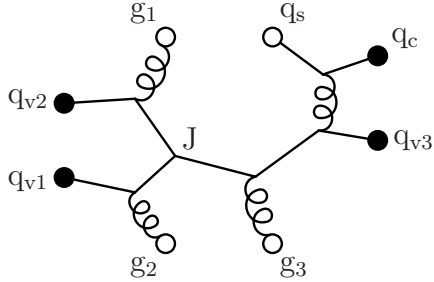

\begin{figure}[t]
\begin{center}
\begin{tabular}{ccc}\hspace*{-.1cm}
\includegraphics*[scale=0.73]{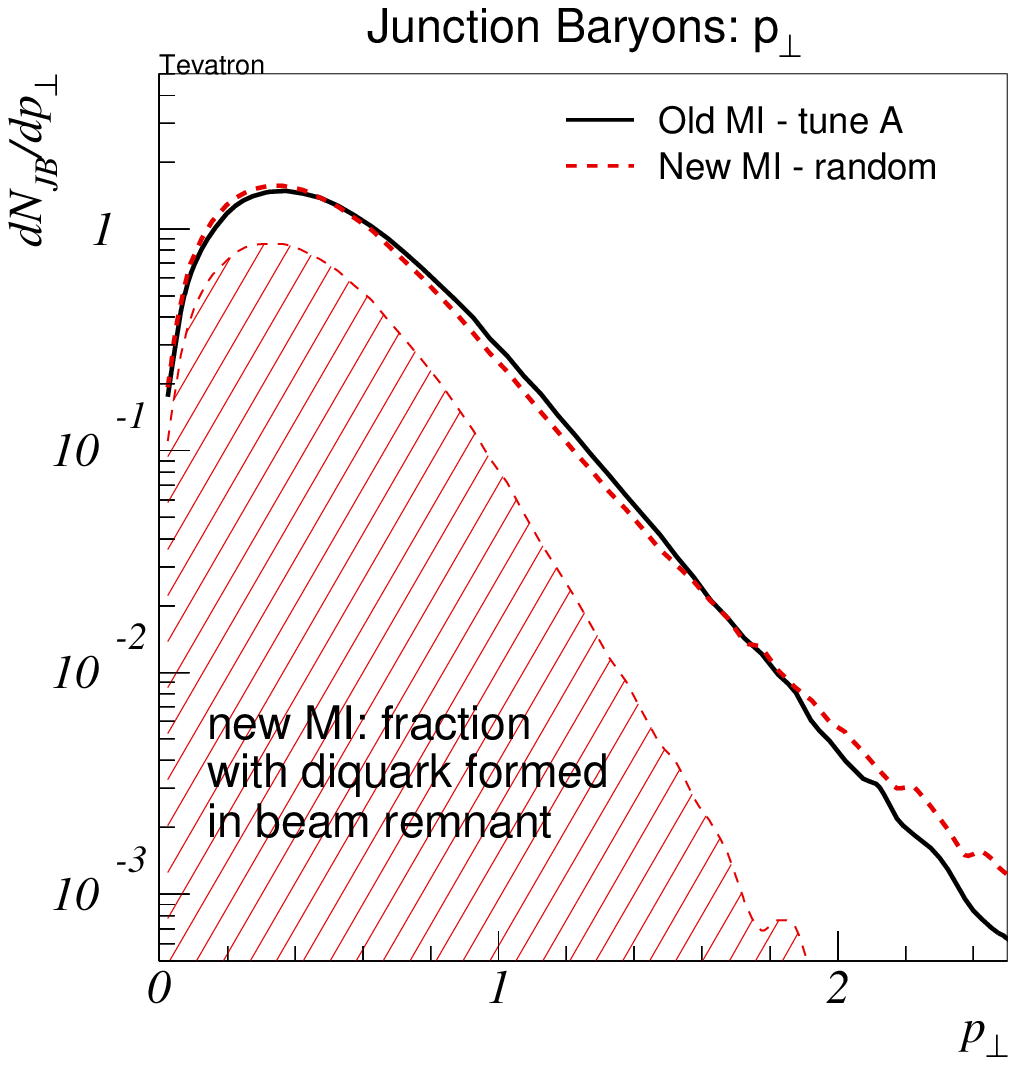}\hspace*{-.1cm}&&\hspace*{-.1cm}
\includegraphics*[scale=0.73]{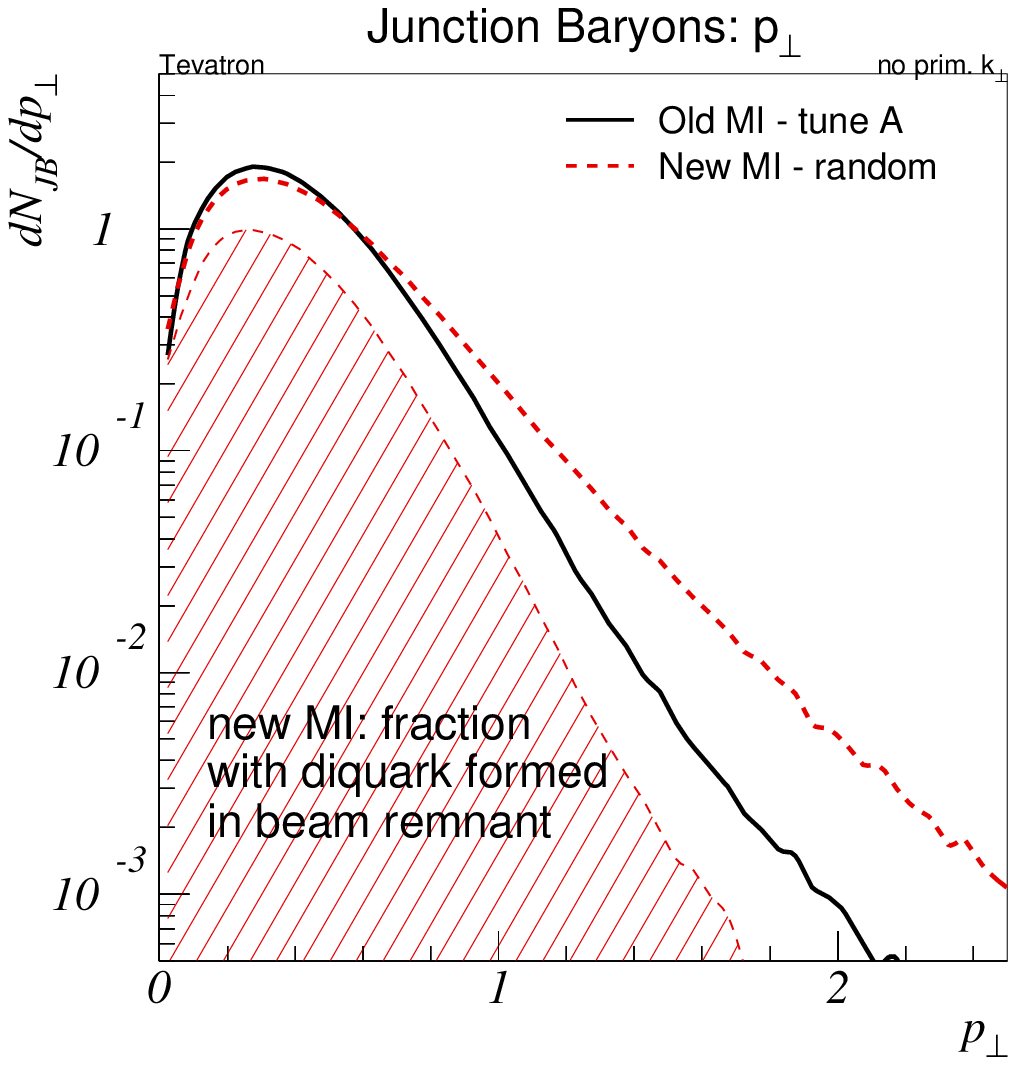} \hspace*{-0.1cm}\\[-2mm]
\it a) & & \it b)\\
\end{tabular}
\caption{$\pT$ spectra for junction baryons, {\it a)} with primordial $\kT$
  switched on, and {\it b)} with primordial $\kT$ switched off. The shaded
  area represents the distribution in the new model of those junction baryons
  which arose by first forming a diquark in the beam remnant, cf.~Section
  \ref{sec:remnants}. \label{fig:connectpt}}
\end{center}
\end{figure}

The simplest solution would be to assume that Nature arranges
these correlations randomly, i.e.~that gluons should be attached
to the initial quark lines in a random order, see
Fig.~\ref{fig:initialstate}. 
In this case, the junction (and
hence the baryon number) would rarely be colour connected directly
to two valence quarks in the beam remnant, even in the quite
common case that two such quarks are actually present (multiple
valence quark interactions are rare). It should be clear that the
migration of the baryon number depends sensitively upon which
partons in the final state the junction ends up being connected to
(see further Section \ref{sec:jufrag}). The conclusion is that if
the connections are \emph{purely} random, as above, then the
baryon number will in general be disconnected from the beam
remnant valence quarks. Hence the formation of a diquark in the
beam remnant would be rare and the baryon number of the initial
state should quite often be able to migrate to small $x_F$ values,
as previously illustrated in Fig.~\ref{fig:compflow}. One
could expect this longitudinal migration to be accompanied by a
migration in the transverse plane, such that the junction baryon
should generally migrate to larger $\pT$ values when the junction
is allowed to `float' more. However, as Fig.~\ref{fig:connectpt}a
illustrates, no large differences in the total $\pT$ spectrum are apparent 
when comparing the new model (thick dashed) with the old Tune A (solid).
 
The reason that no large transverse migration effect is visible, relative to
the old model, is that the latter also has a broader leading--baryon
$\pT$-spectrum than for normal baryons, as follows. 
In the old model, the  beam
remnant diquark (around which the junction baryon forms in the
fragmentation) always receives the full primordial $\kT$ kick from
the hard interaction initiator, by default a Gaussian distribution
with a width of 1\GeV. In the fragmentation process, when the baryon
acquires a fraction of the diquark longitudinal momentum, it obtains the same
transverse momentum fraction. Additional \pT\ will be imparted to the
junction baryon from the newly created quark, at the level 
of 0.36\GeV, but with some dependence on the momentum--space
location of its nearest neighbour in colour space. Essentially,
these two effects combine to yield the solid curve in
Fig.~\ref{fig:connectpt}a.
 
In the new model, we must distinguish between the old--model-like case 
when a diquark is formed in the beam remnant, on one hand,
and those cases where the junction is `free' to migrate, on the other.
 
If a diquark is formed, then it consists of two undisturbed beam
remnant quarks which are colour connected to the junction, and the
situation is indeed very similar to the old model. The baryon
forming around this diquark receives \pT\ from three sources.
Firstly, the diquark will have some intrinsic primordial \kT,
distributed according to eq.~(\ref{eq:ktbroad}). Since the diquark
resides entirely within the beam remnant, this \kT\ will always
be at the level of the fragmentation \pT. Secondly,
primordial \kT\ will be imparted to the diquark by recoil effects
from other beam remnant and initiator partons. In the case that primordial
$\kT$ kicks are compensated for uniformly by all initiator and remnant
partons, it is normally impossible for the diquark to acquire more than a
fraction of the hardest interaction initiator's primordial $\kT$. Even when
$\kT$ compensation is more local, by straightforward combinatorics, the more
initiators  present in the event, the smaller the chances that the initiator
parton(s) closest in colour to the diquark is associated with a scattering at
large $Q^2$. Hence, again according to eq.~(\ref{eq:ktbroad}), it is apparent
that the diquark usually will not receive a very hard primordial \kT\ kick. 
Thus, such a diquark will in general
have a smaller total primordial \kT\ than a diquark
in the old model. As before, the baryon will keep a large fraction of
this diquark \pT\ in the fragmentation process, as well as obtaining extra
fragmentation \pT. The net result is a softer junction baryon transverse
momentum spectrum than in the old model, as can be verified by comparing the
asymptotic slope of the shaded area in Fig.~\ref{fig:connectpt}a with that of
the solid curve. This conclusion is further established by the observation
that, when primordial $\kT$ effects are not included, see 
Fig.~\ref{fig:connectpt}b, indeed the spectrum of the old model becomes
almost identical to that of the shaded region. In addition, it can already
here be recognized that the junction baryon must have larger $\pT$ in those
events where a diquark is \emph{not} formed, by comparing the slopes of the
full junction baryon spectrum (dashed curves) with those of the shaded
regions in either figure. We now study this further.
 
If a diquark is not formed, then the junction may \emph{a priori}
be colour connected to partons going in widely different
directions in the transverse plane. Nonetheless, as was described in Section
\ref{sec:jufrag}, the fragmentation occurs in such a way that the
junction baryon is always the last, i.e.\ normally slowest, hadron to 
be formed in either of the three directions. Hence, while the colour
neighbours of the junction may themselves have large transverse
momenta, this momentum will in general be taken by the leading
hadrons formed in the fragmentation and not by the junction
baryon. Unless two of those partons are going in roughly the
same direction in $\varphi$, the junction baryon itself will still
only obtain a fairly small \pT. The end result is a rather small
\pT\ enhancement, that is masked by the decreased primordial 
\kT, Fig.~\ref{fig:connectpt}. 
 
Thus, the main difference in the new model is that the beam baryon
number can migrate \emph{longitudinally} to a much larger extent
than in the old model. Empirically, it may be desirable to be able to limit 
the degree to which this baryon number stopping occurs, and furthermore both
perturbative and impact-parameter arguments allow much of the
activity to be correlated in `hot spot' regions that leave the
rest of the proton largely unaffected. Therefore a free
suppression parameter is introduced, such that further gluons
more frequently connect to a string piece that has already been
disturbed. In this way, gluons would preferentially be found on
one of the three colour lines to the junction. This will reduce the amount of
baryon number stopping and is an important first
modification, but most likely it is not the \emph{only} relevant 
ordering principle. 
 
\subsubsection{Ordered Colour Correlations}
\begin{figure}
\begin{center}
\vspace*{0.5cm}
\begin{fmffile}{fmfcolcor}
\begin{tabular}{cc}
\begin{fmfgraph*}(200,150)
\fmfset{curly_len}{2.5mm}
\fmfforce{0.4w,0.9h}{g1}
\fmfforce{0.65w,0.25h}{g2}
\fmfforce{0.9w,0.7h}{g3}
\fmfforce{0.1w,0.1h}{br}
\fmfforce{0.0w,0.0h}{O}
\fmfforce{1w,0h}{R}
\fmfforce{0w,1h}{T}
\fmf{plain}{O,T}
\fmf{plain}{O,R}
\fmfv{d.sh=tri,d.siz=10,d.ang=270,lab=$i$,l.ang=-45}{R}
\fmfv{d.sh=tri,d.siz=10,lab=$y$,l.ang=135}{T}
\fmfv{d.sh=circ,d.fil=empty,lab=BR,d.siz=22,l.dist=0}{br}
\fmfv{d.sh=circ,d.fil=empty,lab=I$_1$,d.siz=18,l.dist=0}{g1}
\fmfv{d.sh=circ,d.fil=empty,lab=I$_2$,d.siz=18,l.dist=0}{g2}
\fmfv{d.sh=circ,d.fil=empty,lab=I$_3$,d.siz=18,l.dist=0}{g3}
\fmf{gluon,foreground=(0.8,,0.8,,0.8)}{br,g1}
\fmf{gluon,foreground=(0.8,,0.8,,0.8)}{br,g2}
\fmf{gluon,foreground=(0.8,,0.8,,0.8)}{br,g3}
\fmf{dashes,foreground=(0.,,.8,,0.)}{br,g1}
\fmf{dashes,foreground=(0.,,0.,,1.)}{br,g3}
\fmf{dashes,foreground=(0.,,.8,,0.)}{g3,g2}
\fmf{dashes,foreground=(1.,,0.,,0.)}{g2,g1}
\end{fmfgraph*} &
\begin{fmfgraph*}(200,150)
\fmfset{curly_len}{2.5mm}
\fmfforce{0.4w,0.9h}{g1}
\fmfforce{0.65w,0.25h}{g2}
\fmfforce{0.9w,0.7h}{g3}
\fmfforce{0.1w,0.1h}{br}
\fmfforce{0.0w,0.0h}{O}
\fmfforce{1w,0h}{R}
\fmfforce{0w,1h}{T}
\fmf{plain}{O,T}
\fmf{plain}{O,R}
\fmfv{d.sh=tri,d.siz=10,d.ang=270,lab=$i$,l.ang=-45}{R}
\fmfv{d.sh=tri,d.siz=10,lab=$y$,l.ang=135}{T}
\fmfv{d.sh=circ,d.fil=empty,lab=BR,d.siz=22,l.dist=0}{br}
\fmfv{d.sh=circ,d.fil=empty,lab=I$_1$,d.siz=18,l.dist=0}{g1}
\fmfv{d.sh=circ,d.fil=empty,lab=I$_2$,d.siz=18,l.dist=0}{g2}
\fmfv{d.sh=circ,d.fil=empty,lab=I$_3$,d.siz=18,l.dist=0}{g3}
\fmf{gluon,foreground=(0.8,,0.8,,0.8)}{br,g1}
\fmf{gluon,foreground=(0.8,,0.8,,0.8)}{br,g2}
\fmf{gluon,foreground=(0.8,,0.8,,0.8)}{br,g3}
\fmf{dashes,foreground=(0.,,.8,,0.)}{br,g2}
\fmf{dashes,fore=blue}{br,g3}
\fmf{dashes,foreground=(0.,,.8,,0.)}{g3,g1}
\fmf{dashes,fore=red}{g2,g1}
\end{fmfgraph*}\\[2mm]
\it a) & \it b) \\
\end{tabular}
\end{fmffile}
\caption{Example of initial-state colour correlations in an imagined event
  in which three gluons have been knocked out of an incoming hadron by
  colourless objects (for simplification), with no parton showers.
In case {\it a)} the gluons have been randomly
  attached in colour to each other and to the beam remnant, as indicated by
  the dashed lines, whereas in case
  {\it b)} the connections follow the rapidities of the hard scattering
  systems.
\label{fig:colcor}}
\end{center}
\end{figure}
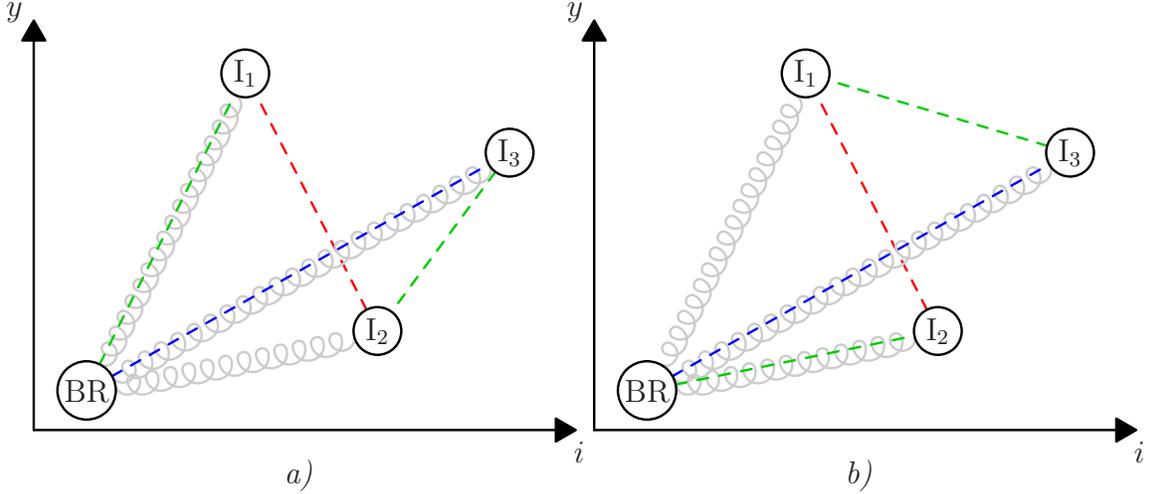

With the gluons connected preferentially along one of the three colour lines
to the junction, we now address the question of their relative order along 
that line. If this order is random, then
strings will in general be stretched criss-cross in the event. This is
illustrated in Fig.~\ref{fig:colcor}a for a very simplified situation. 
However, it is unlikely that such a scenario catches all the relevant
physics. More plausible is that, among all the possible final-state 
colour topologies, those that correspond to the smaller total
string length are favoured, all other aspects being the same. One possible
way of introducing such correlations is illustrated in Fig.~\ref{fig:colcor}b.
 
In case {\it a)} of Fig.~\ref{fig:colcor}  there are four string pieces
  criss-crossing the rapidity range between the systems $I_2$ and
  $I_3$, while in case {\it b)} there are only two string pieces spanning
  this range. Hence the total string length should on average
be smallest for the latter type of correlations.
However, the rapidity distance is not the only variable determining the
  string lengths, also the transverse separations play a role. Moreover, when
  the interactions exchange colour between the colliding objects, there is no
  longer a unique correspondence between the colour flow
  of the initial state and that of the final state.
 
To investigate these effects quantitatively, we consider three different
possibilities for initial-state colour correlations (with the suppression of
attachments breaking up the beam remnant applicable to all cases):
\begin{enumerate}
\item Random correlations, as in Fig.~\ref{fig:colcor}a.
\item Initiator gluons are attached preferentially in those places that order
  the hard scattering systems in rapidity, as in
  Fig.~\ref{fig:colcor}b. The rapidities are calculated at a stage
before primordial $\kT$ is included. Hence, $y = \frac12 \ln \frac{x_1}{x_2}$.
 
For beam remnant partons, the rapidities are not yet known at the stage
discussed here, since the initial-state colour connections are in our
framework made \emph{before} primordial $\kT$ and beam remnant $x$ values are
assigned. 
However, beam remnant partons are almost by definition characterized by
having large longitudinal and small transverse momenta.  Thus, we assign a
fixed, but otherwise
arbitrary, large rapidity to each of the beam remnant partons, in the
direction of its parent hadron. Finally, gluons are attached
sequentially to the initial valence topology, with the attachments ordered by
minimization of the measure
\begin{equation}
\Delta y = \left|y_g-y_1\right|+\left|y_g-y_2\right|,
\end{equation}
where $y_g$ is the rapidity associated with the attached gluon and $y_{1,2}$
are the rapidities associated with the partons it is inserted between. For
those gluons which appear only as parents of sea quark pairs, the rapidity of
the most central of the daughters is used. 

Note that, since the same hard-scattering rapidities are used for both 
beam remnants, the ordering in the two remnants will 
be closely correlated in this scenario, at least as long as only 
gluon--gluon interactions are considered. 
\item Initiator gluons are attached preferentially in those places that
will give rise to the smaller string lengths in the final
state. This is the most aggressive possibility, where the actual momentum
separations of final-state partons, together with the full colour flow
between the two sides of a hadronic collision, is used to determine
which gluon attachment will result in the smallest increase in potential
energy (string length) of the system, with each gluon being attached one
after the other. The measure we use to define the increase in string length,
for a particular attachment, is \cite{BNV,lambda}
\begin{equation}
\Delta\lambda = \ln \left[\frac{2}{m_0^2}\frac{
(p_{c_1} \cdot p_{\bar{c}_1})~(p_{c_2} \cdot p_{\bar{c}_2})}{
(p_{\bar{c}_1}\cdot p_{c_2})}\right],
\label{eq:stringlength}
\end{equation}
where $m_0$ is a normalization constant, which drops out when comparing the
string lengths of two different gluon attachments, and $c_1$ ($\bar{c}_2$)
represents the final-state parton carrying the colour (anticolour) index of
the attached gluon.
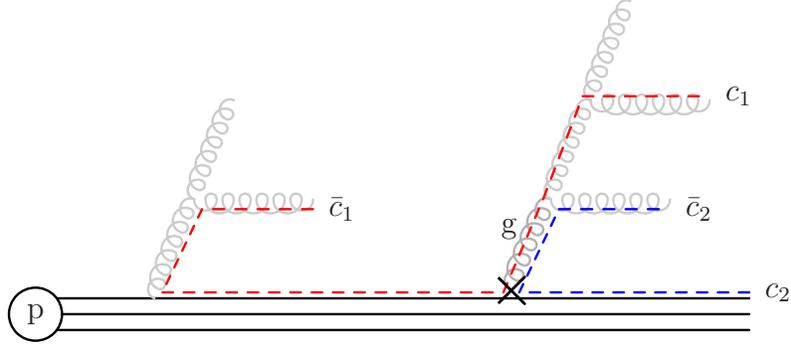
\begin{figure}
\begin{center}
\vspace*{-0.5cm}\begin{fmffile}{fmfcolmom}
\begin{fmfgraph*}(300,150)
\fmfset{curly_len}{2.3mm}
\fmfset{dots_len}{1mm}
\fmfforce{0.05w,0.1h}{ll}
\fmfforce{0.95w,0.1h}{rl}
\fmfforce{0.95w,0.115h}{rl.r}
\fmfforce{0.05w,0.06h}{ll2}
\fmfforce{0.95w,0.06h}{rl2}
\fmfforce{0.05w,0.02h}{ll3}
\fmfforce{0.95w,0.02h}{rl3}
\fmfforce{0.2w,0.1h}{v1}
\fmfforce{0.25w,0.35h}{vgg1}
\fmfforce{0.3w,0.6h}{v1e}
\fmfforce{0.4w,0.35h}{vgg1e}
\fmfforce{0.65w,0.12h}{v2}
\fmfforce{0.7w,0.35h}{vgg2}
\fmfforce{0.85w,0.35h}{vgg2e}
\fmfforce{0.75w,0.6h}{vgg3}
\fmfforce{0.9w,0.6h}{vgg3e}
\fmfforce{0.8w,0.85h}{v2e}
\fmfforce{0.64w,0.115h}{v2.l}
\fmfforce{0.69w,0.35h}{vgg2.l}
\fmfforce{0.74w,0.61h}{vgg3.l}
\fmfforce{0.9w,0.61h}{vgg3e.l}
\fmfforce{0.21w,0.115h}{v1.r}
\fmfforce{0.26w,0.325h}{vgg1.r}
\fmfforce{0.4w,0.325h}{vgg1e.r}
\fmfforce{0.66w,0.115h}{v2.r}
\fmfforce{0.71w,0.325h}{vgg2.r}
\fmfforce{0.85w,0.325h}{vgg2e.r}
\fmfforce{0.75w,0.6h}{vgg3.r}
\fmfforce{0.9w,0.6h}{vgg3e.r}
\fmfforce{0.8w,0.85h}{v2e.r}
\fmf{plain}{ll,rl}
\fmf{plain}{ll2,rl2}
\fmf{plain}{ll3,rl3}
\fmf{gluon,foreground=(0.8,,0.8,,0.8)}{vgg1,v1}
\fmf{gluon,foreground=(0.8,,0.8,,0.8)}{v1e,vgg1}
\fmf{gluon,foreground=(0.8,,0.8,,0.8)}{vgg1,vgg1e}
\fmf{gluon,foreground=(0.8,,0.8,,0.8)}{v2e,vgg3,vgg2}
\fmf{gluon,foreground=(0.7,,0.7,,0.7),label=g}{vgg2,v2}
\fmf{gluon,foreground=(0.8,,0.8,,0.8)}{vgg2,vgg2e}
\fmf{gluon,foreground=(0.8,,0.8,,0.8)}{vgg3,vgg3e}
\fmf{dashes,foreground=(1.0,,0.0,,0.0)}%
{vgg1e.r,vgg1.r,v1.r,v2.l,vgg2.l,vgg3.l,vgg3e.l}
\fmf{dashes,foreground=(0.0,,0.0,,1.0)}{rl.r,v2.r,vgg2.r,vgg2e.r}
\fmfv{d.sh=cross}{v2}
\fmfv{d.sh=circ,d.fill=empty,d.siz=20,label=p,l.dist=0}{ll2}
\fmfv{lab=$\bar{c}_1$,l.ang=0}{vgg1e.r}
\fmfv{lab=$\bar{c}_2$,l.ang=0}{vgg2e.r}
\fmfv{lab=$c_1$,l.ang=0}{vgg3e.l}
\fmfv{lab=$c_2$,l.ang=0}{rl.r}
\fmfv{d.sh=cross}{v2}
\end{fmfgraph*}
\end{fmffile}
\caption{Example showing the colour flow produced by attaching the gluon $\g$
  at the place indicated by the cross.
\label{fig:colmom}}
\end{center}
\end{figure}
To illustrate, Fig.~\ref{fig:colmom} shows the partons that enter the above
expression for a specific example. Before the attachment, a single string
piece is spanned between the final-state partons that carry the colour
indices denoted $\bar{c}_1$ and $c_2$. After the attachment, there are two
string pieces, one that is spanned between $c_1$ and $\bar{c}_1$, the other
between $c_2$ and $\bar{c}_2$, hence the \emph{increase} in string length is
given by the expression eq.~(\ref{eq:stringlength}).
 
As above, however, note that neither primordial $\kT$ nor beam
remnant longitudinal momenta have yet been assigned at this stage.
Simplified kinematics are therefore set up, to be used only for
the purpose of determining the colour connections: the momentum
remaining in the beam remnant on each side is divided evenly among
the respective remnant partons (junctions are here treated
simply as `fictitious partons', receiving the same momentum as the
`real' remnant partons), and primordial $\kT$ effects are ignored.
Thereby, parton pairs involving (at least) two partons in one of
the beam remnants will come to have zero invariant mass, hence the
total $\Delta\lambda$ will be negative infinity for such pairs. Obviously,
this is not desirable; one string piece with vanishing invariant
mass should not affect the comparison, hence we impose a minimum
invariant mass of $m_0$ for each string piece. If knowledge of the
full kinematics of the final state was available a better choice
could of course be made here. However, these two aspects are
intertwined. The kinematics of the final state may depend on the
colour connections assumed for the initial state (see Section
\ref{sec:kT} above), and vice versa. Our choice has been to
determine the initial-state colour connections first, and then
subsequently construct the final-state kinematics, hence some
approximation is necessary at this point.
\end{enumerate}
 
A variable which we have found to be sensitive to the colour connections in
an event is the mean $\pT$ vs.~charged multiplicity, $\langle
\pT\rangle(n_{\mathrm{ch}})$ \cite{Zijl}. 
In scenarios with large string lengths, each
additional interaction would result in a large increase in hadron
multiplicity.  This large multiplicity per interaction means
that, in such scenarios, observed average charged multiplicities are
reproduced with comparatively large values of $p_{\perp 0}$, i.e.~with
only a few parton--parton interactions taking place per event. Hence,
correspondingly little perturbative $\pT$ is generated. On the other hand, in
scenarios with smaller string lengths, comparatively more interactions would
be required to produce the same multiplicity, hence more perturbative $\pT$
would be generated per charged particle, bringing $\langle
\pT\rangle(n_{\mathrm{ch}})$ up.
 
\begin{figure}
\begin{center}
\includegraphics*[scale=0.9]{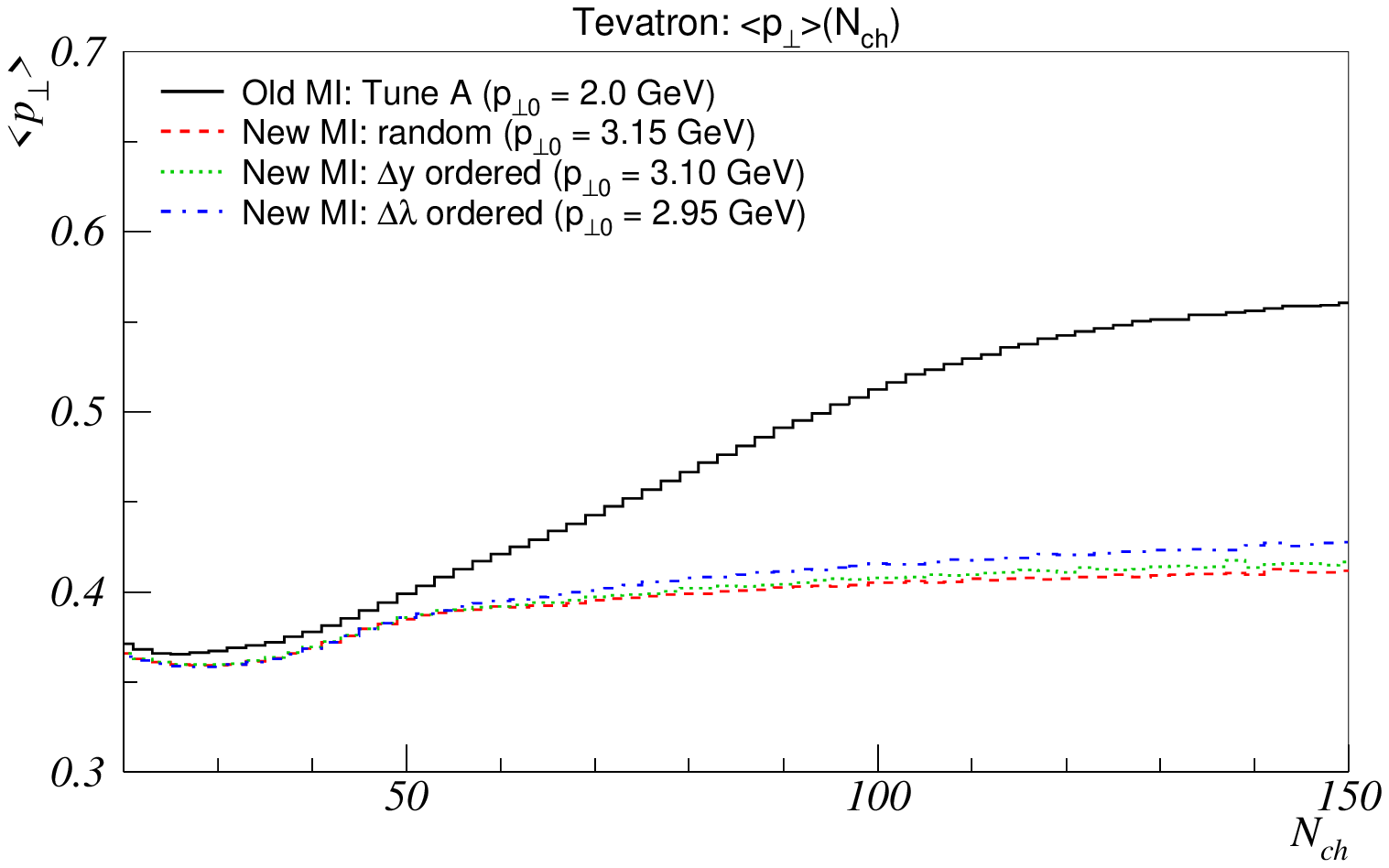}\vspace*{-.2cm}
\caption{$\left<p_{\perp}\right>$ vs.~$n_{\mathrm{ch}}$ at the Tevatron
for Tune A (solid lines), and for the new model with
random (dashed lines), rapidity ordered (solid lines), and
string length ordered (dotted lines) correlations in the initial
state. Note that the origo of the plot is \emph{not} at (0,0).
For each of the new MI scenarios, $\pTo$ was selected to give the same
average charged multiplicity as Tune A, with the same impact parameter
dependence as Tune A (i.e.~a double Gaussian matter distribution).
\label{fig:initialptvsnch}}
\end{center}
\end{figure}

In Fig.~\ref{fig:initialptvsnch}, we show the $\left<\pT\right>$
vs.~$n_{\mathrm{ch}}$ distribution for each of the possibilities described
above.
In all cases, $p_{\perp 0}$ was first selected so as to give
identical average multiplicities, corresponding to the multiplicity obtained
with Tune A.
Since the $\Delta\lambda$ ordering results in the largest average
number of interactions for a given multiplicity, it has the largest average
$\pT$ per particle of the new scenarios, while the random ordering results in
the smallest number of interactions.
 
One also notes that Tune A, which more or less agrees
with recent experimental data \cite{CDFglobal,CDFped,meanptfornch}, 
shows an even steeper rise with $n_{\mathrm{ch}}$ than any of the
new scenarios. This tune of the old model is such that the 
partons produced by subsequent scatterings will almost always be
hooked up to the existing configuration in the way that minimizes
the total string length. This is more or less like the
$\Delta\lambda$ ordering described above, but with the essential
difference that the $\Delta\lambda$ ordering only concerns
the colour lines that are present in the \emph{initial state}, while the
ordering of parton attachments in the old model occurs in the
\emph{final state}, without any attempt at constructing a consistent colour
flow in the event.
 
From these observations, an interesting inference can be made. By
the failure of even the $\Delta\lambda$ ordering of the colour
lines in the initial state to describe the $\langle
\pT\rangle(n_{\mathrm{ch}})$ distribution, it appears that the
colour flow in physical events cannot be correctly described by
merely arranging the colour lines present in the initial state.
We imagine two possible causes for this. Firstly, the initial-state 
showers associated with each scattering are constructed by
backwards DGLAP evolution of each scattering initiator separately,
down to the shower cutoff scale. This does not take into account
the possibility that the showers could be intertwined, i.e.\ that a
parton at low virtuality, but above the shower cutoff scale, could
have branched to give rise to \emph{two} higher--virtuality
scattering initiators. Secondly, one or more mechanisms causing
colour exchanges between the showers may be active, both in the
initial state as well as in the final state. Below we present some first
studies related to the topic of colour exchanges, well aware that no simple
solutions are to be expected.
 
\subsection{Final-State Colour (Re-)Connections \label{sec:fsi}}
To investigate how much more we need to `mess around with the
colours', we study a crude model of colour exchanges in the final
state. Essentially, we rearrange the colour connections between
the final-state partons in a manner that, taken to the extreme,
will converge on that string configuration which has the smallest total
`string length', according to the $\lambda$ measure introduced above. This
corresponds roughly to a minimization of the multiplicity produced
when the parton system hadronizes. Note that we only apply this procedure to
events where at least two interactions have occurred, to avoid the
reconnections leading to large central rapdity gaps, i.e.~diffractive
topologies.  

For a given configuration of final-state partons in momentum and
colour space, we apply an iterative procedure that successively brings down 
the total string length of the system, by the following steps:
\begin{enumerate}
\item First, we have assigned Les Houches Accord style colour tags
  \cite{lha1} to all partons, so that each colour tag in the final state is
  matched by exactly one corresponding anticolour tag in the final
  state, with one string piece spanned between them. 
  Junctions are special, since colour lines end
  there. In the following, we do not consider string pieces ending on
  junctions. 
\item Secondly, we decide on a fraction, $F$, of the colour tags present in
  the event for which we will attempt to make a reassignment. Note that $F$
  can be larger than one, since several different reassignments 
($n(n-1)/2$ for $n$ colour tags, neglecting junctions) are normally possible. 
\item Next, we select two colour tags at random, $c_1$ and $c_2$. Denoting 
the final-state parton carrying $c_1$ colour (anticolour) by $i_1$
($j_1$) and 
the one carrying $c_2$ colour (anticolour) by $i_2$ ($j_2$), we
compute the combined string length, $\lambda$, for the two string pieces
$i_1$---$j_1$, $i_2$---$j_2$:
\begin{equation}
\lambda = \ln\left(\frac{2p_{i_1}\cdot p_{j_1}}{m_0^2}\frac{2p_{i_2}\cdot
  p_{j_2}}{m_0^2}\right).
\end{equation}
By swapping e.g.~the anticolours, a different string topology arises,
$i_1$---$j_2$, $i_2$---$j_1$, with length
\begin{equation}
\lambda' =  \ln\left(\frac{2p_{i_1}\cdot p_{j_2}}{m_0^2}\frac{2p_{i_2}\cdot
  p_{j_1}}{m_0^2}\right).
\end{equation}
If $\lambda'<\lambda$, the colour reassignment is
accepted, otherwise the original assigments are kept. If the fraction of
colour tags tried so far is smaller than $F$, a new pair of random colour
tags is selected. 
\item Once the fraction $F$ of colour tags has been tried, two
  things can happen. If at least one reconnection was made, then the colour
  topology now looks different, and the entire iteration is restarted. If no
  reconnection was made, the iteration ends.
\end{enumerate}
Briefly summarized, we thus introduce the fraction $F$ as a free parameter 
that controls the strength of colour reconnections in the final state.

\begin{figure}
\begin{center}
\includegraphics*[scale=0.9]{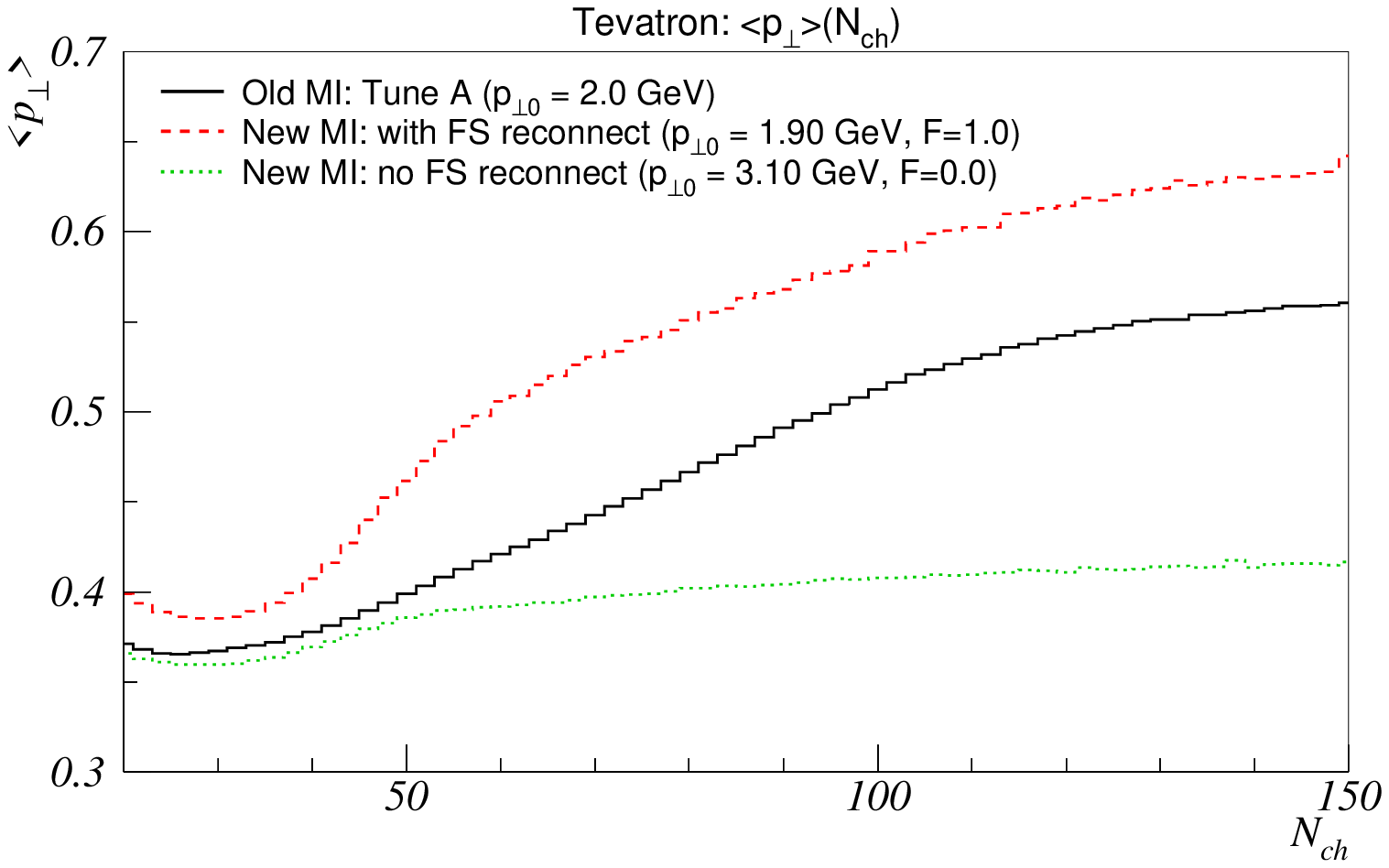}\vspace*{-1mm}
\caption{
$\left<p_{\perp}\right>$ vs.~$n_{\mathrm{ch}}$ at the Tevatron
for Tune A (solid line), and for the new model with (dashed line) and
without (dotted line) final-state reconnections allowed. Both of the new
models use rapidity ordering of the colour lines in the initial state and 
give the same average charged multiplicity as Tune A, with the same impact
parameter dependence as Tune A.
\label{fig:reconnect}} 
\end{center}
\end{figure}

As stated, this method is very crude and should not be
interpreted as representing physics \emph{per se}, but it does allow us to
study whether a significant effect can be achieved by manipulating the colour
correlations to reduce the string lengths. As illustrated by the dashed
histogram in Fig.~\ref{fig:reconnect}, this is very much the case. Here, we
have allowed a large amount of reconnections to occur, $F=1$, adjusting
$\pTo$ down so as to reproduce the average charged multiplicity of Tune
A. 
As could be expected with this somewhat extreme choice of parameters, the
new model now lies well above the data. (We shall return to more realistic
tunings in Section \ref{sec:tests}.)

However, this result should not be taken as evidence for the existence of 
colour reconnections in
physical events. Rather it allows us to infer that, by changing the colour
structure of events, it should be possible to obtain agreement with the data
within our framework. It is encouraging that, by studying and attempting to
describe this distribution, we may learn interesting lessons concerning the
highly non--trivial issue of colour flow in hadronic interactions. We
therefore plan to go further, to construct more physically motivated models
for colour rearrangement between partons, both in the initial state and in
the final state, and also to allow for the possibility of intertwining the
initial-state showers. 

Although all of these issues appear almost hopelessly
complicated from the point of view of pure QCD, the salient features of the
resulting physics may not be all that hard to penetrate. For instance, we
imagine that the probability for two hard-scattering initiators to have
originated from a common branching should be proportional to the probability
that their spatial wavefunctions overlap, i.e.~two very high--virtuality
initial-state partons associated with different scatterings 
are most likely uncorrelated, since they only resolve very small
distance scales in their parent hadron. On the other hand, two
low--virtuality partons, even though associated with different scatterings,
may very well have come from one and the same parent parton, since their
wavefunctions are comparatively much larger. We will explore such and other
ideas in a future study.

\section{Model Studies\label{sec:tests}}
In this Section we concentrate on illustrating the properties of
models that, as a baseline, roughly reproduce the charged
multiplicity distribution of Tune A in $\p\pbar$ collisions at a
centre-of-mass energy of 1800\GeV. Our studies only concern inelastic
nondiffractive events, i.e.~essentially the same as the experimentally
defined (trigger-dependent) ``min-bias'' event sample; we will here use
the two concepts
interchangeably. 

A general feature of the new multiple interactions modeling is that the
added parton showers and the less efficient string energy minimization result 
in a higher
  multiplicity per interaction than Tune A. In order to 
arrive at the same average hadron multiplicity as Tune A, without too much
colour reconnection required in the final state, 
a generally larger $\pTo$ cutoff should be used. Table
\ref{tab:models} lists three different tunes of the new framework, with
successively smaller $\pTo$ values and with different schemes for the
initial-state colour correlations.
\begin{table}[t]
\begin{center}
\begin{tabular}{lccc|ccc|ccc|}
Model&IS Colour            &  $\pTo$     &    &
\multicolumn{3}{c|}{Tevatron} &  \multicolumn{3}{c|}{LHC}\\[-1mm] 
name&\multicolumn{1}{l}{Ordering}&[\GeV] &$F$ & 
$\langle n_{\mathrm{int}}\rangle$ & 
$\langle n_{\mathrm{part}}\rangle$ & 
$\langle n_{\mathrm{rcp}}\rangle$ & 
$\langle n_{\mathrm{int}}\rangle$ & 
$\langle n_{\mathrm{part}}\rangle$ & 
$\langle n_{\mathrm{rcp}}\rangle$ \\ 
\toprule
Ran & random    & 2.50 & 0.55 & 3.3 & 21.5 & 18 & 4.2 & 40.2 & 45 \\
Rap & $\Delta y$& 2.40 & 0.55 & 3.6 & 22.8 & 19 & 4.5 & 43.5 & 49 \\
Lam & $\Delta\lambda$&2.30&0.65& 3.9& 24.5 & 20 & 4.8 & 45.8 & 52 \\
\cmidrule{1-10}
Tune A & --     & 2.00 & --   & 5.7 & 19.2 & -- & 6.9 & 27.7 & -- \\
\bottomrule
\end{tabular}
\caption{Parameters of the three models investigated in the text, and for
  Tune A where applicable. Also shown for each model is the average number of
  parton--parton interactions, $\langle n_{\mathrm{int}}\rangle$, the average
  number of final-state partons, $\langle n_{\mathrm{part}}\rangle$,
  and the average number of colour reconnections taking place, $\langle
  n_{\mathrm{rcp}}\rangle$, per min-bias 
collision at the Tevatron and at the LHC. \label{tab:models}}    
\end{center}
\end{table}
\begin{itemize}
\item The ``Ran'' model is based on a random ordering of the initial-state
  colour correlations, with a fairly large suppression of initiator
  gluon  attachments to colour lines wholly within the beam remnant. Since
  only a minimal ordering of the colour correlations in the initial state are
  thus imposed, each additional interaction will \emph{ab initio} 
  give rise to a relatively large increase in hadron multiplicity. Therefore,
  a comparatively large cutoff $\pTo$ is used, and the $F$ parameter ---
  controlling the amount of final-state reconnections --- is likewise chosen
  fairly large, so as to get the correct average charged multiplicity. 
\item The ``Rap'' model uses the $\Delta y$ measure introduced above to order
  the intial-state colour connections. $\pTo$ can
  thus here be slightly smaller, allowing more interactions on the average
  (with the same $F$ fraction) for the same average charged multiplicity.
\item The ``Lam'' model employs the $\Delta \lambda$ ordering of the
  initial-state colour correlations. In principle, this model should provide
  the most 
  ordered initial state of the three, and thus allow a smaller $\pTo$ and/or
  $F$. Unfortunately, the earlier-mentioned limitations, that the beam
  remnant kinematics is not fully fixed when the minimization is performed,
  leads to final string lengths which are not significantly shorter than for
  the $\Delta y$ ordering. Choosing a smaller $\pTo$ for this tune, the $F$
  fraction is consequently also required to be slightly higher, in order to
  reproduce the Tune A average charged multiplicity.
\end{itemize}
Observe that all three models have a significant number of reconnections per
event, cf.~$\langle n_{\mrm{rcp}}\rangle$ 
in Table \ref{tab:models}. As a fraction
of the total number of potential colour rearrangements it is below the 10\%
level, but one should keep in mind that several clusters of partons appear in
reasonably collimated jets, where reconnections would not be expected
anyway. In this perspective, the amount of reconnections is quite significant.

Common for the new models is a rather smooth overlap profile ExpOfPow(1.8),
as compared to the more peaked double Gaussian of Tune A, this to better
reproduce the shape of the Tune A multiplicity distribution. 
In addition, all of
the new models assume that primordial $\kT$ kicks are compensated uniformly
among all other initiator partons, that composite objects can only be formed in
the beam remnant by valence quarks, and that initial-state colour connections
breaking up the beam remnant are suppressed, so that the relative
probability of attaching a gluon between two remnant partons (i.e.~breaking
up the remnant) as compared to an attachment where at least one `leg' is
outside the remnant is 0.01, whenever the latter type of attachment is
possible. 

\begin{figure}[tp]
\begin{center}
\begin{tabular}{c}
\includegraphics*[scale=1]{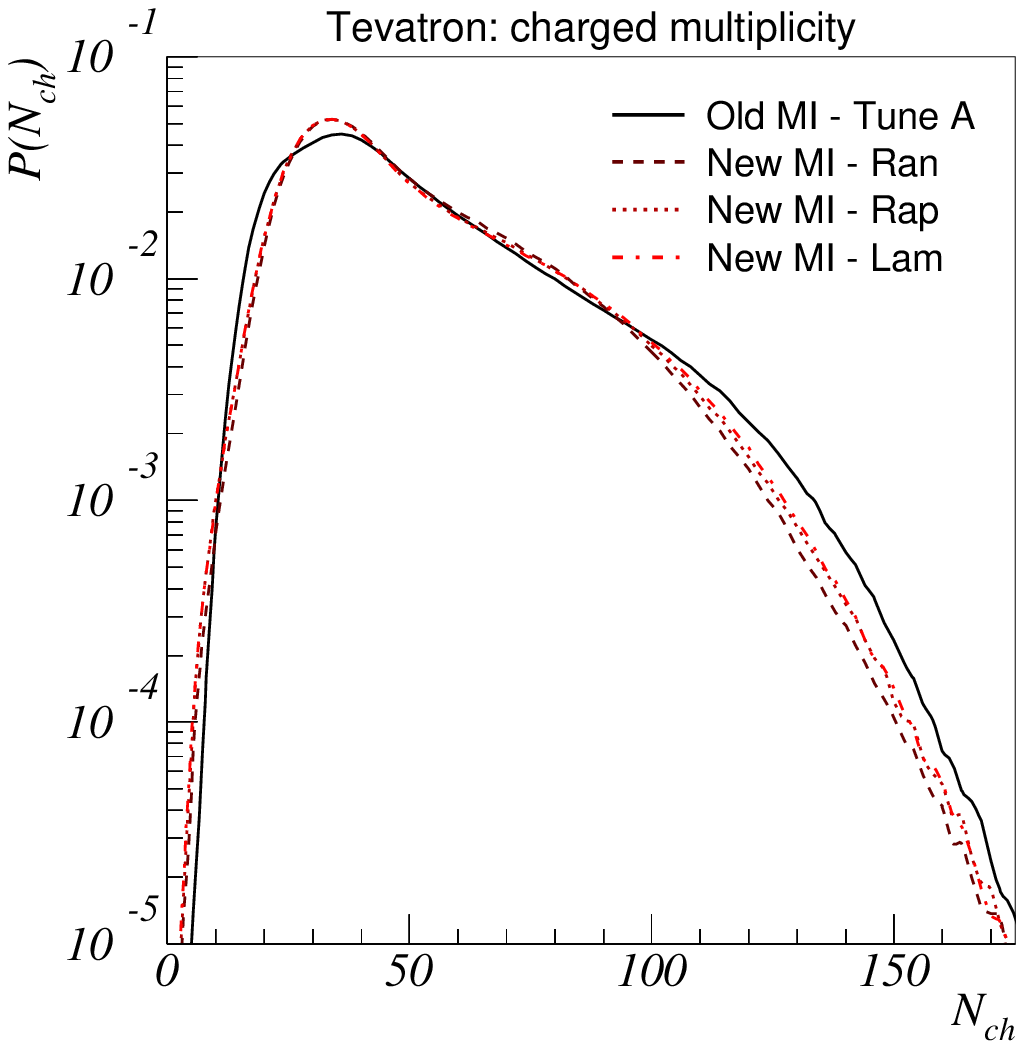}\\[-1mm]
\end{tabular}
\caption{Multiplicity distributions
  for the Tevatron as obtained with Tune A (solid), and the Ran (dashed), Rap
  (dotted), and Lam (dash-dotted) models defined in 
  Table \ref{tab:models}. In all cases, the average charged multiplicity is
  49.5 (within $\pm 0.5$). 
\label{fig:mtest20}}
\end{center}
\end{figure}

Fig.\  \ref{fig:mtest20} shows the Tevatron multiplicity
distributions of these models, as compared to Tune A.
It is apparent that, while the average charged multiplicity is the same, 
the shape of the Tune A multiplicity distribution is not
exactly reproduced by any of the models here investigated. This
should not be taken too seriously; our aim is not to present full-fledged
tunes, rather it is to explore the general properties of the new
framework, and how these compare with those of Tune A. 

Below, we first present comparisons for $\p\pbar$ min-bias 
collisions at 1.8\TeV\ CM energy, highlighting the differences (and
similarities) between the new models and Tune A. Thereafter, 
we apply the same models to the case of $\p\p$
min-bias events at 14\TeV\ CM energy. 

\subsection{Comparisons at the Tevatron}
\begin{figure}[tp]
\begin{center}
\includegraphics*[scale=0.8]{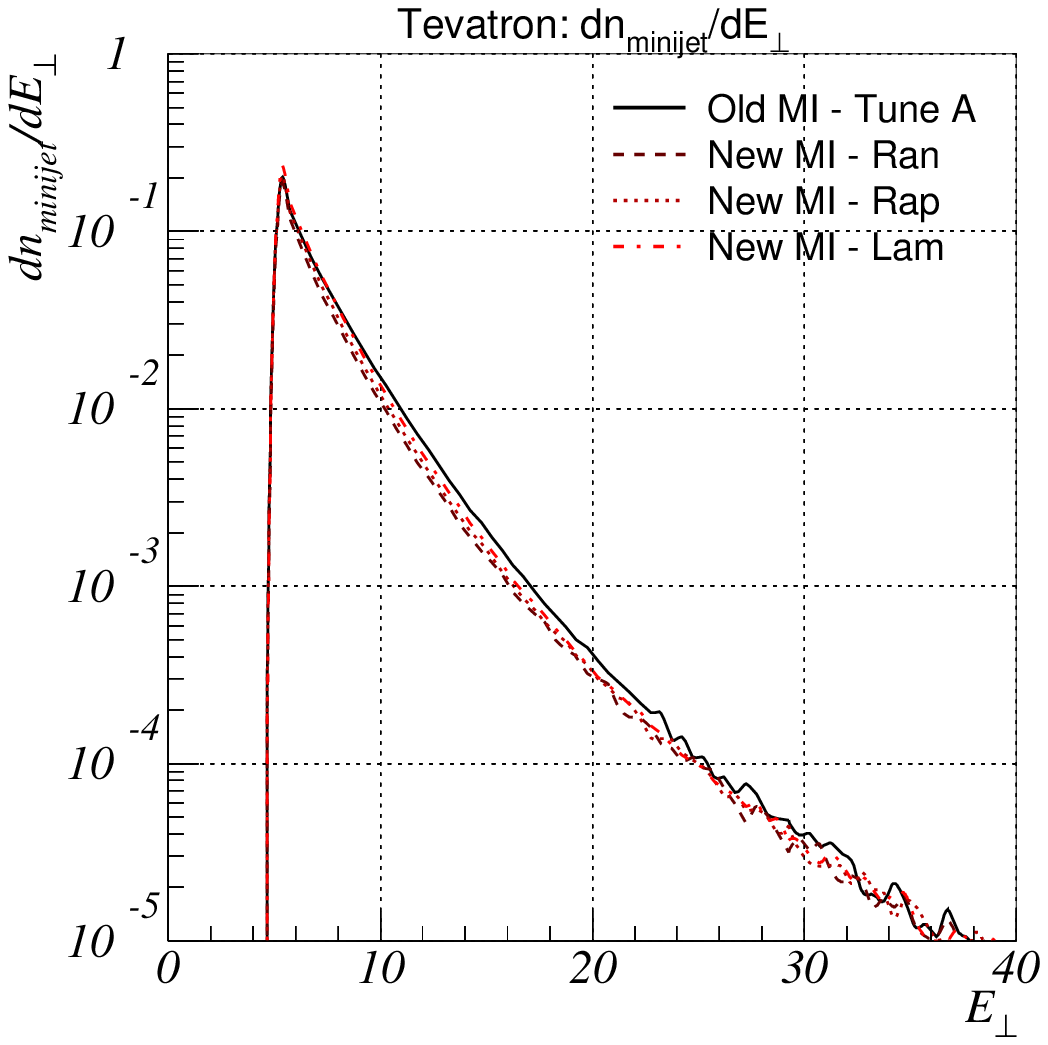}\\[-1mm]
\caption{The number of jets as a function of jet $E_\perp$ (for
  $E_\perp>5\GeV$) in min-bias collisions at the Tevatron.
Results are shown for Tune A (solid), Ran (dashed), Rap (dotted), and
  Lam (dash--dotted), as defined in Table \ref{tab:models}.
\label{fig:minijets}}
\end{center}
\end{figure}

\begin{figure}[tp]
\begin{center}
\begin{tabular}{ccc}\hspace*{-.1cm}
\includegraphics*[scale=0.73]{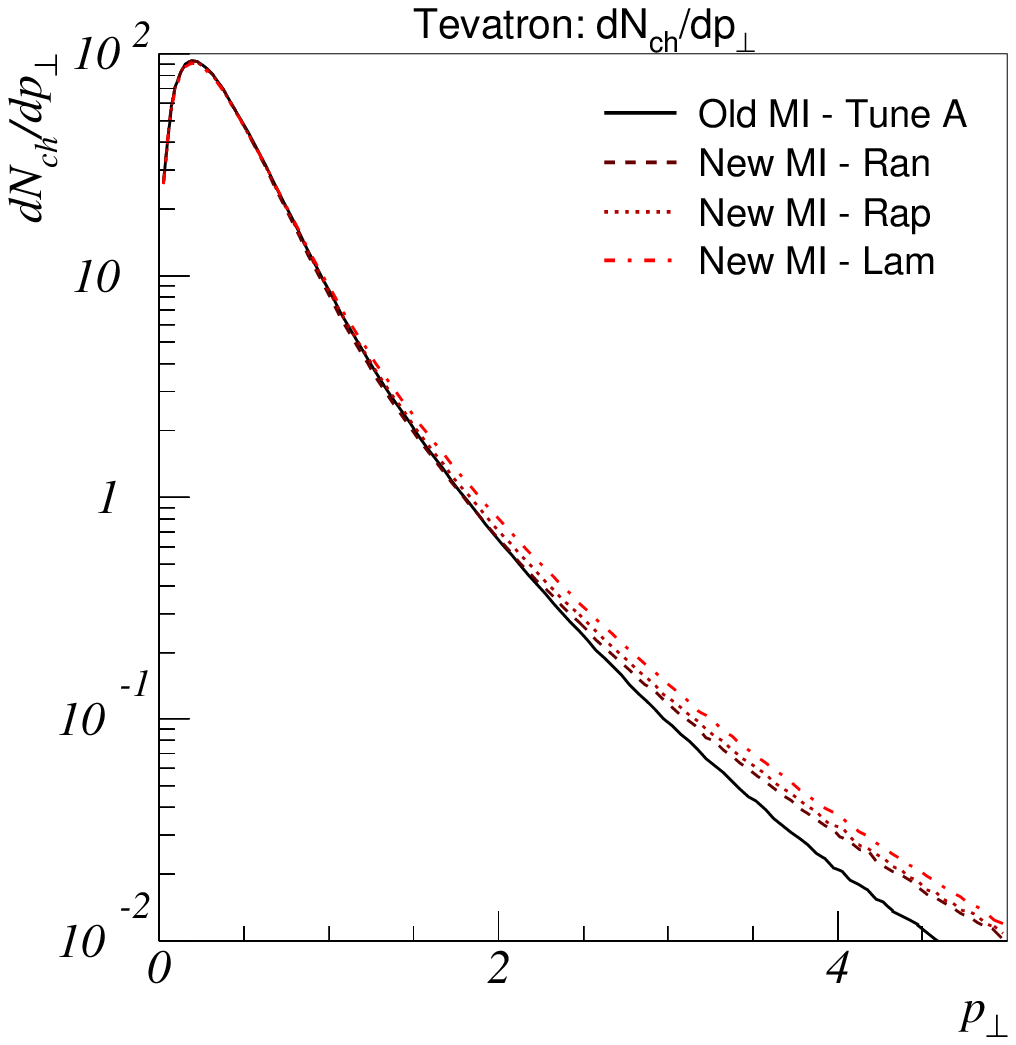}\hspace*{-.1cm}&&\hspace*{-.1cm}
\includegraphics*[scale=0.73]{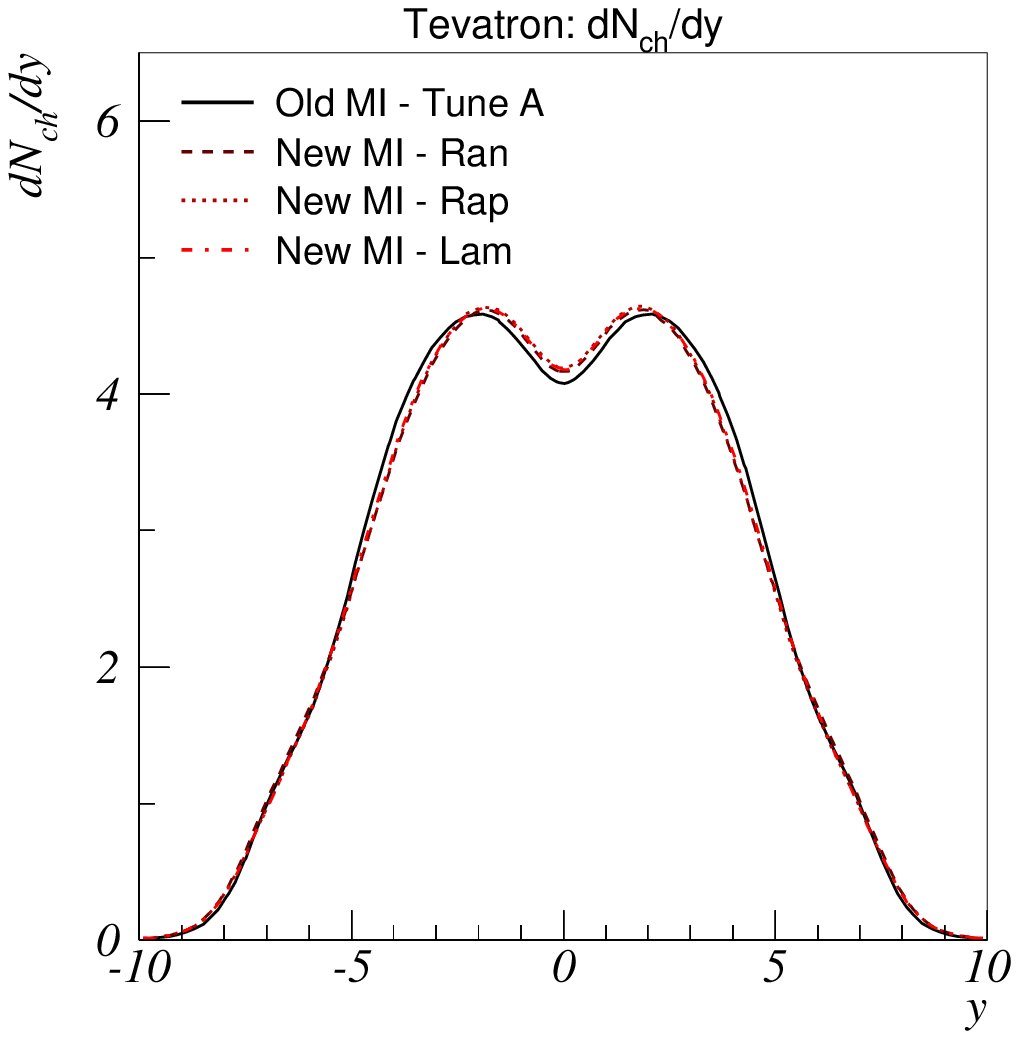} \hspace*{-0.1cm}\\[-1mm]
\it a) & & \it b)\\
\end{tabular}
\caption{Charged particle $\pT$ and $y$ spectra at the Tevatron.
Results are shown for Tune A (solid), Ran (dashed), Rap (dotted), and
  Lam (dash--dotted), as defined in Table \ref{tab:models}.
\label{fig:tevinclusive}}
\end{center}
\end{figure}

Despite the differences between the old and new frameworks, the starting point
in both cases is still that of a perturbative sequence of $\pT$-ordered
scatterings. Especially for the hardest partons there should thus be next to 
no difference between the old and new frameworks. An illustration of this is
given in Fig.\  \ref{fig:minijets}, where the probability of finding a jet 
with transverse energy $E_\perp$ is plotted against $E_\perp$. 
A simple cone algorithm with cone size $\Delta
R=\sqrt{\Delta\eta^2+\Delta\phi^2}=0.7$ 
has been used to cluster the jets, and only particles with
$\left|\eta\right|<2.5$ are included.  
As can be observed, there is hardly any difference between Tune A and the new
models here. Further, the models exhibit similar charged particle spectra
both in transverse momentum and in rapidity, 
Fig.\  \ref{fig:tevinclusive}, with
a slightly harder $\pT$ spectrum and, consequently, a slightly more
central $y$ one in the new models. 

\begin{figure}[tp]
\begin{center}\vspace*{-7mm}
\begin{tabular}{ccc}\hspace*{-.1cm}
\includegraphics*[scale=0.73]{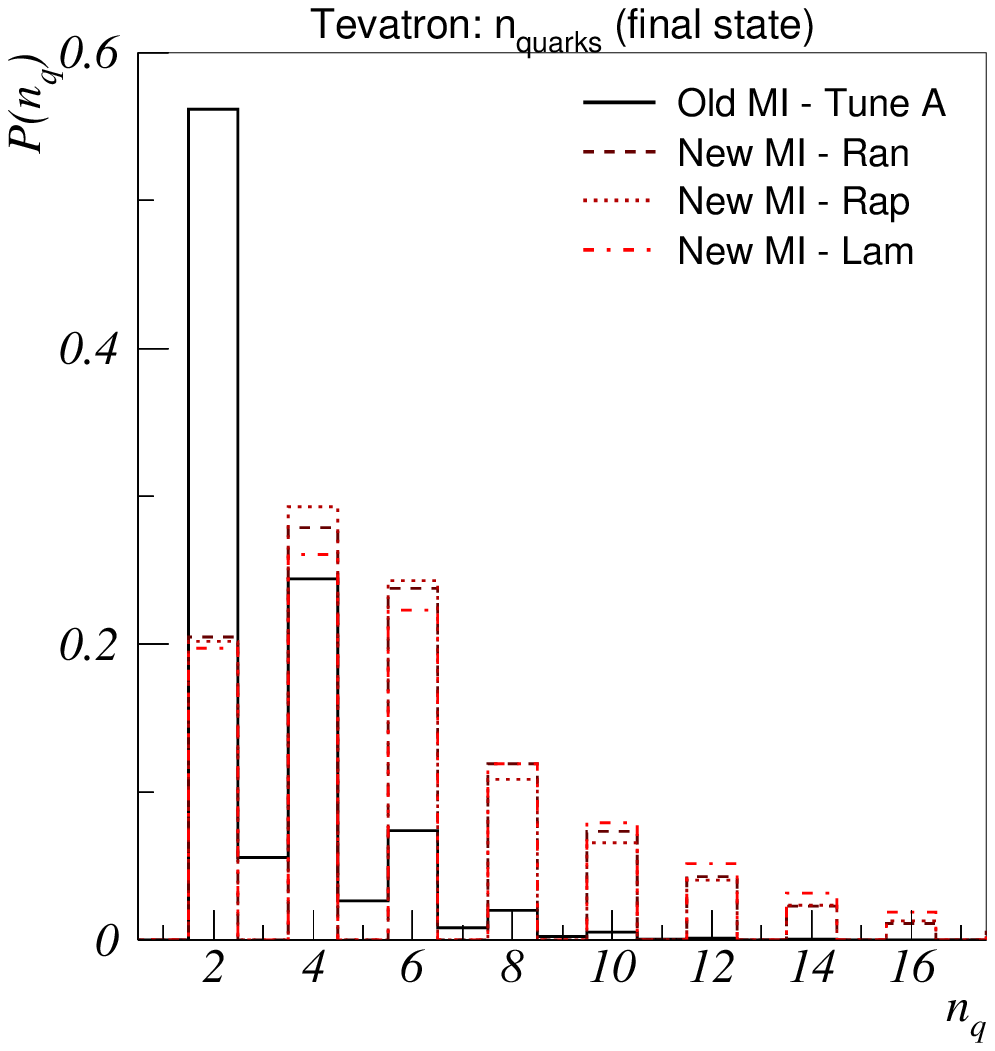}\hspace*{-.1cm}&&\hspace*{-.1cm}
\includegraphics*[scale=0.73]{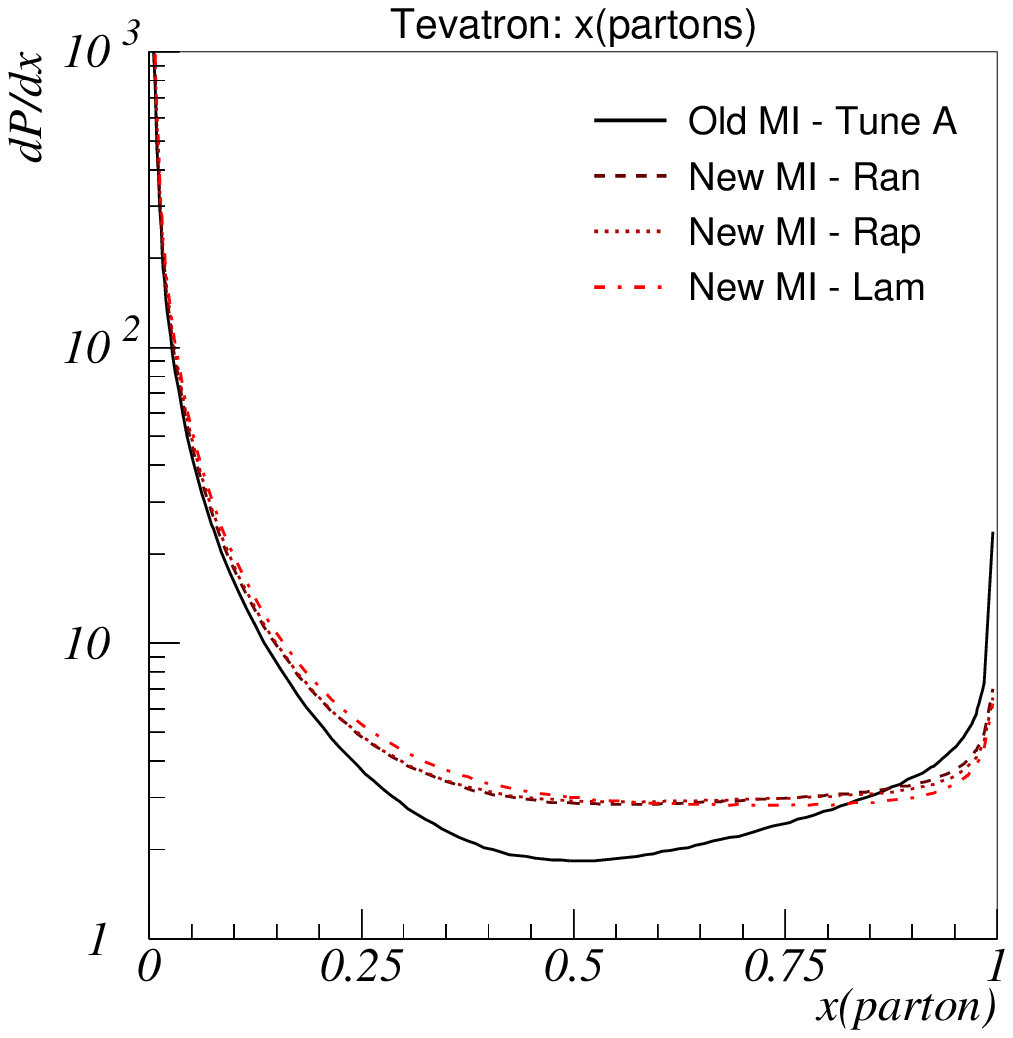} \hspace*{-0.6cm}\\[-1mm]
\it a) & & \it b)\\
\end{tabular}
\caption{{\it a)} Number of final-state quarks (not counting diquarks) 
and {\it b)} final-state parton
  $x$ values at the Tevatron.
Results are shown for Tune A (solid), Ran (dashed), Rap (dotted), and
  Lam (dash--dotted), as defined in Table \ref{tab:models}.
\label{fig:nquarks}}
\end{center}
\end{figure}

Having thus convinced ourselves that the overall features of the models agree, 
we turn to the aspects in which the new and old scenarios are
expected to differ. One significant change is the possibility 
to knock out several valence quarks from the beam hadron. To
quantify, the number of quarks (excluding diquarks) in the final
state is illustrated in Fig.\  \ref{fig:nquarks}a. 
As can readily be observed, final-state quarks are much more abundant in the
new scenarios. Valence quark interactions are, however, not the only cause of
this. Diquarks are not included in Fig.\ 
\ref{fig:nquarks}a, and (as discussed in Section \ref{sec:remnants}) 
these are less
frequently formed in the new scenarios, hence more of the quark content here 
appears as individual quarks in the final state. Further
illustration of this is given by Fig.\  \ref{fig:nquarks}b, where the
$x=2E/\sqrt{s}$ values of all final-state partons, including diquarks, 
are shown. The peak towards low values
comes mainly from gluons and is the same for Tune A and the new
models. However,  
some of the content of the $x=1$ peak in Tune A has vanished in the new
models and has been replaced by a larger plateau at intermediate $x$,
since a fraction of large-$x$ diquarks has been split up into
individual valence--like quarks.

\begin{figure}[tp]
\begin{center}
\begin{tabular}{ccc}\hspace*{-.3cm}
\includegraphics*[scale=0.73]{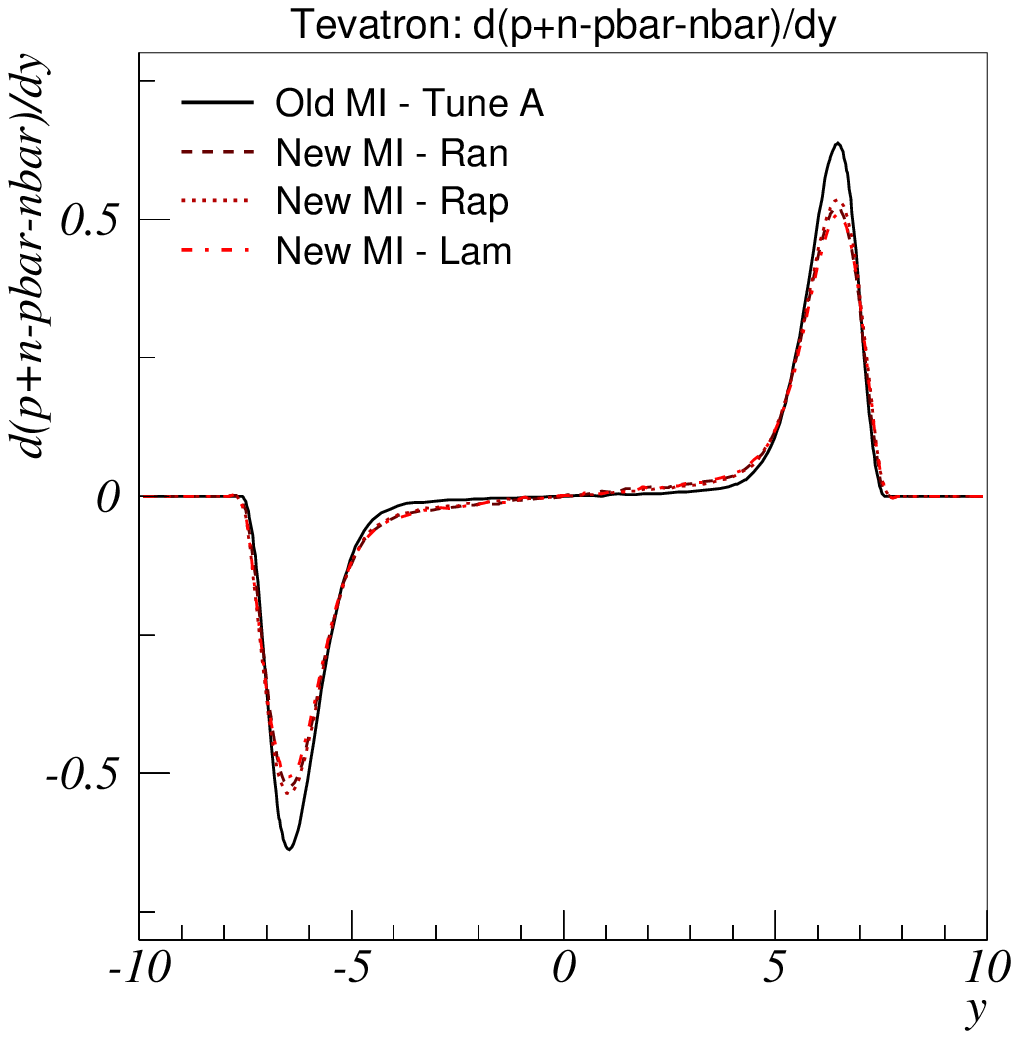}&&\hspace*{-.3cm}
\includegraphics*[scale=0.73]{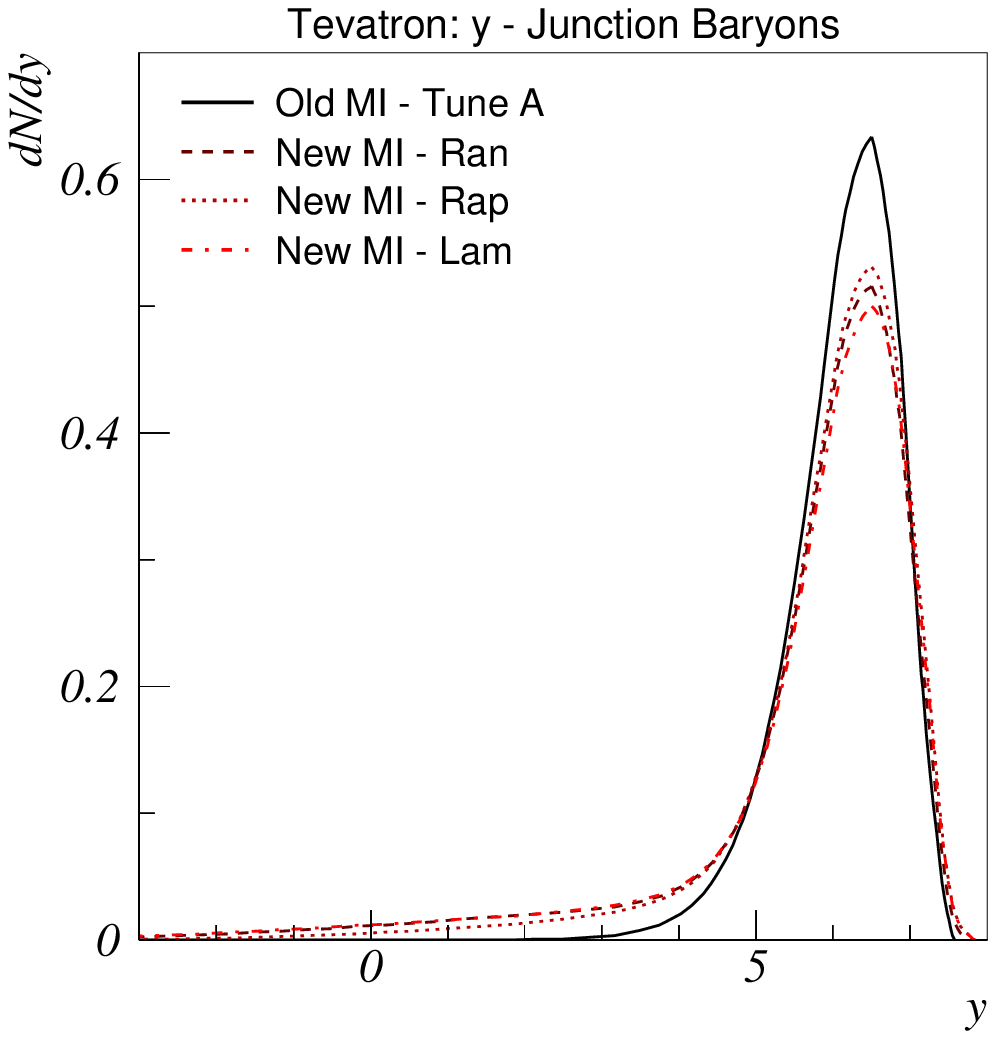}\\[-1mm]
\it a) & & \it b)\\
\end{tabular}
\caption{{\it a)} $n(\p)+n(\n)-n(\pbar)-n(\nbar)$ and {\it b)} junction
  baryon rapidity distributions at the Tevatron.
Results are shown for Tune A (solid), Ran (dashed), Rap (dotted), and
  Lam (dash--dotted), as defined in Table \ref{tab:models}.
\label{fig:tevbaryon}}
\end{center}
\end{figure}

Another point where the models differ is in the treatment of the beam
remnant, especially concerning the flow of baryon number. Fig.\ 
\ref{fig:tevbaryon}a shows the distribution of baryons minus the
distribution of antibaryons as a function of rapidity, illustrating one way
of experimentally probing the location of the beam baryon numbers in the
final state.  
As expected, the distribution is more peaked in the old scenario, where the
beam baryon number is `locked' inside the escaping remnant diquark. Further
illustration of the baryon number stopping is given by Fig.\ 
\ref{fig:tevbaryon}b, which shows the rapidity distribution of the
final-state baryon carrying the baryon number of the incoming proton.  
Note especially the long tail
in the new models, even extending to negative rapidities. The height of this
tail depends on a set of model parameters, such as the choice of initial- and
final-state colour correlations and the rules for diquark formation,
cf.~Fig.~\ref{fig:compflow} fore more extreme scenarios.

\begin{figure}[tp]
\begin{center}
\includegraphics*[scale=0.8]{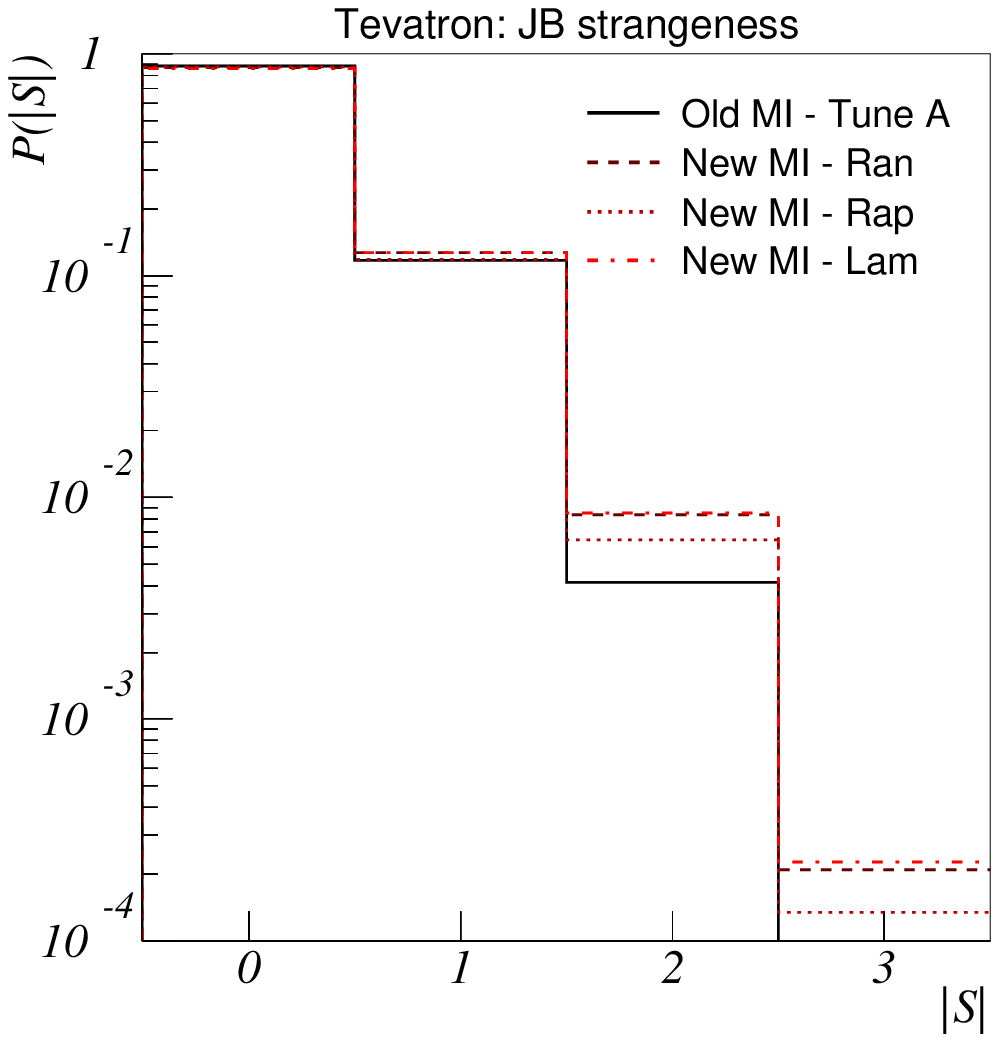}\\[-1mm]
\caption{Junction baryon strangeness.
Results are shown for Tune A (solid), Ran (dashed), Rap (dotted), and
  Lam (dash--dotted), as defined in Table \ref{tab:models}.
\label{fig:tevstrange}}
\end{center}
\end{figure}

The fact that the baryon number is not so closely associated with
the valence quarks in the new model has an interesting side effect: apart
from migrating in physical space, the initial-state baryon number can also
`migrate' in flavour to a much larger extent than before. When the junction is 
resolved, the flavour selection rules of ordinary string fragmentation take
over in determining the flavour composition of the junction baryon. Thereby, 
it becomes possible for the junction baryon to have a larger
strangeness number $S$ than before, as shown in Fig.\  \ref{fig:tevstrange}.
The last bin, $\left|S\right|=3$, is actually empty for Tune A, since 
in the old model it is impossible to produce an
$\Omega^-$ from the incoming beam baryon number. We note that heavy-ion
experiments do observe a ratio $\Omega^-/\Omega^+ > 1$ \cite{Omega}, which is
in contradiction with the old string model but which would be 
more in line with expectations based on the junction scenario introduced here. 

\begin{figure}[tp]
\begin{center}
\includegraphics*[scale=0.8]{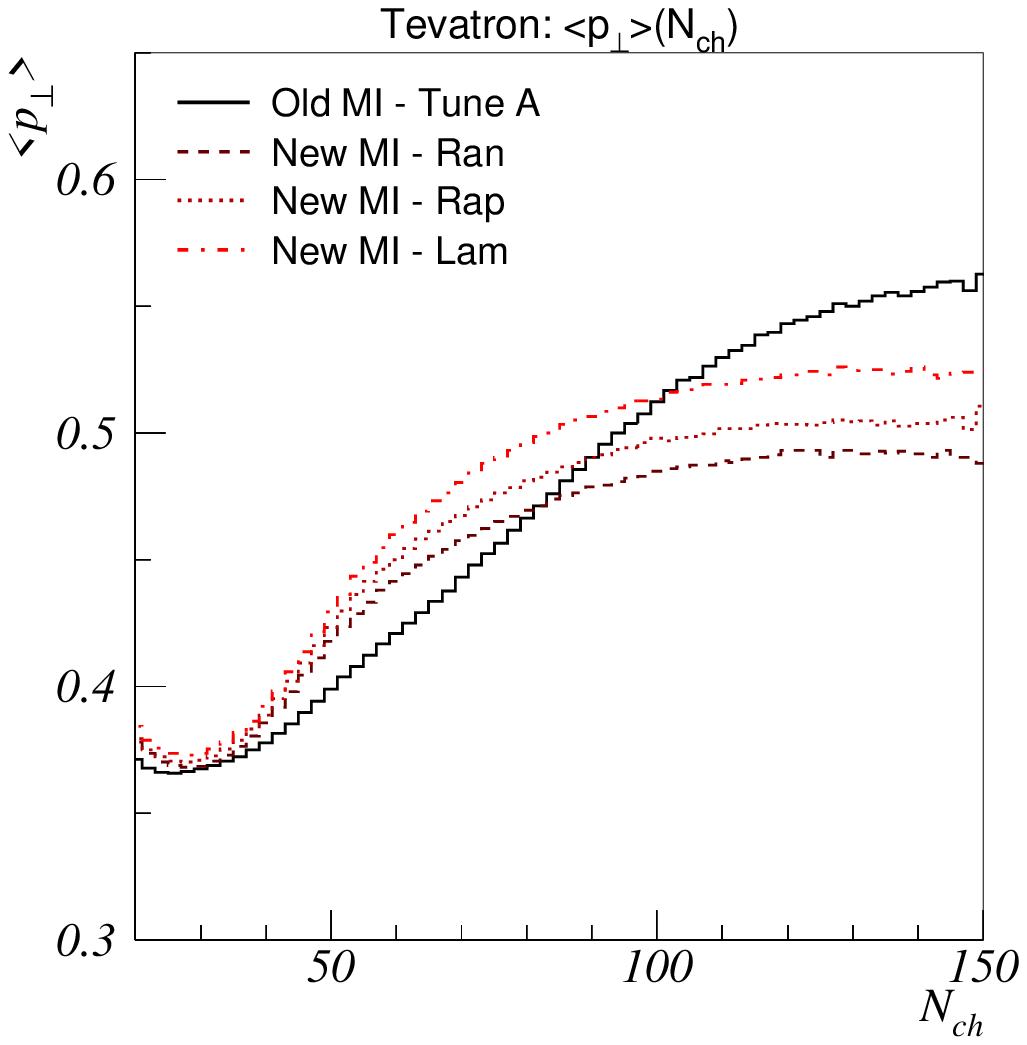}\\[-1mm]
\caption{Average $\pT$ as a function of charged multiplicity for
min-bias collisions at the Tevatron.
Results are shown for Tune A (solid), Ran (dashed), Rap (dotted), and
  Lam (dash--dotted), as defined in Table \ref{tab:models}.
\label{fig:tevpt}}
\end{center}
\end{figure}
Finally, Fig.\  \ref{fig:tevpt} shows the $\langle \pT\rangle(n_{\mathrm{ch}})$
distributions. As previously noted, Fig.~\ref{fig:reconnect},
this distribution is very sensitive
to the colour correlations present at the hadronization stage. Since this is
one of the major open issues remaining, it is not surprising that the
agreement here is far from perfect: the new models exhibit a too 
early rise to a too low plateau, as compared to Tune A. While it is possible to
obtain a much better agreement by varying the scheme adopted for the
colour reconnections in the final state, our attempts in this direction
have so far led to poorer descriptions of other distributions, the charged
multiplicity distribution in particular. Moreover, the colour reconnection
scheme adopted here is meant only as an instructive example, not as
a model of the physics taking place. Our plan is to continue the study of
colour correlations in more depth. In the context of such studies, it is
encouraging to note that the acute sensitivity of the $\langle
\pT\rangle(n_{\mathrm{ch}})$ distribution to these aspects makes its proper
description a prime test for any physical model of the colour flow.

\subsection{Comparisons at the LHC}
Turning now to the situation at the LHC, the lack of experimental constraints 
increases the uncertainties. We here focus on only a subset of these,
assuming the same energy scaling of the $\pTo$ cutoff as for Tune A,
proportional to $E_{\mrm{cm}}^{0.25}$, and using the same parton distributions
in all cases. We note that these aspects do constitute important sources of
uncertainty in our ability to 
make trustworthy `forecasts' for the LHC. Here, however, our aim is merely
to compare alternative scenarios under identical boundary conditions.

\begin{figure}[tp]
\begin{center}\hspace*{-0.3cm}%
\begin{tabular}{c}
\includegraphics*[scale=1]{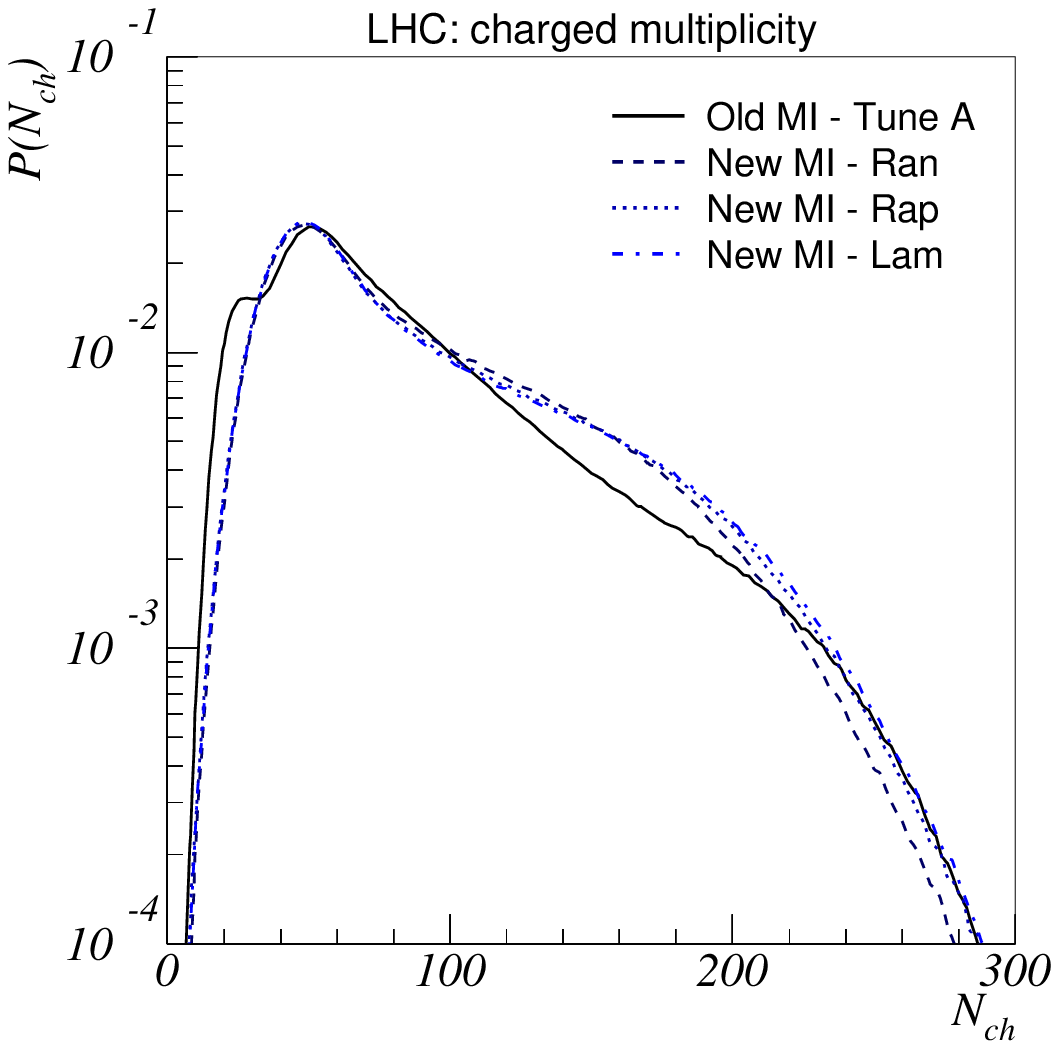}\\[-1mm]
\end{tabular}
\caption{Multiplicity distributions
  for the LHC as obtained with Tune A (solid), and the Ran (dashed), Rap
  (dotted), and Lam (dash-dotted) models defined in 
  Table \ref{tab:models}. For Tune A, the average charged multiplicity is 
$\langle n_{\mathrm{ch}}\rangle=81$, whereas for the new models it is in the
  range 87--89.
\label{fig:lhc20}}
\end{center}
\end{figure}

Fig.\  \ref{fig:lhc20} shows the LHC charged multiplicity distribution for
the same models as used for the Tevatron, but with the $\pTo$ cutoff scaled
to the LHC CM energy.
The new models exhibit average multiplicities of 6--8 more charged
particles per event than the Tune A value, $\langle
  n_{\mathrm{ch}}\rangle=81$. This illustrates a general effect in the new
  models, which is due to the increased shower activity arising from
  associating also the sub-leading interactions with initial- and final-state
  cascades. The larger available phase space  
at higher energies implies that showers are more important at LHC, cf.~the
average number of final-state partons, $\langle n_{\mrm{part}}\rangle$, in
Table \ref{tab:models}. All else being equal this causes the multiplicity
to increase more rapidly with energy than in the old model. (The
strange bump on the Tune A distribution at low multiplicities is merely an
artifact of the way parton distributions at low $Q^2$ are handled in that
model.) 

\begin{figure}[tp]
\begin{center}
\includegraphics*[scale=0.8]{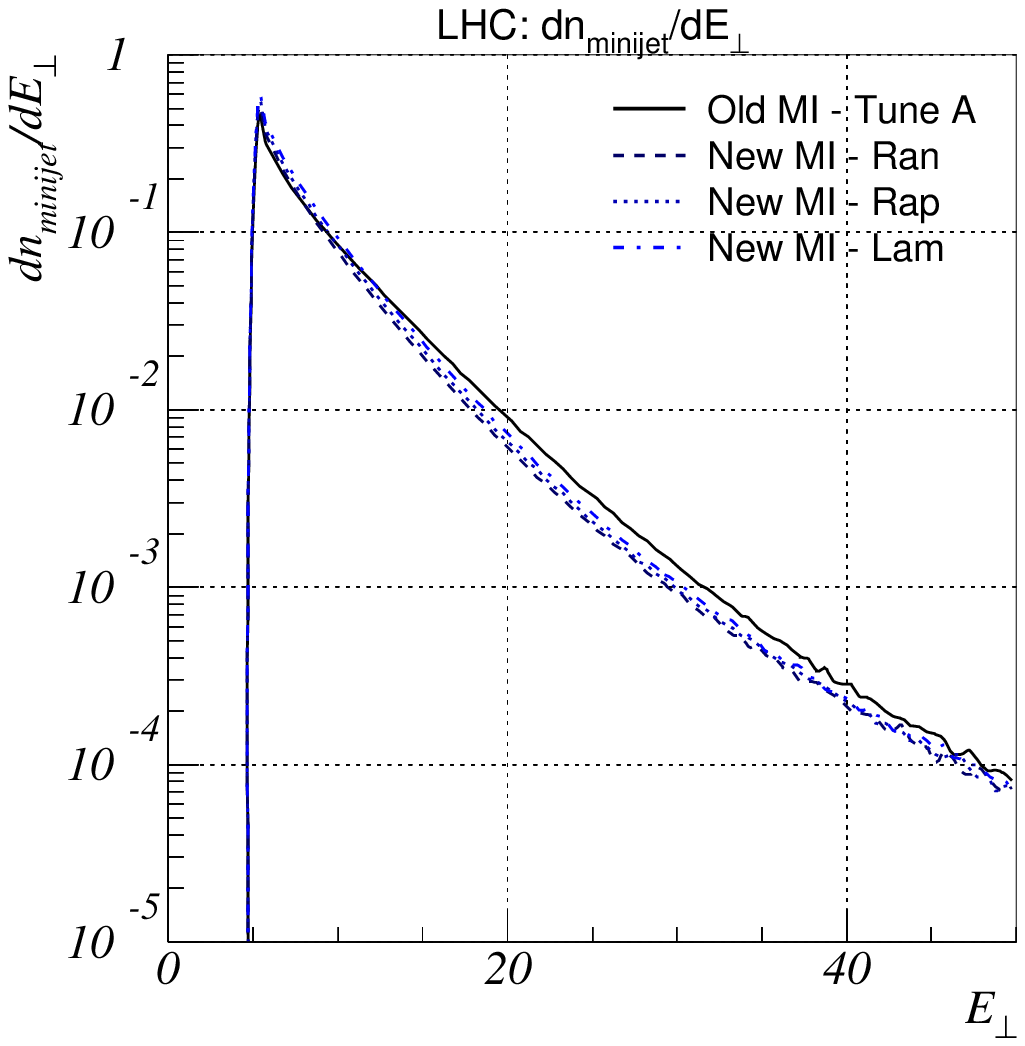}\\[-1mm]
\caption{The number of jets as a function of jet $E_\perp$ (for
  $E_\perp>5\GeV$) in min-bias collisions at the LHC.
Results are shown for Tune A (solid), Ran (dashed), Rap (dotted), and
  Lam (dash--dotted), as defined in Table \ref{tab:models}.
\label{fig:lhcminijets}}
\end{center}
\end{figure}

The addition of parton showers also increases the total amount of partonic
transverse energy, but owing to a partial cancellation of the effects of
radiation in the initial state
(boosting some partons to larger $\pT$) and in the final state (jet
broadening), the jet rates come out similar, as depicted in 
Fig.\  \ref{fig:lhcminijets} (adopting the same cone
algorithm and $\left|\eta\right|<2.5$ region as before).

\begin{figure}[tp]
\begin{center}
\begin{tabular}{ccc}\hspace*{-.3cm}
\includegraphics*[scale=0.73]{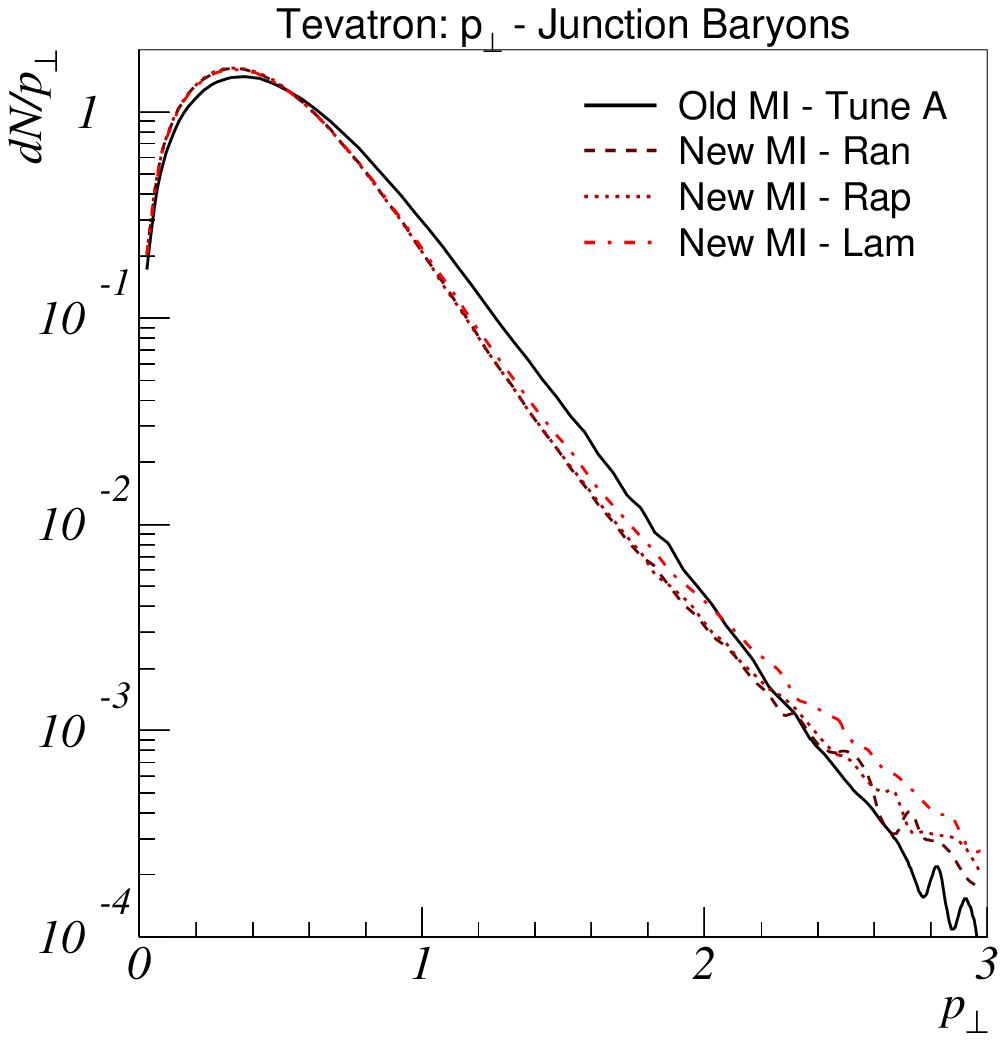}&&\hspace*{-.3cm}
\includegraphics*[scale=0.73]{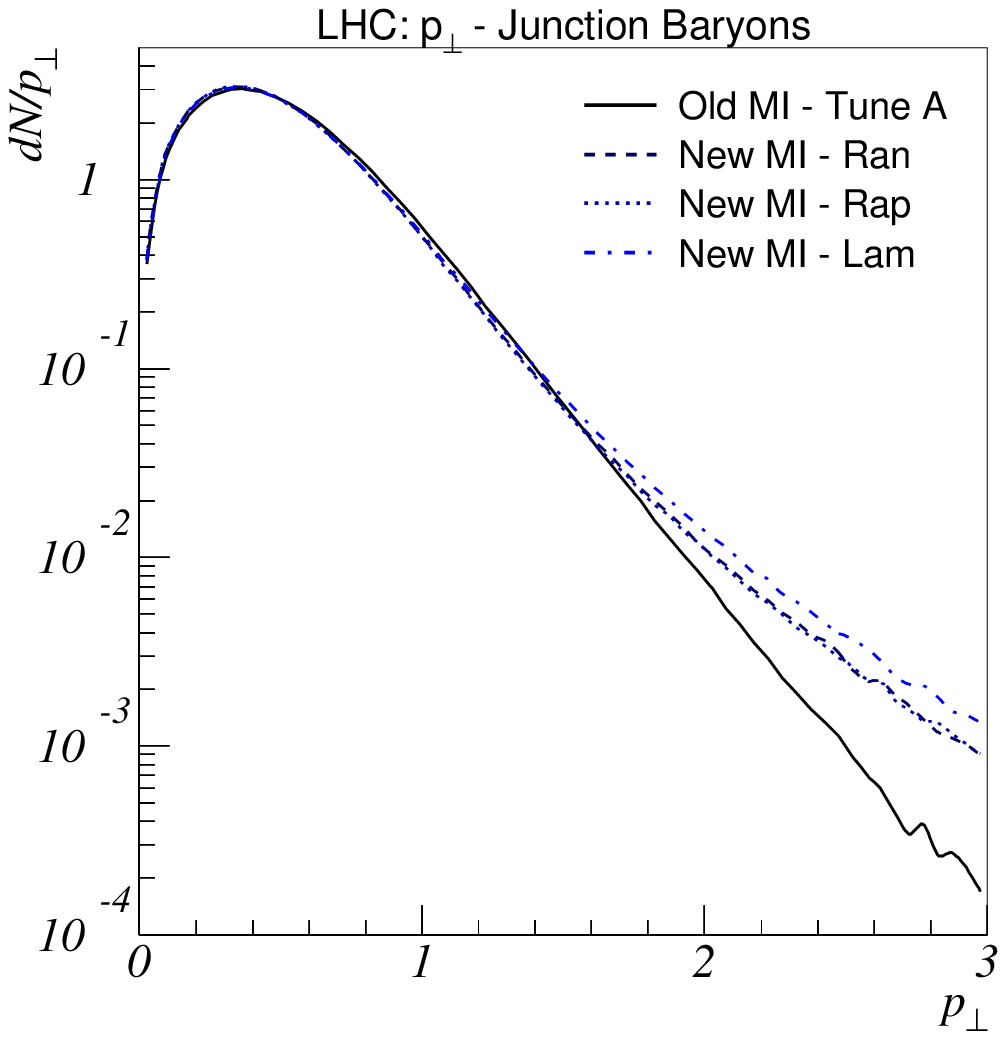} \\[-1mm]
\it a) & & \it b)\\
\end{tabular}
\caption{Junction baryon $\pT$ spectrum at {\it a)} the Tevatron and {\it b)}
  the LHC. Results are shown for Tune A (solid), Ran (dashed), Rap (dotted),
  and Lam (dash--dotted), as defined in Table
  \ref{tab:models}.\label{fig:jbpt}} 
\end{center}
\end{figure}

Fig.\  \ref{fig:jbpt} compares the junction baryon $\pT$ 
distributions at the Tevatron (left plot) and at
the LHC (right plot), for Tune A and the new models. An interesting
difference is that the junction baryon can be significantly harder in
$\pT$ at the LHC than at the Tevatron in the new models,
whereas Tune A exhibits spectra which are almost identical between
the two energies.  
This is due to the intrinsic difference between the way primordial $\kT$ is
treated in the two frameworks. In the old model, the width of the primordial
$\kT$ distribution for the parton initiating the hardest scattering is
fixed, to 1\GeV\ by default, hence there is no mechanism that would allow the 
junction baryon spectrum to depend on the CM energy (at sufficiently high
energies that energy--momentum conservation effects can be neglected). 
In the new model, the amount of primordial $\kT$ given to initiators
depends on the $Q^2$ of their associated hard scattering. With the increased
phase space at the LHC, more primordial $\kT$ is thus imparted by recoil
effects to the junction baryon than at the Tevatron, hence the $\pT$ spectrum
becomes harder.

\begin{figure}[tp]
\begin{center}
\includegraphics*[scale=0.8]{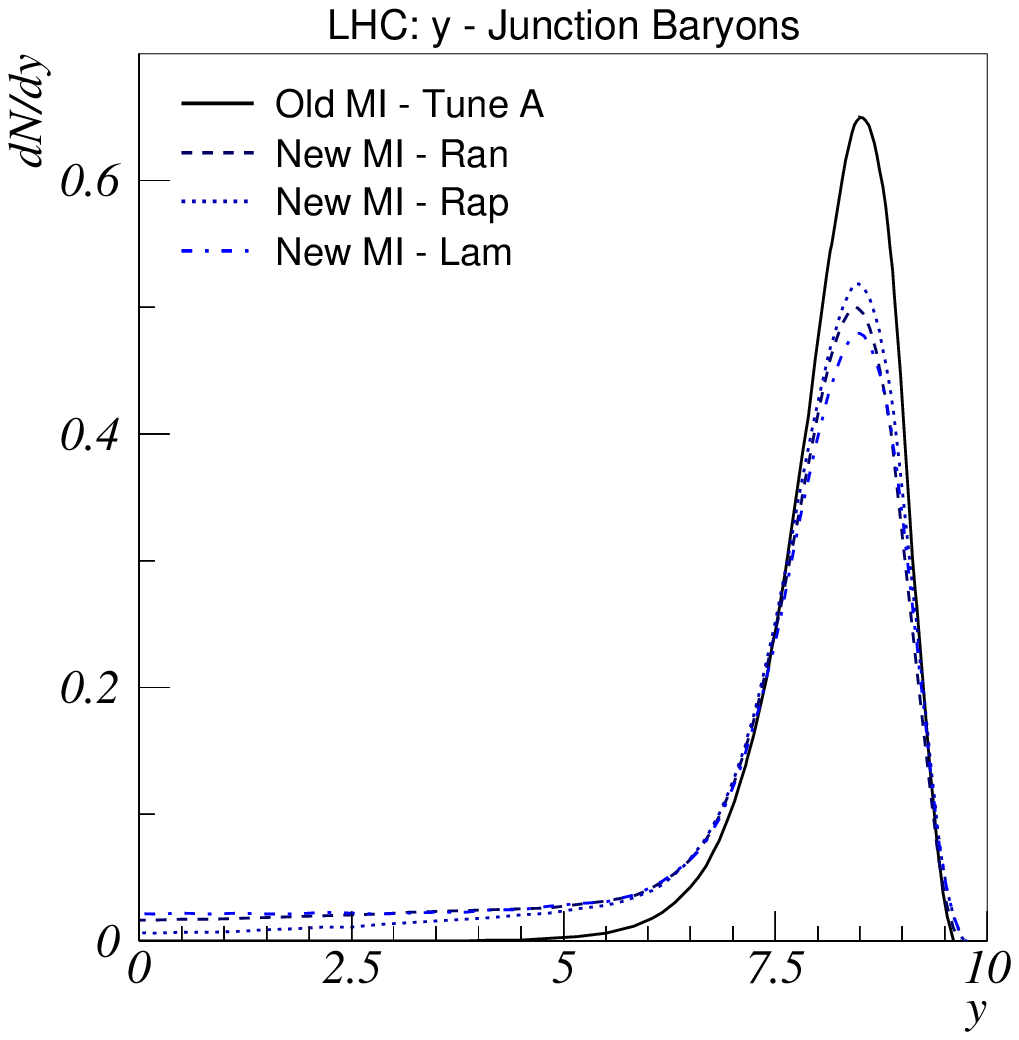}\\[-1mm]
\caption{Junction baryon rapidity distributions at the
  LHC. Note: at the
LHC \emph{both} beam baryon numbers are included in the figure, whereas in
the Tevatron plots, Fig.\  \ref{fig:tevbaryon}, the antibaryon number is
  not. Results are shown for Tune A (solid), Ran
  (dashed), Rap (dotted), and Lam (dash--dotted), as defined in Table
  \ref{tab:models}.\label{fig:lhcbaryon}}  
\end{center}
\end{figure}

Also the junction baryon longitudinal migration shows some
difference. Comparing the Tevatron junction baryon rapidity distribution,
Fig.\  \ref{fig:tevbaryon}b above, with the LHC one, Fig.\ 
\ref{fig:lhcbaryon}, we may distinguish two components. 
One is the peak at
large rapidities, which corresponds to an (effective) diquark fragmentation
and which is only shifted outwards in rapidity relative to the Tevatron by
the increased energy. The other is the tail to central rapdities, which
corresponds to baryon stopping. This tail \emph{does} increase with energy,
following the increase in the average number of interactions.

\begin{figure}[tp]
\begin{center}
\includegraphics*[scale=0.8]{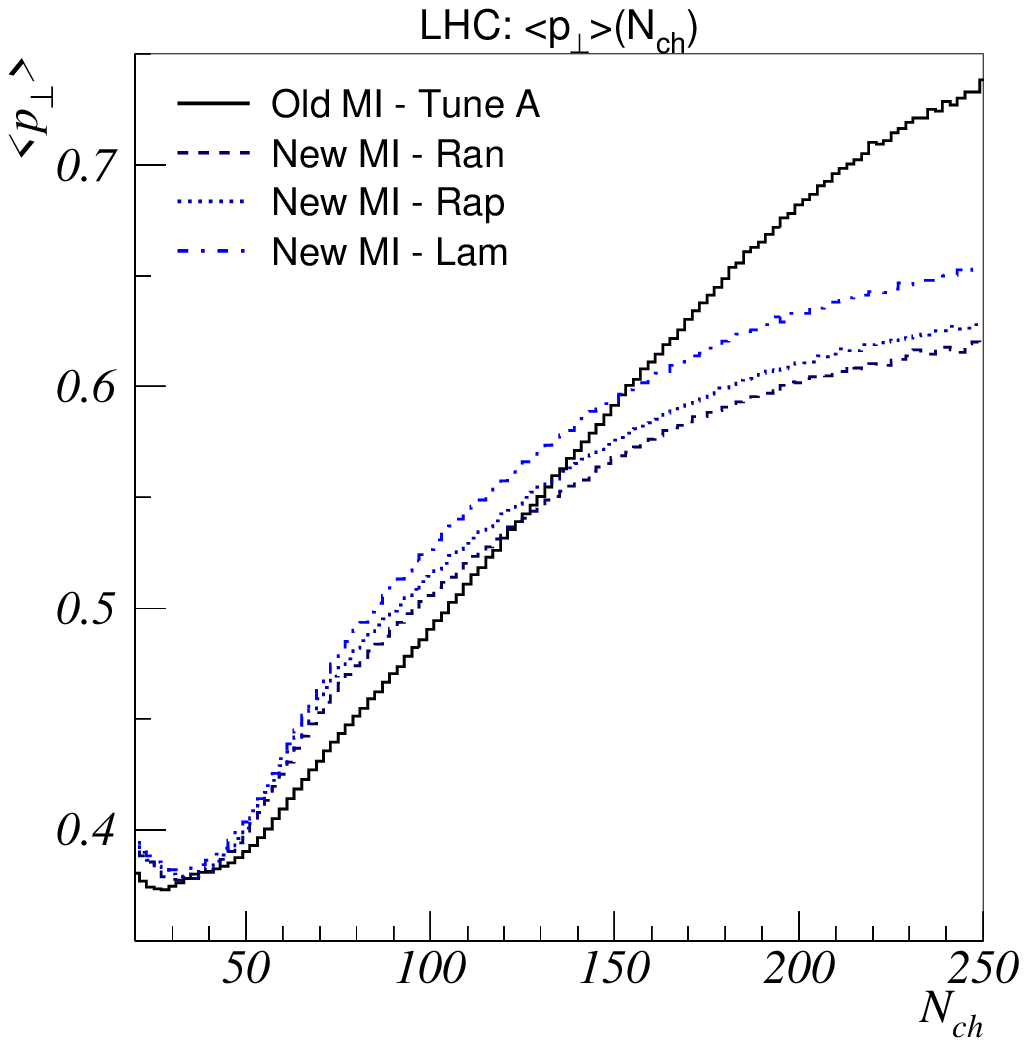}\\[-1mm]
\caption{Average $\pT$ as a function of charged multiplicity for
min-bias collisions at the LHC.
Results are shown for Tune A (solid), Ran (dashed), Rap (dotted), and
  Lam (dash--dotted), as defined in Table \ref{tab:models}.
\label{fig:lhcpt}}
\end{center}
\end{figure}

Finally, we show the $\langle \pT\rangle(n_{\mathrm{ch}})$ distributions in
Fig.\  \ref{fig:lhcpt}. 
The same qualitative behaviour as at the Tevatron is apparent: 
the new models exhibit an earlier rise to a lower plateau, as compared to
Tune A. Again, it is premature to draw any strong conclusions, in view of the
still simple-minded description of the colour flow that we have included
here. Further and more detailed studies of possible colour correlation
mechanisms in hadronic collisions 
will be required in order to fully understand these aspects. 

\section{Conclusion and Outlook}

Only in the last few years have multiple interactions gone from
being a scientific curiosity, by most assumed relevant only for
some rare topologies of four-jet events, to being accepted as 
the key element for understanding the structure of underlying 
events. However, this leaves a lot of questions to be addressed,
such as:\\
\textit{(i)} What is the detailed mechanism and functional form of the 
dampening of the perturbative cross section at small $p_{\perp}$?\\ 
\textit{(ii)} What is the energy dependence of the mechanism(s) involved?\\
\textit{(iii)} How is the internal structure of the proton reflected in 
an impact-parameter-dependent multiple interactions rate, as manifested
e.g.\ in jet pedestal effects?\\
\textit{(iv)} How can the set of colliding partons from a hadron be
described in terms of correlated multiparton distribution functions
of flavours and longitudinal momenta?\\
\textit{(v)} How does a set of initial partons at some low perturbative 
cutoff scale evolve into such a set of colliding partons?
Is standard DGLAP evolution sufficient, or must BFKL/CCFM effects be 
taken into account?\\ 
\textit{(vi)} How would the set of initiators correlate with the flavour 
content of, and the longitudinal momentum sharing inside, the left-behind 
beam remnant?\\
\textit{(vii)} How are the initiator and remnant partons correlated by 
confinement effects, e.g.\ in primordial $k_{\perp}$?\\
\textit{(viii)} How are all produced partons, both the interacting and 
the beam-remnant ones, correlated in colour? Is the large 
number-of-colours limit relevant, wherein partons can be hooked up into 
strings representing a linear confinement force?\\
\textit{(ix)} How is the original baryon number of an incoming proton 
reflected in the colour topology?\\ 
\textit{(x)} To what extent would a framework with independently 
fragmenting  string systems, as defined from the colour topology, be 
modified by the space--time overlap of several strings?

Tentative answers to some of the questions are provided by the Tune A
of the \textsc{Pythia} multiple interactions framework. Thus we now
believe that:\\
$\bullet$ The matter overlap when two hadrons collide can be described 
by an impact-parameter dependence more spiked than a Gaussian but 
less so than an exponential.\\ 
$\bullet$ The $\pTo$ regularization scale does increase with energy.\\
$\bullet$ The colours of final-state partons are not random but 
correlated, somehow, to give a reduced string length.

This still leaves many questions unanswered. Worse, existing event
generators would not even address many of the relevant issues, at least 
not in a deliberate or realistic fashion. In this article we have 
therefore tried to take the next step towards a better understanding 
of the structure of a hadronic event, addressing several of the points 
above. This in particular has concerned 
the correlations between initiator and remnant partons in the hadron 
beams, in terms of flavour, longitudinal and transverse momenta. 
Colour correlations have also been studied, and here it appears that 
the final-state partons need be involved as well. The complexity of the 
colour issues is tremendous, however, and we do not consider the studies 
finished in this area. 

A specific new topic addressed is that of baryon number flow. Data from 
hadronic collisions, and even more from heavy-ion ones, show large 
excesses of baryon over antibaryon production in the central rapidity 
region of events \cite{barexcess}, suggesting a significant influx 
of baryon number from the high-rapidity colliding beams, more than 
would be expected from standard quark/diquark fragmentation models. 
When the junction is introduced as a topological feature of the colour 
field in the baryon, however, the fate of the baryon number of an 
incoming beam particle may partly or wholly decouple from that of the 
valence quarks \cite{Zijl,barexcessthy}.  We have here demonstrated that
the junction topology in combination with multiple interactions can 
induce quite large rapidity shifts, of the desired kind. 

The problem may actually be the opposite, i.e.\ not to move the baryon 
number by too much. To this end, we have assumed a suppression of 
interactions that affect several of the three colour chains that 
connect the valence quarks to the junction. This could be an 
impact-parameter-related effect, that not the whole proton is 
involved in the hard processes. If so, the suppression should be less
pronounced in heavy-ion collisions, where the interactions of a proton 
with several nucleons in the other nucleus could occur at different 
positions in the transverse plane and thereby affect different chains. 
Obviously it would be a major undertaking to construct a complete model 
for heavy-ion collisions to study these ideas, but we hope in the future 
to be able to present a simple study of the baryon number flow. 

Another open issue is that of intertwined initial-state showers, 
whereby two seemingly unrelated partons, each undergoing a hard
scattering,  reconstruct back to come from a common shower ancestor.
With the new $\pT$-ordered showers now being implemented in 
\textsc{Pythia} \cite{newshowers} we intend to introduce enough 
flexibility that such issues could be addressed.
 
This will also further constrain the initial-state colour flow.
The possibility of final-state colour reconnections remains, 
however, and has been proposed as a mechanism to introduce 
diffractive topologies in a number of processes \cite{ingelman}.
One here needs to better understand how much reconnections
are allowed/required, and of what character.

We see that much work remains, before the physics of the underlying
event is truly understood. Progress will not be possible without a
constructive dialogue between theory and experiment. We have
frequently had reason to mention Tune A as a role model here,
because it offers a convenient reference that more sophisticated 
models can be tested against, without the need to know the details
of the CDF detector. However, only a few distributions 
went into the tune, and so we do not know what to aim for in 
many other respects. 

To give one specific example, it would be valuable to have 
information on the `lumpiness' of the underlying event, such as 
$n$-jet rates as a function of some jet resolution parameter, 
similarly to $\e^+\e^-$-annihilation QCD analyses. One would 
there hope for an intermediate resolution region, between the coarse 
one that is dominated by the perturbative QCD structure and the fine 
one that mainly is sensitive to hadronization details, where the 
structure of the multiple interactions would play a key role. An
understanding of this lumpiness is related to the fluctuations in
the jet pedestal, and thereby to the smearing of jet energies in
SUSY searches, say. It all hangs together \ldots

In summary, striving for a better understanding of the physics of 
the underlying event is both interesting and useful. Interesting
because it forces us to consider many issues normally swept under 
the carpet, and to confront dramatically different scenarios. 
Useful because it ties in with so many other physics analyses at
hadron colliders. So there is plenty of interesting and useful
work ahead of us before the picture has clarified completely! 

\subsection*{Acknowledgments}
 
The authors gratefully acknowledge the stimulating atmosphere at the
Les Houches 2003 Workshop on Physics at TeV Colliders, the CERN
Workshop on Monte Carlo tools for the LHC, and the Collider Physics 2004 
Workshop at KITP, UCSB. We have benefitted from
discussions with R.D. Field, J. Huston, A. Moraes, and many more.
We are also grateful to the NorduGRID project, for use of 
computing resources.
 
This research was supported in part by 
the National Science Foundation under Grant No.~PHY99-07949, 
and by 
The Royal Physiographic Society in Lund.


\begin{thebibliography}{99}
 
\bibitem{Zijl}
T. Sj\"ostrand and M. van Zijl, Phys. Rev. {\bf D36} (1987) 2019.
 
\bibitem{higgsvertex}
C. Seez, CMS TN/93--115;\\
CMS Collaboration, Technical Proposal, CERN/LHCC 94--38, p. 50,
ECAL Technical Design Report, CERN/LHCC 97--33, p. 330;\\
U. Egede, Ph.D. Thesis, Lund University LUNFD6/(NFFL--7150) 1997,
ISBN 91--628--2804--5;\\
ATLAS Collaboration, Detector and Physics Performance Technical
Design Report, Vol. I, CERN/LHCC 99--14, p. 228
 
\bibitem{BNV}
T. Sj\"ostrand and P.Z. Skands, Nucl. Phys. {\bf B659} (2003) 243
 
\bibitem{string}
B. Andersson, G. Gustafson, G. Ingelman and T. Sj\"ostrand,
Phys. Rep. {\bf 97} (1983) 31;\\
B. Andersson, `The Lund Model' (Cambridge University Press, 1998)
 
\bibitem{briefdesc}
P.~Skands and T.~Sj\"ostrand, in the proceedings 
of HEP 2003, Eur.\ Phys. J.\ C Direct, DOI: 10.1140/epjcd/s2003-03-520-7 
[hep-ph/0310315]; \\ 
T.~Sj\"ostrand and P.~Skands [hep-ph/0401060], in the proceedings 
of 3rd Les Houches Workshop on Physics at TeV Colliders, M.~Dobbs et
al.\ [hep-ph/0403100]

\bibitem{CTEQ5}
CTEQ Collaboration, H.L. Lai et al., Eur. Phys. J. {\bf C12} (2000) 375
 
\bibitem{DL}
A. Donnachie and P.V. Landshoff, Phys. Lett. {\bf B296} (1992) 227
 
\bibitem{Dischler}
J. Dischler and T. Sj\"ostrand, EPJdirect {\bf C2} (2001) 1
 
\bibitem{Pythia}
T. Sj\"ostrand, P. Ed\'en, C. Friberg, L. L\"onnblad, G. Miu, S. Mrenna
and E. Norrbin, Computer Phys. Commun. {\bf 135} (2001) 238;\\
T. Sj\"ostrand, L. L\"onnblad, S. Mrenna and P. Skands, LU TP 03-38
[hep-ph/0308153]
 
\bibitem{Schuler}
G.A. Schuler and T. Sj\"ostrand, Phys. Rev. {\bf D49} (1994) 2257
 
\bibitem{Sudakov}
V.V. Sudakov, Zh.E.T.F. {\bf 30} (1956) 87 (Sov. Phys. J.E.T.P.
{\bf 30} (1956) 65)

\bibitem{pythiagamma}
G.A. Schuler and T. Sj\"ostrand, Nucl. Phys. {\bf B407} (1993) 539,
Z. Phys. {\bf C73} (1997) 677;\\
C. Friberg and T. Sj\"ostrand, JHEP {\bf 09} (2000) 10

\bibitem{DTU}
H.-M. Chan et al., Nucl. Phys. {\bf B86} (1975) 479, {\bf B92} (1975) 13;\\
G.F. Chew and C. Rosenzweig, Phys. Rep. {\bf 41} (1978) 263
 
\bibitem{DPM}
A. Capella, U. Sukhatme, C.-I. Tan and J. Tran Thanh Van,
Phys. Lett. {\bf 81B} (1979) 68, Phys. Rep. {\bf 236} (1994) 225;\\
H. Minaka, Phys. Rev. {\bf D20} (1979) 1656;\\
G. Cohen-Tannoudji et al., Phys. Rev. {\bf D21} (1980) 2699;\\
A. Capella and J. Tran Thanh Van, Z. Phys. {\bf C10} (1981) 249,
Phys. Lett {\bf 114B} (1982) 450;\\
K. Fia\/lkowski and A. Kotanski, Phys. Lett. {\bf 107B} (1981) 132,
{\bf 115B} (1982) 425;\\
P. Aurenche and F.W. Bopp, Phys. Lett. {\bf 114B} (1982) 363;\\
A.B. Kaidalov, Phys. Lett. {\bf 116B} (1982) 459;\\
A.B. Kaidalov and K.A. Ter Martirosyan, Phys. Lett. {\bf 117B} (1982) 247;\\
P. Aurenche, F.W. Bopp and J. Ranft, Z. Phys. {\bf C23} (1984) 67,
{\bf C26} (1984) 279, Phys. Rev. {\bf D33} (1986) 1867;\\
V.M. Chudakov and V.V. Lugovoi, Z. Phys. {\bf C59} (1993) 511
 
\bibitem{diagrammatic}
L.V. Gribov, E.M. Levin and M.G. Ryskin, Phys. Rept. {\bf 100} (1983) 1;\\
E.M. Levin and M.G. Ryskin, Phys. Rep. {\bf 189} (1990) 267
 
\bibitem{AKG}
V.N. Gribov, Sov. Phys. JETP {\bf 26} (1968) 414;\\
V.A. Abramovski, O.V. Kancheli and V.N. Gribov,
Soviet J. Nucl. Phys.{\bf 18} (1974) 308;\\
G. Veneziano, Nucl. Phys. {\bf B74} (1974) 365,
{\bf B117} (1976) 519;\\
J. Bartels and M.G. Ryskin, Z. Phys. {\bf C76} (1997) 241
 
\bibitem{Isajet}
H. Baer, F.E. Paige, S.D. Protopopescu, X. Tata, hep-ph/0001086
 
\bibitem{Dtujet}
P. Aurenche, F.W. Bopp, A. Capella, J. Kwiecinski, M. Maire, J. Ranft
and J. Tran Thanh Van, Phys. Rev. {\bf D45} (1992) 92;\\
P. Aurenche, F.W. Bopp, R. Engel, D. Pertermann, J. Ranft and S. Roesler,
Computer Phys. Commun. {\bf 83} (1994) 107
 
\bibitem{Phojet}
R. Engel, Z. Phys.  {\bf C66} (1995) 203;\\
R. Engel, J. Ranft, Phys.Rev. {\bf D54} (1996) 4244
 
\bibitem{Dpmjet}
J. Ranft, Phys. Rev. {\bf D51} (1995) 64;\\
S. Roesler, R. Engel and J. Ranft, hep-ph/0012252
 
\bibitem{twoint}
P.V. Landshoff and J.C. Polkinghorne,
Phys. Rev. {\bf D18} (1978) 3344;\\
C. Goebel, F. Halzen and D.M. Scott,
Phys. Rev. {\bf D22} (1980) 2789;\\
M. Mangano, Z. Phys. {\bf C42} (1989) 331
 
\bibitem{manyint}
F. Takagi, Phys. Rev. Lett. {\bf 43} (1979) 1296;\\
N. Paver and D. Treleani, Nuovo Cimento {\bf 70A} (1982) 215,
Phys. Lett. {\bf 146B} (1984) 252, Z. Phys. {\bf C28} (1985) 187;\\
B. Humpert, Phys. Lett. {\bf 131B} (1983) 461;\\
B. Humpert and R. Odorico, Phys. Lett. {\bf 154B} (1985) 211;\\
G. Pancheri and Y.N. Srivastava, Phys. Lett. {\bf B182} (1986) 199;\\
N. Brown, Mod. Phys. Lett. {\bf A4} (1989) 2447
 
\bibitem{manyintdiff}
R.E. Ecclestone and D.M. Scott, Z. Phys. {\bf C19} (1983) 29;\\
B. Humpert, Phys. Lett. {\bf 135B} (1984) 179;\\
M. Mekhfi, Phys. Rev. {\bf D32} (1985) 2371;\\
F. Halzen, P. Hoyer and W.J. Stirling, Phys. Lett. {\bf B188} (1987) 375;\\
R.W. Robinett, Phys. Lett. {\bf B230} (1989) 153;\\
R.M. Godbole, S. Gupta and J. Lindfors, Z. Phys. {\bf C47} (1990) 69;\\
M. Drees and T. Han, Phys. Rev. Lett. {\bf 77} (1996) 4142;\\
A. Del Fabbro and D. Treleani, Phys. Rev. {\bf D61} (2000) 077502;\\
A. Kulesza and W.J. Stirling, Phys. Lett. {\bf B475} (2000) 168
 
\bibitem{sigmaeikonal}
T.T. Chou and C.N. Yang, Phys. Rev. {\bf 170} (1968) 1591,
{\bf D19} (1979) 3268;\\
C. Bourrely, J. Soffer and T.T. Wu,
Nucl.Phys. {\bf B247} (1984) 15; Eur. Phys. J. {\bf C28} (2003) 97\\
P. L'Heureux, B. Margolis and P. Valin,
Phys. Rev. {\bf D32} (1985) 1681
 
\bibitem{sigmaminijet}
L. Durand and P. Hong, Phys. Rev. Lett. {\bf 58} (1987) 303;\\
A. Capella, J. Kwiecinski and J. Tran Thanh Van,
Phys. Rev. Lett. {\bf 58} (1987) 2015;\\
M.M. Block, F. Halzen and B. Margolis,
Phys. Rev. {\bf D45} (1992) 839;\\
R.M. Godbole and G. Pancheri, Eur. Phys. J. {\bf C19} (2001) 129
 
\bibitem{hadronpcorr}
J. Kuti and V.F. Weisskopf, Phys. Rev. {\bf D4} (1971) 3418;\\
K. Konishi, A. Ukawa and G. Veneziano,
Nucl. Phys. {\bf B157} (1979) 45;\\
R. Kirschner, Phys. Lett. {\bf 84B} (1979) 266;\\
H.R. Gerhold, Nuovo Cimento {\bf 59A} (1980) 373;\\
V.P. Shelest, A.M. Snigirev and G.M. Zinovjev,
Phys. Lett. {\bf 113B} (1982) 325;\\
A.M. Snigirev, Phys. Rev. {\bf D68} (2003) 114012
 
\bibitem{hadronccorr}
M. Mekhfi, Phys. Rev. {\bf D32} (1985) 2380;\\
M. Mekhfi and X. Artru, Phys. Rev. {\bf D37} (1988) 2618
 
\bibitem{hadronbcorr}
A. Bia\/las and E. Bia\/las, Acta. Phys. Pol. {\bf B5} (1974) 373;\\
T.T. Chou and C.N. Yang, Phys. Rev. {\bf D32} (1985) 1692;\\
S. Barshay, Z. Phys. {\bf C32} (1986) 513;\\
W.R. Chen and R.C. Hwa, Phys. Rev. {\bf D36} (1987) 760
 
\bibitem{saturation}
A.H. Mueller and J. Qiu, Nucl. Phys. {\bf B268} (1986) 427
 
\bibitem{Herwig}
G. Corcella, I.G. Knowles, G. Marchesini, S. Moretti, K. Odagiri,
P. Richardson, M.H. Seymour and B.R. Webber,
JHEP {\bf 0101} (2001) 010, hep-ph/0210213
 
\bibitem{UA5gen}
UA5 Collaboration, G.J. Alner et al.,
Nucl. Phys. {\bf B291} (1987) 445
 
\bibitem{Jimmy}
J.M. Butterworth, J.R. Forshaw and M.H. Seymour,
Z. Phys. {\bf C72} (1996) 637
 
\bibitem{Ivan}
I. Borozan and M.H. Seymour, JHEP {\bf 0209} (2002) 015
 
\bibitem{BFKL}
E.A. Kuraev, L.N. Lipatov and V.S. Fadin,
Sov. Phys. JETP {\bf 45} (1977) 199;\\
I.I. Balitsky and L.N. Lipatov,
Sov. J. Nucl. Phys. {\bf 28} (1978) 822
 
\bibitem{CCFM}
M. Ciafaloni, Nucl. Phys. {\bf B296} (1988) 49;   \\
S. Catani, F. Fiorani and G. Marchesini, Nucl. Phys. {\bf B336}
(1990) 18
 
\bibitem{LDC}
B. Andersson, G. Gustafson and J. Samuelsson,
Nucl. Phys. {\bf B467} (1996) 443;\\
B. Andersson, G. Gustafson and H. Kharraziha,
Phys. Rev. {\bf D57} (1998) 5543;\\
H. Kharraziha and L. L\"onnblad, JHEP {\bf 03} (1998) 006
 
\bibitem{Miukt}
G. Gustafson, L. L\"onnblad and G. Miu,
JHEP {\bf 09} (2002) 005
 
\bibitem{Miu}
G. Gustafson, L. L\"onnblad and G. Miu,
Phys. Rev. {\bf D67} (2003) 034020
 
\bibitem{heavyion}
K. Kajantie, P.V. Landshoff and J. Lindfors,
Phys. Rev. Lett. {\bf 59} (1987) 2527;\\
K.J. Eskola, K. Kajantie and J. Lindfors,
Nucl. Phys. {\bf B323} (1989) 37;\\
K.J. Eskola, K. Kajantie and K. Tuominen,
Nucl. Phys. {\bf A700} (2002) 509
 
\bibitem{AFSmiev}
AFS Collaboration, T. {\AA}kesson et al., Z. Phys. {\bf C34} (1987) 163
 
\bibitem{UA2miev}
UA2 Collaboration, J. Alitti et al., Phys. Lett. {\bf B268} (1991) 145
 
\bibitem{CDFmiev}
CDF Collaboration, F. Abe et al., Phys. Rev. {\bf D47} (1993) 4857
 
\bibitem{CDFmievnew}
CDF Collaboration, F. Abe et al., Phys. Rev. Lett. {\bf 79} (1997) 584,
Phys. Rev. {\bf D56} (1997) 3811
 
\bibitem{ZEUSmiev}
ZEUS Collaboration, C. Gwenlan, Acta Phys. Polon. {\bf B33} (2002) 3123
 
\bibitem{D0miev}
D0 Collaboration, V.M. Abazov et al., Phys. Rev. {\bf D67} (2003) 052001
 
\bibitem{UA1minijet}
UA1 Collaboration, C.-E. Wulz, in proceedings of the 22nd Rencontres de
Moriond, Les Arcs, France, 15-21 March 1987
 
\bibitem{KNO}
Z. Koba, H.B. Nielsen and P. Olesen, Nucl. Phys. {\bf B40} (1972) 317
 
\bibitem{multdistu}
UA5 Collaboration, G.J. Alner et al., Phys. Lett. {\bf 138B} (1984) 304;\\
UA5 Collaboration, R.E. Ansorge et al., Z. Phys. {\bf C43} (1989) 357
 
\bibitem{multdiste}
E735 Collaboration, T. Alexopoulos et al.,
Phys. Lett. {\bf B435} (1998) 453
 
\bibitem{eeKNO}
E.D. Malaza and B.R. Webber, Nucl. Phys. {\bf B267} (1986) 702;\\
E.D. Malaza, Z. Phys. {\bf C31} (1986) 143
 
\bibitem{flavnonuniv}
ZEUS Collaboration, M. Derrick et al., Z. Phys. {\bf C68} (1995) 29;\\
H1 Collaboration, S. Aid et al., contribution 479 to EPS HEP95, Brussels;\\
T. Sj\"ostrand, J. Phys. {\bf G22} (1996) 709;\\
H1 Collaboration, Abstract 1102, submitted to ICHEP02, Amsterdam
 
\bibitem{Schleinstring}
R608 Collaboration, A.M. Smith et al., Phys. Lett. {\bf B163} (1985) 267
 
\bibitem{fbcorr}
UA5 Collaboration, R.E. Ansorge et al., Z. Phys. {\bf C37} (1988) 191;\\
E735 Collaboration, T. Alexopoulos et al.,
Phys. Lett. {\bf B353} (1995) 155
 
\bibitem{UA1ped}
UA1 Collaboration, C. Albajar et al., Nucl. Phys. {\bf B309} (1988) 405
 
\bibitem{CDFpedjet}
CDF Collaboration, T. Affolder et al., Phys. Rev. {\bf D65} (2002) 092002
 
\bibitem{CDFped}
R.D. Field (CDF Collaboration), in the proceedings of Snowmass 2001,
eConf C010630 (2001) P501 [hep-ph/0201192], CDF Note 6403,
talks available from webpage
\ttt{http://www.phys.ufl.edu/}$\sim$\ttt{rfield/cdf/};\\
V. Tano (CDF Collaboration), to appear in the proceedings of the
37th Rencontres de Moriond, Les Arcs, France, 16--23 March 2002
[hep-ex/0205023]
 
\bibitem{H1ped}
H1 Collaboration, S. Aid et al., Z. Physik {\bf C70} (1996) 17
 
\bibitem{meanptfornch}
A. Breakstone et al., Phys. Lett. {\bf 132B} (1983) 463;\\
UA1 Collaboration, C. Albajar et al., Nucl. Phys. {\bf B335} (1990) 261;\\
E735 Collaboration, T. Alexopoulos et al.,
Phys. Lett. {\bf B336} (1994) 599;\\
CDF Collaboration, D. Acosta et al., Phys. Rev. {\bf D65} (2002) 072005
 
\bibitem{jetshape}
UA1 Collaboration, Phys. Lett. {\bf 132B} (1983) 214;\\
D0 Collaboration, S. Abachi et al.,  Phys. Lett. {\bf B357} (1995) 500;\\
ZEUS Collaboration, J. Breitweg et al., Eur. Phys. J. {\bf C1} (1998) 109
 
\bibitem{oldglobal}
CDF Collaboration, F. Abe et al., Phys. Rev. {\bf D59} (1999) 032001;\\
P. Bartalini et al., in  the proceedings of the Workshop on Standard Model
physics (and more) at the LHC, CERN 2000-004, p.293
 
\bibitem{CDFglobal}
R.D. Field, presentations at the `Matrix Element and Monte Carlo Tuning
Workshop', Fermilab, 4 October 2002 and 29--30 April 2003, talks available
from webpage \texttt{http://cepa.fnal.gov/CPD/MCTuning/}, and further
recent talks available from 
\ttt{http://www.phys.ufl.edu/}$\sim$\ttt{rfield/cdf/}
 
\bibitem{D0global}
A. Kharchilava, presentation at the `Matrix Element and Monte Carlo Tuning
Workshop', Fermilab, 4 October 2002, talk available from webpage\\
\texttt{http://cepa.fnal.gov/CPD/MCTuning/}
 
\bibitem{ATLASglobal}
A. Moraes, C. Buttar, I. Dawson and P. Hodgson, ATLAS internal notes;
notes and talks available from webpage
\ttt{http://amoraes.home.cern.ch/amoraes/}

\bibitem{Butterworth:2002ts}
J.~M.~Butterworth and S.~Butterworth,
Comput.\ Phys.\ Commun.\  {\bf 153} (2003) 164;\\ 
see also \texttt{http://jetweb.hep.ucl.ac.uk}

\bibitem{Landdeb}
P.V. Landshoff, Nucl. Phys. Proc. Suppl. {\bf 99A } (2001) 311
 
\bibitem{DKMT}
Yu.L. Dokshitzer, V.A.Khoze, A.H. Mueller and S.I. Troyan,
`Basics of Perturbative QCD' (Editions Fronti\`eres, 1991)
 
\bibitem{backw}
T. Sj\"ostrand, Phys. Lett. {\bf 157B} (1985) 321; \\
M. Bengtsson, T. Sj\"ostrand and M. van Zijl, Z. Phys. {\bf C32}
(1986) 67
 
\bibitem{Huston}
C. Bal\'azs, J. Huston and I. Puljak, Phys. Rev. {\bf D63} (2001) 014021
 
\bibitem{Erik}
E. Thom\'e, master's thesis, Lund University, LU TP 04--01 [hep-ph/0401121];\\
J. Huston, I. Puljak, T. Sj\"ostrand, E. Thom\'e, LU TP 04--07 
[hep-ph/0401145]

\bibitem{EMC}
EMC Collaboration, M. Arneodo et al., Z. Physik {\bf C36} (1987) 527;\\
L. Apanasevich et al., Phys. Rev. {\bf D59} (1999) 074007
 
\bibitem{NCinfinite}
G. 't Hooft, Nucl. Phys. {\bf B72} (1974) 461
 
\bibitem{Jadestring}
B. Andersson, G. Gustafson and T. Sj\"ostrand,
Phys. Lett. {\bf B94} (1980) 211;\\
JADE Collaboration, W. Bartel et al., Phys. Lett. {\bf 101B} (1981) 129,
Z. Phys. {\bf C21} (1983) 37, Phys. Lett. {\bf 134B} (1984) 275;\\
TPC/2$\gamma$ Collaboration, H. Aihara et al., Z. Phys. {\bf C28}
(1985) 31;\\
TASSO Collaboration, M. Althoff et al., Z. Phys. {\bf C29} (1985) 29
 
\bibitem{mrv80}
L. Montanet, G.C. Rossi and G. Veneziano, Phys.~Rep.\
{\bf 63} (1980) 149

\bibitem{artrustring}
X. Artru, Nucl. Phys. {\bf B85} (1975) 442

\bibitem{earlyjunc}
M. Imachi, S. Otsuki and F. Toyoda, Progr. Theor. Phys. {\bf 54}
(1975) 280;\\
Y. Igarashi et al., Progr. Theor. Phys. Suppl. {\bf 63} (1978) 49;\\
G.C. Rossi and G. Veneziano, Nucl. Phys. {\bf B123} (1977) 507

\bibitem{confjunction}
G. {'}t Hooft, Physica Scripta {\bf 25} (1982) 133

\bibitem{BNVsmall}
P. Skands, in Proceedings of the 10th International Conference on
Supersymmetry and Unification of Fundamental Interactions, DESY, Hamburg,
17--23 Jun 2002, eds. P. Nath, P.M. Zerwas and C. Grosche (Hamburg, DESY, 
2002), vol. 2, p. 831 [hep-ph/0209199]

\bibitem{HansUno}
H.-U. Bengtsson, Computer Physics Commun. {\bf 31} (1984) 323
 
\bibitem{crosstalkexp}
N. van Remortel, hep-ex/0305014, to appear in the proceedings of 38th
Rencontres de Moriond on QCD and High-Energy Hadronic Interactions,
Les Arcs, Savoie, France, 22-29 Mar 2003;\\
A. Straessner, hep-ex/0304031, ibid.
 
\bibitem{ppbarBE}
UA1 Collaboration, C.~Albajar et al., Phys.~Lett. {\bf B226} (1989) 410;\\
E735 Collaboration, T.~Alexopoulos et al., Phys.~Rev. {\bf D48} (1993) 1931
 
\bibitem{lambda}
B. Andersson, G. Gustafson and B. S\"oderberg, Z. Phys. {\bf C20}
(1983) 317;\\
B. Andersson, S. Mohanty and F. S\"oderberg,
Nucl. Phys. {\bf B646} (2002) 102
 
\bibitem{lha1}
E.~Boos et al., in the proceedings of the Workshop on Physics at 
TeV Colliders, 21 May -- 1 June 2001, Les Houches, France,
eds. P. Aurenche et al. [hep-ph/0109068]

\bibitem{Omega}
WA97 Collaboration, F. Antinori et al., Nucl. Phys. {\bf A661} (1999) 130;\\
STAR Collaboration, C. Suire et al., Nucl. Phys. {\bf A715} (2003) 470;\\
NA57 Collaboration, F. Antinori et al., J. Phys. {\bf G30} (2004) S199

\bibitem{barexcess}
G. Belletini et al., Nuovo Cimento {\bf A42} (1977) 85;\\
B. Alper et al., Nucl. Phys. {\bf B100} (1975) 237;\\
A. Breakstone et al., Z. Physik {\bf C28} (1985) 335;\\
H1 Collaboration, paper 556, submitted to ICHEP98, Vancouver,
July 1998, and in preparation;\\
STAR Collaboration, C. Adler et al., Phys. Rev. Lett. {\bf 86} (2001) 
4778;\\
PHENIX Collaboration, K. Adcox et al., Phys. Rev. Lett. {\bf 89} (2002)
092302;\\
PHOBOS Collaboration, B.B. Back et al., Phys. Rev. {\bf C67} (2003) 
021901;\\
BRAHMS Collaboration, I.G. Bearden et al., Phys. Rev. Lett. {\bf 90} 
(2003) 102301
 
\bibitem{barexcessthy}
D. Kharzeev, Phys. Lett. {\bf B378} (1996) 238;\\
B. Kopeliovich and B. Povh, Z. Physik {\bf C75} (1997) 693;\\
S.E. Vance, M. Gyulassy and X.-N. Wang, Phys. Lett. {\bf B443} (1998) 45;\\
G.T. Garvey, B.Z. Kopeliovich, B. Povh, Comments Mod. Phys.
{\bf A2} (2001) 47

\bibitem{newshowers}
T. Sj\"ostrand, LU TP 04-05, submitted to the proceedings of the 
Workshop on Physics at TeV Colliders, Les Houches, France, 
26 May -- 6 June 2003 [hep-ph/0401061] 

\bibitem{ingelman}
R. Enberg, G. Ingelman and N. Timneanu,
Phys. Rev. {\bf D64} (2001) 114015
 
\end{thebibliography}
\end{document}